\documentclass[12pt]{article}

\usepackage{enumerate}
\usepackage{amsfonts}
\usepackage{amsmath}
\usepackage{amssymb}
\usepackage{amsthm}

\def\A{\mathfrak{A}}
\def\D{\mathcal{D}}
\def\H{\mathcal{H}}

\def\M{\mathbb{M}}

\def\S{\mathfrak{S}}
\def\C{\mathfrak{C}}
\def\N{\mathbb{N}}
\def\F{\mathfrak{F}}
\def\T{\mathfrak{T}}
\def\B{\mathfrak{B}}

\newcommand{\supp}{\mathrm{supp}}
\newcommand{\rank}{\mathrm{rank}}
\newcommand{\id}{\mathrm{Id}}
\newcommand{\Tr}{\mathrm{Tr}}

\newcommand{\shs}{\hspace{1pt}}

\newcounter{defin}  \newcounter{lemma}  \newcounter{theorem}
\newcounter{proposition} \newcounter{corol}  \newcounter{remark} \newcounter{example}

\newenvironment{lemma}{\par\refstepcounter{lemma}
     \textbf{Lemma \thelemma.} }{\rm\par}
\newenvironment{theorem}{\par\refstepcounter{theorem}
     \textbf{Theorem \thetheorem.}\ }{\rm\par}
\newenvironment{proposition}{\par\refstepcounter{proposition}
     \textbf{Proposition \theproposition.}\ }{\rm\par}
\newenvironment{corollary}{\par\refstepcounter{corol}
     \textbf{Corollary \thecorol.} }{\rm\par}
\newenvironment{definition}{\par\refstepcounter{defin}
     \textbf{Definition \thedefin.}\ }{\rm\par}
\newenvironment{remark}{\par\refstepcounter{remark}
     \textbf{Remark \theremark.}}{\rm\par}
\newenvironment{example}{\par\refstepcounter{example}
     \textbf{Example \theexample.}}{\rm\par}

\begin{document}

\title{Continuity of characteristics of composite quantum systems}
\date{}
\author{M.E. Shirokov\footnote{email:msh@mi.ras.ru}\\Steklov Mathematical Institute, Moscow, Russia}

\maketitle
\vspace{-20pt}

\begin{abstract}
General methods of quantitative and qualitative continuity analysis of characteristics of composite quantum systems are described.
Several modifications of the Alicki-Fannes-Winter method are considered, which make it applicable to a wide class of characteristics in both finite-dimensional and infinite-dimensional cases. A new approximation method for obtaining local continuity conditions for various characteristics of quantum systems is proposed and described in detail. This method allows us to prove several general results (Simon-type dominated convergence theorem, the theorem about preserving continuity under convex mixtures, etc.).

Uniform continuity bounds and local continuity conditions for basic characteristics of composite quantum systems are presented.
Along with the results obtained earlier by different authors, a number of new results proved by the proposed methods are described.
\end{abstract}

%
%

%
%
%
\tableofcontents

\section{Introduction}

When studying the information abilities of quantum systems, various  characteristics of states of these systems play an important role. Mathematically, these characteristics are functions of states or ensembles of states of these systems.  Naturally, the question arises about the continuity and uniform continuity of such characteristics. This question is especially relevant in the study of quantum systems of infinite dimension, the main characteristics of which have singular properties (discontinuity on the whole set of states, infinite or indefinite values, etc.) In this case, it is necessary to investigate questions about the correctness of definition of a given characteristic and its continuity or uniform continuity on certain subsets of quantum states.

In practice, two types of problems arise related to the continuity properties of a given characteristic $f$ of a quantum system:
\begin{enumerate}[A)]
  \item proof of the uniform continuity of a function $f$ on a certain subset $\S$ of states of this system with obtaining a possibly more accurate upper bound on the quantity
 \begin{equation}\label{GCB-def}
   \sup\left\{|f(\rho)-f(\sigma)|\,|\,\rho,\sigma\in\S,d(\rho,\sigma)\leq\varepsilon\right\}
 \end{equation}
   for any given $\varepsilon>0$, where $d$ is some metric on $\S$, that are traditionally called \emph{(uniform) continuity bounds};
   \item obtaining easily verifiable \emph{local continuity conditions}, i.e. conditions under which
   \begin{equation}\label{f-L}
    \lim_{n\to+\infty}f(\rho_n)=f(\rho_0)
   \end{equation}
   for a given sequence $\{\rho_n\}$ of states of this system
   converging to a state $\rho_0$.
\end{enumerate}

The importance of the first of the above problems  is confirmed by the (very non-complete) list of articles \cite{A&F}-\cite{RHA} published in the last two decades
which are entirely dedicated to getting uniform continuity  bounds for different characteristics of quantum systems and channels. Note that most of these articles are devoted to the study of characteristics of finite-dimensional systems and channels, the uniform continuity of which directly follows from their definitions. The main problem in this case is to obtain explicit upper bound on the quantity  in (\ref{GCB-def}) which is sharp or close-to-sharp in a particular sense.


The first decision of the above task A in the context of quantum systems should probably be considered the Fannes continuity bound for the von Neumann entropy obtained in \cite{Fannes}. This continuity bound and its optimized version obtained by Audenaert in \cite{Aud} are widely used in quantum information theory. This is explained by the basic role of the von Neumann entropy in description of quantum systems and their informational abilities. The second basic example is the Alicki-Fannes continuty bound for the quantum conditional entropy \cite{A&F}, which is refined recently by Winter \cite{W-CB} (and by Wilde for states of a special type \cite{Wilde-CB}). The Fannes-Audenaert and Alicki-Fannes-Winter  continuity bounds are extremely useful tools of analysis of quantum systems, they  allows us to get continuity bounds for many other characteristics of finite-dimensional quantum systems and channels, including basic correlation and entanglement measures in composite systems and basic capacities of quantum channels \cite{L&S,Nielsen,CHI,W-CB,Wilde,C&W,O&C}.

The significance of Alicki-Fannes' work also lies in the fact that the  method used in this work is quite universal and
allows to obtain uniform continuity bounds for a wide class of functions on the set of quantum states.  This method was analysed and improved by different authors \cite{SR&H,W-CB,M&H}, its optimal form is proposed by Winter in \cite{W-CB}. In Section 3.1 we  describe the Alicki-Fannes-Winter method in a general form (what hasn't been done anywhere before). It is our basic tool for quantitative continuity analysis of characteristics of composite finite-dimensional quantum systems.

At the same time, there are many other methods developed for getting continuity bounds for basic  characteristics  of finite-dimensional quantum systems: the quantum relative
entropy and its generalizations \cite{A&E-I,A&E-II,Hanson, Ras-q-re-I, Ras-q-re-II}, the Renyi and Tsallis entropy \cite{Aud,RCE,FYK,Ras-u-ent,Woods,Rag,Zhang}, the Renyi conditional entropy and its generalizations \cite{RCE,R-ce-II,R-ce-III,R-ce-I}, etc. A quite universal method of getting continuity bounds was proposed and used by Hanson and Datta in \cite{Hanson&Datta-1} to obtain many new results in this directions.

An important new step  was made by Winter in 2015, who obtained continuity bounds for the von Neumann entropy and the quantum conditional entropy in infinite-dimensional quantum systems under the energy-type constraint by using two-step approach consisting of finite-dimensional approximation of states with bounded "energy" followed by the Alicki-Fannes-Winter technique. This work stimulated research in this direction. As a result, a series of papers appeared where both new general methods of quantitative analysis of characteristics of infinite-dimensional quantum systems and the continuity bounds for specific characteristic were obtained \cite{Datta-2,CHI,AFM,MCB,CBM,B&D}. In particular, Becker, Datta and Jabbour constructed an optimal continuity bound for the von Neumann entropy under the energy constraint in one-mode quantum oscillator  that refines the Winter continuity bound in this specific case \cite{Datta-2}. The universal method of getting continuity bounds under the energy-type constraints based on initial purification of states followed by the Alicki-Fannes-Winter technique was proposed in \cite{AFM}.  In Section 3.2 we describe this method and the Winter two-step approach in a general form (which makes them  more universal) as well as other methods of getting continuity bounds for characteristics of infinite-dimensional composite quantum systems.

As for task B (local continuity conditions), the first results in this direction are the famous Simon convergence theorems for von Neumann entropy $S$ presented in
the Appendix in \cite{Ruskai} (where they are used in the proof of the basic result of quantum information theory!). These theorems give several conditions under which (\ref{f-L}) holds with $f=S$ for a  sequence $\{\rho_n\}$ of trace class operators (in particular, quantum states) converging to an operator $\rho_0$ (in different senses). The other important and widely used result of this kind is the claim about continuity of the von Neumann entropy
on the set of states with bounded energy provided that the Hamiltonian satisfies the Gibbs condition (condition (\ref{H-cond}) in Section 2) presented in \cite{W}. The above results are strengthened and generalized in \cite{SSP}, where a criterion of local continuity of the von Neumann entropy is proved.
Several local continuity conditions for specific characteristics of quantum systems and channels were obtained and used in \cite{ESP,EM,Kuz,Sh-H-ED,CMI}.

Since there is no way to describe in one review all the methods for solving the above tasks A and B and the obtained results, we have narrowed the field of study by considering only the characteristics of composite quantum systems, the definition of which is based on the von Neumann entropy. The main aim of this review is to describe
two general methods of solving these  tasks in finite and infinite dimensions: the Alicki-Fannes-Winter method in a general form (including its modifications and infinite-dimensional generalizations) and the special approximation technique based on the general result called Dini-type lemma (since the main step of its proof  resembles the proof of the classical Dini lemma).
In the last part of the review we consider concrete results obtained by these and other methods for basic characteristics of composite quantum systems.
Conditions of local continuity of these characteristics are always expressed via local continuity of one or several marginal entropies, since
the conditions of local continuity of the von Neumann entropy are well known \cite{Ruskai,W,SSP}. Along with the results obtained earlier by different authors,
a number of new results that have not been published before are presented.

\section{Preliminaries}

\subsection{Basic notation}

Let $\H$ be a finite-dimensional or separable infinite-dimensional Hilbert space. Denote by
$\B(\H)$ the algebra of all bounded operators on $\H$ with the operator norm $\|\cdot\|$ and by $\T(\H)$ the
Banach space of all trace-class operators on $\H$  with the trace norm $\|\!\cdot\!\|_1$. Let
$\S(\H)$  be the set of quantum states (operators from the positive cone $\T_{+}(\H)$ of $\T(\H)$
with unit trace) \cite{H-SCI,N&Ch,Wilde}.

Denote by $I_{\H}$ the unit operator on a Hilbert space
$\H$ and by $\id_{\H}$ the identity
transformation of the Banach space $\T(\H)$.

We will use Greek letters to denote operators in $\T(\H)$ and Latin letters for other operators, in particular, unbounded operators on a Hilbert
space $\H$. The \emph{support} $\mathrm{supp}\rho$ of an operator $\rho$ in  $\T_{+}(\H)$ is the closed subspace spanned by the eigenvectors of $\rho$ corresponding to its positive eigenvalues.  The dimension of $\mathrm{supp}\rho$ is called \emph{rank} of $\rho$ and denoted by $\rank\rho$.

If quantum systems $A_1,..,A_n$ are described, respectively, by Hilbert spaces  $\H_{A_1},..,\H_{A_n}$ then the composite system $A_1...A_n$ is described by the tensor product of these spaces, i.e. $\H_{A_1..A_n}\doteq\H_{A_1}\otimes...\otimes\H_{A_n}$. We will denote by $\rho_{A_k}$ the marginal state $\Tr_{A_1..A_n\setminus A_k}\rho$  of $\rho\in \S(\H_{A_1..A_n})$ corresponding to the
system $A_k$ \cite{H-SCI,Wilde}.

The \emph{fidelity} between quantum states $\rho$ and $\sigma$ in $\S(\H)$ can be defined as
\begin{equation*}
  F(\rho,\sigma)=\|\sqrt{\rho}\sqrt{\sigma}\|^2_1.
\end{equation*}
By Uhlmann theorem $F(\rho,\sigma)=\max_{\psi}|\langle\varphi|\psi\rangle|^2$, where $\varphi$ is a given purification of $\rho$ in $\S(\H\otimes\H_R)$, $\H_R$ is a separable Hilbert space, and the
maximum is over all purifications $\psi$ of $\sigma$ in $\S(\H\otimes\H_R)$ \cite{Uhl}. Thus,
\begin{equation}\label{F-Tn-eq}
\min_{\psi}\textstyle\frac{1}{2}\||\varphi\rangle\langle\varphi|-|\psi\rangle\langle\psi|\|_1=\sqrt{1-\max_{\psi}|\langle\varphi|\psi\rangle|^2}=\sqrt{1-F(\rho,\sigma)}.
\end{equation}
The following relations  between the fidelity and the  trace-norm distance hold (cf.\cite{F-ref})
\begin{equation}\label{F-Tn-ineq}
1-\sqrt{F(\rho,\sigma)}\leq\textstyle\frac{1}{2}\|\rho-\sigma\|_1\leq\sqrt{1-F(\rho,\sigma)}.
\end{equation}
We will use the trace norm distance and the fidelity as measures of closeness of quantum states (the trace norm is basic, but the fidelity is more convenient in some cases).

The \emph{von Neumann entropy} of a quantum state
$\rho \in \S(\H)$ is  defined by the formula
$S(\rho)=\Tr\eta(\rho)$, where  $\eta(x)=-x\ln x$ if $x>0$
and $\eta(0)=0$. The function $S(\rho)$ is lower semicontinuous on the set~$\S(\H)$ and takes values in~$[0,+\infty]$, it is concave and satisfies the inequality
\begin{equation}\label{S-LAA-2}
S(p\rho+(1-p)\sigma)\leq pS(\rho)+(1-p)S(\sigma)+h_2(p)
\end{equation}
valid for any states  $\rho$ and $\sigma$ in $\S(\H)$ and $p\in[0,1]$
with possible values $+\infty$ in both sides, where $\,h_2(p)=\eta(p)+\eta(1-p)\,$ is the binary entropy \cite{L-2,W,H-SCI,N&Ch,Wilde}.

The \emph{quantum relative entropy} for two quantum states $\rho$ and
$\sigma$ in $\S(\H)$ is defined as
$$
D(\rho\shs\|\shs\sigma)=\sum\langle
i|\,\rho\ln\rho-\rho\ln\sigma\,|i\rangle,
$$
where $\{|i\rangle\}$ is the orthonormal basis of
eigenvectors of the state $\rho$ and it is assumed that
$D(\rho\shs\|\sigma)=+\infty$ if $\,\mathrm{supp}\rho\shs$ is not
contained in $\shs\mathrm{supp}\shs\sigma$ \cite{L-2,H-SCI,Wilde}.

A finite or countable set  $\{\rho_{k}\}$ of quantum states
with a  probability distribution $\{p_{k}\}$ is called (discrete) \emph{ensemble} and denoted by $\{p_k,\rho_k\}$. The state $\bar{\rho}=\sum_{k} p_k\rho_k$ is called  the \emph{average state} of  $\{p_k,\rho_k\}$.  The Holevo quantity of an ensemble
$\{p_k,\rho_k\}$ defined as
\begin{equation*}
\chi(\{p_k,\rho_k\})= \sum_{k} p_k D(\rho_k\|\bar{\rho})=S(\bar{\rho})-\sum_{k} p_kS(\rho_k),
\end{equation*}
where the second formula holds provided that $S(\bar{\rho})$ is finite, gives a upper bound on the amount of classical information that can be obtained from  quantum measurements over the ensemble \cite{H-73}.

A \emph{quantum channel} $\Phi$ from a system $A$ to a system $B$ is a linear completely positive trace preserving map
from $\T(\H_A)$ to $\T(\H_B)$ \cite{Wilde,H-SCI}.

The \emph{constrained Holevo capacity} of a channel $\Phi$ at a state $\rho$
in $\S(\H_A)$ is defined as
\begin{equation}\label{CHI-QC-def}
\bar{C}(\Phi,\rho)=\sup_{\sum_k p_k\rho_k=\rho}\chi(\{p_k,\Phi(\rho_k)\}),
\end{equation}
where the supremum is over all ensembles  $\{p_k,\rho_k\}$ of states in $\S(\H_A)$ with the average state $\rho$.
This quantity is related to the classical (unassisted) capacity of a channel \cite{H-SCI,H-Sh-2}.

The \emph{mutual information} $I(\Phi,\rho)$ of a  quantum channel $\Phi:A\to B$ at a state $\rho$ in $\S(\H_A)$
can be defined as
\begin{equation}\label{MI-Ch-def}
I(\Phi,\rho)=I(B\!:\!R)_{\Phi\otimes\id_R(\bar{\rho})}\doteq D(\Phi\otimes\id_R(\bar{\rho})\|\Phi(\rho)\otimes\bar{\rho}_R),
\end{equation}
where $\bar{\rho}$ is pure state in $\S(\H_A\otimes\H_R)$ such that $\Tr_R\bar{\rho}=\rho$ and
$I(B\!:\!R)$ denotes the quantum mutual information of a state in $BR$ \cite{H-SCI,Wilde}.
This quantity is related to the classical entanglement-assisted capacity of a channel \cite{BSST,H-c-ch}.

\subsection{The set of quantum states with bounded values of energy-type functional}

Let $H$ be a positive (semi-definite)  operator on a Hilbert space $\mathcal{H}$ (we will always assume that positive operators are self-adjoint). Denote by $\mathcal{D}(H)$ the domain of $H$. For any positive operator $\rho\in\T(\H)$ we will define the quantity $\Tr H\rho$ by the rule
\begin{equation}\label{H-fun}
\Tr H\rho=
\left\{\begin{array}{l}
        \sup_n \Tr P_n H\rho\;\; \textrm{if}\;\;  \supp\rho\subseteq {\rm cl}(\mathcal{D}(H))\\
        +\infty\;\;\textrm{otherwise}
        \end{array}\right.
\end{equation}
where $P_n$ is the spectral projector of $H$ corresponding to the interval $[0,n]$ and ${\rm cl}(\mathcal{D}(H))$ is the closure of $\mathcal{D}(H)$. If
$H$ is a Hamiltonian (energy observable) of a quantum system described by the space $\H$ then
$\Tr H\rho$ is the mean energy of a state $\rho$.

For any positive operator $H$ the set
$$
\C_{H,E}=\left\{\rho\in\S(\H)\,|\,\Tr H\rho\leq E\right\}
$$
is convex and closed (since the function $\rho\mapsto\Tr H\rho$ is affine and lower semicontinuous). It is nonempty if $E> E_0$, where $E_0$ is the infimum of the spectrum of $H$.

The von Neumann entropy is continuous on the set $\C_{H,E}$ for any $E> E_0$ if and only if the operator $H$ satisfies  the \emph{Gibbs condition}
\begin{equation}\label{H-cond}
  \Tr\, e^{-\beta H}<+\infty\quad\textrm{for all}\;\,\beta>0
\end{equation}
and the supremum of the entropy on this set is attained at the \emph{Gibbs state}
\begin{equation}\label{Gibbs}
\gamma_H(E)\doteq e^{-\beta(E) H}/\Tr e^{-\beta(E) H},
\end{equation}
where the parameter $\beta(E)$ is determined by the equation $\Tr H e^{-\beta H}=E\Tr e^{-\beta H}$ \cite{W}. Condition (\ref{H-cond}) can be valid only if $H$ is an unbounded operator having  discrete spectrum of finite multiplicity. It means, in Dirac's notation, that
\begin{equation}\label{H-form}
H=\sum_{k=0}^{+\infty} E_k |\tau_k\rangle\langle\tau_k|,
\end{equation}
where
$\mathcal{T}\doteq\left\{\tau_k\right\}_{k=0}^{+\infty}$ is the orthonormal
system of eigenvectors of $H$ corresponding to the \emph{nondecreasing} unbounded sequence $\left\{E_k\right\}_{k=0}^{+\infty}$ of its eigenvalues
and \emph{it is assumed that the domain $\D(H)$ of $H$ lies within the closure $\H_\mathcal{T}$ of the linear span of $\mathcal{T}$}. In this case
\begin{equation*}
\Tr H \rho=\sum_i \lambda_i\|\sqrt{H}\varphi_i\|^2
\end{equation*}
for any operator $\rho$ in $\T_+(\H)$ with the spectral decomposition $\rho=\sum_i \lambda_i|\varphi_i\rangle\langle\varphi_i|$ provided that
all the vectors $\varphi_i$ lie in $\D(\sqrt{\H})=\{ \varphi\in\H_\mathcal{T}\,| \sum_{k=0}^{+\infty} E_k |\langle\tau_k|\varphi\rangle|^2<+\infty\}$.


We will use the function
\begin{equation}\label{F-def}
F_{H}(E)\doteq\sup_{\rho\in\C_{H,E}}S(\rho)=S(\gamma_H(E)).
\end{equation}
This is a strictly increasing concave function on $[E_0,+\infty)$ \cite{W-CB,EC}. It is easy to see that $F_{H}(E_0)=\ln m(E_0)$, where $m(E_0)$ is the multiplicity of $E_0$. By Proposition 1 in \cite{EC} the Gibbs condition (\ref{H-cond}) is equivalent to the following property
\begin{equation}\label{H-cond-a}
  F_{H}(E)=o\shs(E)\quad\textrm{as}\quad E\rightarrow+\infty
\end{equation}
that means a restriction on the growth of the function $F_H$ with increasing $E$.

We will often need a stronger restriction on the growth of the function $F_H$. For this purpose we will strengthen the Gibbs condition (\ref{H-cond}) as follows\footnote{Condition (\ref{H-cond}) means that $\lim_{k\rightarrow\infty}E_k/\ln k=+\infty$, where $\{E_k\}$ is the sequence  of eigenvalues of $H$, condition (\ref{H-cond+}) holds  if $\;\liminf_{k\rightarrow\infty} E_k/\ln^p k>0\,$ for some $\,p>2$ \cite[Proposition 1]{AFM}.}
\begin{equation}\label{H-cond+}
  \lim_{\beta\rightarrow0^+}\left[\Tr\, e^{-\beta H}\right]^{\beta}=1.
\end{equation}
By Lemma 1 in \cite{AFM} condition (\ref{H-cond+}) is valid for a positive operator $H$ if and only if
\begin{equation}\label{H-cond+a}
  F_{H}(E)=o\shs(\sqrt{E})\quad\textrm{as}\quad E\rightarrow+\infty.
\end{equation}

It is essential that condition (\ref{H-cond+}) holds for the Hamiltonian of a multi-mode quantum
oscillator (described in Section 3.2.4) which is a basic model in "continuous variable" quantum information theory, since in this case
\begin{equation}\label{B-D-cond-a}
  F_{H}(E)=O\shs(\ln E)\quad\textrm{as}\quad E\rightarrow+\infty.
\end{equation}

Becker and Datta proved recently that asymptotic  property (\ref{B-D-cond-a}) holds provided that there exists
\begin{equation}\label{B-D-cond}
\lim_{E\rightarrow+\infty}N_{\shs\uparrow}[H](E)/N_{\downarrow}[H](E)=a>1,
\end{equation}
where
$$
\!N_{\shs\uparrow}[H](E)\doteq \sum_{k,j: E_k+E_j\leq E} E_k^2\quad \textrm{and}\quad N_{\downarrow}[H_A](E)\doteq \sum_{k,j: E_k+E_j\leq E} E_kE_j
$$
for any $E>E_0$. Namely, if condition (\ref{B-D-cond}) holds then Theorem 3 in \cite{B&D} states that
$$
F_{H}(E)=(a-1)^{-1} (\ln E)(1+o(1))\quad\textrm{as}\quad E\rightarrow+\infty.
$$
Thus, condition (\ref{B-D-cond}) is stronger than  condition (\ref{H-cond+}). In \cite{B&D}
the validity of condition (\ref{B-D-cond}) for Hamiltonians of different quantum systems is shown.

We will often assume that
\begin{equation}\label{star}
  E_0\doteq\inf\limits_{\|\varphi\|=1}\langle\varphi|H|\varphi\rangle=0.
\end{equation}
In this case the concavity and nonnegativity of $F_H$ imply that (cf.\cite[Corollary 12]{W-CB})
\begin{equation}\label{W-L}
  xF_H(E/x)\leq yF_H(E/y)\quad  \forall y>x>0.
\end{equation}

\section{General methods of quantitative continuity\\ analysis}

\subsection{The Alicki-Fannes-Winter (AFW) method for bounded functions on $\Delta$-invariant subsets of $\S(\H)$ and their images under affine maps}

\subsubsection{General results.} In this subsection we describe in a general form an universal method of constructing  uniform continuity bounds
for a function $f$ on a convex subset $\S_0$ of $\S(\H)$ satisfying the inequalities
\begin{equation}\label{LAA-1}
  f(p\rho+(1-p)\sigma)\geq pf(\rho)+(1-p)f(\sigma)-a_f(p)
\end{equation}
and
\begin{equation}\label{LAA-2}
  f(p\rho+(1-p)\sigma)\leq pf(\rho)+(1-p)f(\sigma)+b_f(p),
\end{equation}
for all states $\rho$ and $\sigma$ in $\S_0$ and any $p\in[0,1]$, where $a_f(p)$ and $b_f(p)$ are vanishing functions as $p\rightarrow+0$ (depending of $f$).
These  inequalities can be treated, respectively, as weakened forms of concavity and convexity. We will call functions
satisfying both inequalities (\ref{LAA-1}) and (\ref{LAA-2}) \emph{locally almost affine} (briefly, \emph{LAA functions}), since for any such function $f$ the quantity
$\,|f(p\rho+(1-p)\sigma)-p f(\rho)-(1-p)f(\sigma)|\,$ tends to zero as $\,p\rightarrow 0^+$ uniformly on $\,\S_0\times\S_0$.  For technical simplicity we will assume that
\begin{equation}\label{a-b-assump}
 \textrm{ the functions}\;\; a_f\;\; \textrm{and}\;\; b_f\;\; \textrm{are non-decreasing on}\;\; \textstyle[0,\frac{1}{2}].
\end{equation}

The method we what to describe was originally used by Alicki and Fannes to obtain a continuity bound for the quantum conditional entropy
in a finite-dimensional bipartite system \cite{A&F}. Then this method was analysed and improved by different authors \cite{SR&H,M&H,W-CB}. We will describe and use its optimal form proposed by Winter, who applied it to get \emph{tight} (sharp, in a sense) continuity bounds for the quantum conditional entropy and for the bipartite relative entropy of entanglement \cite{W-CB}.

For arbitrary different states $\rho$ and $\sigma$ in $\S(\H)$ introduce the states
\begin{equation}\label{delta-oper}
\Delta^-(\rho,\sigma)=\varepsilon^{-1}[\shs\rho-\sigma\shs]_- \quad \textrm{and} \quad \Delta^+(\rho,\sigma)=\varepsilon^{-1}[\shs\rho-\sigma\shs]_+,
\end{equation}
where $\varepsilon=\frac{1}{2}\|\shs\rho-\sigma\shs\|_1$ and $[A]_-$ and $[A]_+$ denote, respectively, the negative and positive parts
of a Hermitian operator $A$. We will need the following\smallskip
\begin{definition}\label{S-inv} A subset $\S_0$ of $\S(\H)$ is called $\Delta$-\emph{invariant} if
\begin{equation}\label{Inv-P}
\Delta^{\pm}(\rho,\sigma)\in\S_0\quad \forall \rho,\sigma\in\S_0\;(\rho\neq\sigma).
\end{equation}
\end{definition}
The simplest example of $\Delta$-invariant set is $\S(\H)$.  Other basic examples are:
\begin{itemize}
  \item the convex hull of a finite or countable set of mutually orthogonal states in $\S(\H)$;
  \item $\{\rho\in\S(\H)\,|\,\sum_kP_k\rho P_k=\rho\}$, where $\{P_k\}$ is a set of mutually orthogonal projectors;
  \item the sets of quantum-classical and classical-quantum states in $\S(\H_{AB})$.
\end{itemize}

\begin{theorem}\label{AFW-1} \emph{Let $\S_0$ be a $\Delta$-invariant convex subset of $\S(\H)$. A function $f$ on $\S_0$ satisfying inequalities (\ref{LAA-1}) and (\ref{LAA-2})
is uniformly continuous on $\S_0$ if and only if
\begin{equation*}
C^\bot_f=\sup_{\rho,\sigma\in\S_0, \Tr\rho\sigma=0}|f(\rho)-f(\sigma)|<+\infty.
\end{equation*}
If this condition holds then
\begin{equation}\label{AFW-1+}
C^\bot_f\varepsilon-D_f(\varepsilon)\leq\sup_{\rho,\sigma\in\S_0,\frac{1}{2}\|\shs\rho-\sigma\shs\|_1\leq\varepsilon}|f(\rho)-f(\sigma)|\leq C^\bot_f\varepsilon+D_f(\varepsilon)
\end{equation}
for any $\varepsilon\in(0,1]$, where $\,D_f(\varepsilon)=\displaystyle(1+\varepsilon)(a_f+b_f)\!\left(\frac{\varepsilon}{1+\varepsilon}\right)$. The inequalities in (\ref{AFW-1+}) remain valid with the supremum taken over all $\rho,\sigma\in\S_0$ such that $\frac{1}{2}\|\shs\rho-\sigma\shs\|_1=\varepsilon$.}
\end{theorem}\smallskip

\emph{Proof.} Assume that $C^\bot_f<+\infty$. Let $\varepsilon\in(0,1]$ be arbitrary. Let $\rho$ and $\sigma$ be states in $\S_0$ such  that $\frac{1}{2}\|\shs\rho-\sigma\shs\|_1=\varepsilon$. By the assumption the states $\tau_-=\Delta^{-}(\rho,\sigma)$ and $\tau_+=\Delta^{+}(\rho,\sigma)$ defined in (\ref{delta-oper}) lie in $\S_0$. So, we have
\begin{equation}\label{omega-star}
\frac{1}{1+\varepsilon}\,\rho+\frac{\varepsilon}{1+\varepsilon}\,\tau_-=\omega_*=
\frac{1}{1+\varepsilon}\,\sigma+\frac{\varepsilon}{1+\varepsilon}\,\tau_+,
\end{equation}
where $\omega_*$ is a state in $\S_0$. By applying inequalities (\ref{LAA-1}) and (\ref{LAA-2}) to these decompositions of $\omega_*$  we get
\begin{eqnarray}\label{1-b-ineq}
(1-p)(f(\rho)-f(\sigma))&\leq& p
(f(\tau_+)-f(\tau_-))+a_f(p)+b_f(p),\\
\label{2-b-ineq} (1-p)(f(\sigma)-f(\rho))&\leq& p
(f(\tau_-)-f(\tau_+))+a_f(p)+b_f(p),
\end{eqnarray}
where $p=\varepsilon/(1+\varepsilon)$. By combining these two inequalities we obtain
\begin{equation}\label{A-ineq}
p A-a_f(p)-b_f(p)\leq (1-p)|f(\rho)-f(\sigma)|\leq p A+a_f(p)+b_f(p),
\end{equation}
where $A=|f(\tau_+)-f(\tau_-)|$. This implies the right inequality in (\ref{AFW-1+}), since the r.h.s. of (\ref{AFW-1+}) is a non-decreasing function of $\varepsilon$ due to the assumption (\ref{a-b-assump}).

To prove the left inequality in (\ref{AFW-1+}) it suffices, by the first inequality in (\ref{A-ineq}), to show that for arbitrary
orthogonal states $\tau_+$ and $\tau_-$ in $\S_0$ and arbitrary $\varepsilon\in(0,1]$ there exist states $\rho$ and $\sigma$ in $\S_0$
such that $\tau_-=\Delta^{-}(\rho,\sigma)$, $\tau_+=\Delta^{+}(\rho,\sigma)$ and $\frac{1}{2}\|\shs\rho-\sigma\shs\|_1=\varepsilon$.
It is easy that see that the states
\begin{equation}\label{U-states}
  \rho=\textstyle \frac{1}{2}(1+\varepsilon)\tau_++\frac{1}{2}(1-\varepsilon)\tau_-\quad\textrm{and}\quad\sigma=\frac{1}{2}(1-\varepsilon)\tau_++\frac{1}{2}(1+\varepsilon)\tau_-
\end{equation}
have the required properties.

Assume that $C^\bot_f=+\infty$. Then for each natural $n$ there exists orthogonal states $\tau_+$ and $\tau_-$ in $\S_0$
such that $|f(\tau_+)-f(\tau_-)|\geq n$. By taking the states  $\rho$ and $\sigma$ defined in (\ref{U-states}) with some $\varepsilon>0$ and using the first inequality in (\ref{A-ineq}) it is easy to show that the function $f$ is not uniformly continuous on $\S_0$. $\square$\smallskip

Theorem \ref{AFW-1} gives universal and quite accurate continuity bound for a wide class of functions. The accuracy is illustrated by the following\smallskip

\begin{example} Let $\S_0=\S(\H)$, $d\doteq\dim\H<+\infty$, and $f(\rho)=S(\rho)$ be the von Neumann entropy of a state $\rho$. The concavity of $S$ and inequality (\ref{S-LAA-2})
show that this function satisfies inequalities (\ref{LAA-1}) and (\ref{LAA-2}) with $a_f=0$ and $b_f=h_2$. It is clear that $C^\bot_S=\ln(d-1)$. Thus, Theorem \ref{AFW-1}
implies that
\begin{equation}\label{AFW-1+S}
|S(\rho)-S(\sigma)|\leq \varepsilon\ln(d-1)+g(\varepsilon)
\end{equation}
for any states $\rho$ and $\sigma$ in $\S(\H)$ such that $\frac{1}{2}\|\shs\rho-\sigma\shs\|_1\leq\varepsilon$, where
\begin{equation}\label{g-def}
 g(x)=(x+1)h_2(x/(x+1))=(x+1)\ln(x+1)-x\ln x.
\end{equation}
Continuity bound (\ref{AFW-1+S}) is close to the optimal continuity bound for the entropy obtained by Audenaert in \cite{Aud}. The only difference
is the appearance of the term $g(\varepsilon)$ instead of $h_2(\varepsilon)$, which is not too essential for small $\varepsilon$, since $h_2(\varepsilon)\sim g(\varepsilon)$ as $\varepsilon\to0^+$.
\end{example}\smallskip

\begin{corollary}\label{AFW-1-c+}
\emph{If $f$ is a finite affine function on $\Delta$-invariant convex set $\S_0\subseteq\S(\H)$ then}
\begin{equation}\label{AFW-1-c++}
\sup_{\rho,\sigma\in\S_0,\frac{1}{2}\|\shs\rho-\sigma\shs\|_1=\varepsilon}|f(\rho)-f(\sigma)|=C^\bot_f\varepsilon\qquad\forall\varepsilon\in(0,1].
\end{equation}
\end{corollary}

The claim of Corollary \ref{AFW-1-c+} is valid for any affine map $f$ from $\S_0$ into a normed vector space $X$. One should only replace
$|\cdot|$ by $\|\cdot\|_X$ in (\ref{AFW-1-c++}) and in the definition of  $C^\bot_f$.\smallskip

\begin{example} Let $\S_0=\S(\H)$ and $f(\rho)=\Tr A\rho$, where $A$ is a bounded Hermitian operator on $\H$ with the spectrum $\mathrm{Sp}(A)$. Corollary \ref{AFW-1-c+} implies that
\begin{equation}\label{AFW-1+S}
\sup_{\rho,\sigma\in\S(\H),\frac{1}{2}\|\shs\rho-\sigma\shs\|_1=\varepsilon}|\Tr A(\rho-\sigma)|=\varepsilon(\sup \mathrm{Sp}(A)-\inf \mathrm{Sp}(A)).
\end{equation}
\end{example}

The continuity bound given by Theorem  \ref{AFW-1} can be refined by using the following\smallskip
\begin{corollary}\label{AFW-1-c} \emph{Let the conditions of Theorem  \ref{AFW-1} hold and $\,\{\S_\lambda\}_{\lambda\in\Lambda}$  a family of convex subsets of the set $\S_0$ with the following property
\begin{equation}\label{G-Delta-P}
\textrm{if }\rho,\sigma\in\S_\lambda, \lambda\in\Lambda,\textrm{ then }\Delta^{\pm}(\rho,\sigma)\in \S_{\lambda'}\textrm{ for some }\lambda'\in\Lambda
\end{equation}
Assume that $g$ is a  function on $\S_0$ satisfying inequalities (\ref{LAA-1}) and (\ref{LAA-2})
with the functions $a_g$ and $b_g$ such that $\,a_g+b_g<a_f+b_f$ and
$\,f(\rho)-f(\sigma)=g(\rho)-g(\sigma)\,$ provided that $\rho$ and $\sigma$ lie in the same subset of the family $\{\S_\lambda\}$. Then
\begin{equation*}
\sup_{\rho,\sigma\in\S_\lambda,\frac{1}{2}\|\shs\rho-\sigma\shs\|_1\leq\varepsilon}|f(\rho)-f(\sigma)|\leq C^\bot_f\varepsilon+D_g(\varepsilon)
\end{equation*}
for any $\lambda\in\Lambda$ and $\varepsilon\in(0,1]$, where $\,D_g(\varepsilon)=\displaystyle(1+\varepsilon)(a_g+b_g)\!\left(\frac{\varepsilon}{1+\varepsilon}\right)$.}
\end{corollary}

\emph{Proof.} Note first that (\ref{1-b-ineq}),  (\ref{2-b-ineq}) and (\ref{A-ineq}) hold with $f$ replaced by $g$. Then we may replace
$|g(\rho)-g(\sigma)|$ and $|g(\tau_+)-g(\tau_-)|$ with $|f(\rho)-f(\sigma)|$ and $|f(\tau_+)-f(\tau_-)|$ correspondingly. $\Box$\smallskip

A typical example of the family possessing property (\ref{G-Delta-P}) is the following: $$
\S_0=\S(\H_{AB}),\; \Lambda=\S(\H_B),\; \S_{\lambda}=\{\omega\in \S(\H_{AB})\,|\,\omega_B=\lambda\}.
$$

Corollary \ref{AFW-1-c} is used in the proof of Lemma 2 in \cite{UFA} (implicitly) and in the proof of Proposition \ref{QD-CB-1} in Section 5.3.1 below.\smallskip

We also present a strengthened condition of uniform continuity of LAA functions.\smallskip

\begin{theorem}\label{AFW-2} \emph{Let $\S_0$ be a convex subset of $\S(\H)$ s.t. $\S_0=\Lambda(\S'_0)=\Lambda(\S'_*)$, where
\begin{itemize}
  \item $\S'_0$ is a convex subset of $\S(\H')$;
  \item $\S'_*$ is a subset of $\S'_0$ such that $\Delta^{\pm}(\rho',\sigma')\in\S'_0$ for all $\rho',\sigma'\in\S'_*$, $(\rho'\neq\sigma')$;
  \item $\Lambda$ is an affine map from $\S'_0$ onto $\S_0$ such that the inverse (multi-valied) map $\Lambda^{-1}$
  is selective uniform continuous in the following sense:
$$
r(\varepsilon)\doteq\sup_{\rho,\sigma\in\S_0,\,\|\rho-\sigma\|_1\leq2\varepsilon}\inf\left\{\textstyle{\frac{1}{2}}\|\rho'-\sigma'\|_1\,\left|\,\rho'\in \mathfrak{L}(\rho),\, \sigma'\in \mathfrak{L}(\sigma)\right.\right\}=o(1)\quad \textrm{as}\quad \varepsilon\to0,
$$
where $\mathfrak{L}(\rho)=\Lambda^{-1}(\rho)\cap\S'_*$ and $\mathfrak{L}(\sigma)=\Lambda^{-1}(\sigma)\cap\S'_*$.
\end{itemize}}
\emph{Let $f$ be a function on the set $\S_0$ satisfying inequalities (\ref{LAA-1}) and (\ref{LAA-2})
such that $\,C_f=\sup_{\varrho,\varsigma\in\S_0}|f(\varrho)-f(\varsigma)|<+\infty$.
Then $f$ is uniformly continuous on $\S_0$ and
\begin{equation}\label{AFW-2+}
|f(\rho)-f(\sigma)|\leq C_fr(\varepsilon)+(1+r(\varepsilon))(a_f+b_f)\!\left(\frac{r(\varepsilon)}{1+r(\varepsilon)}\right)
\end{equation}
for any states $\rho$ and $\sigma$ in $\S_0$ such that $\frac{1}{2}\|\shs\rho-\sigma\shs\|_1\leq\varepsilon$.}

\emph{If the set $\S_0'$ is compact then the claim about the uniform continuity of the function $f$ on $\S_0$ holds
without the condition of selective uniform continuity of the map $\Lambda^{-1}$.}
\end{theorem}\smallskip

\emph{Proof.} Assume that $\rho$ and $\sigma$ are states in $\S_0$ such that $\frac{1}{2}\|\shs\rho-\sigma\shs\|_1\leq\varepsilon$. For given arbitrary $\delta>0$ one can choose states $\rho'$ and $\sigma'$ in $\S'_*$ such that $\varepsilon'\doteq\frac{1}{2}\|\shs\rho'-\sigma'\shs\|_1\leq r(\varepsilon)+\delta$ and $\rho=\Lambda(\rho')$, $\sigma=\Lambda(\sigma')$.
Then relation (\ref{omega-star}) holds with  $\rho$, $\sigma$, $\tau_\pm$ and $\varepsilon$ replaced by $\rho'$, $\sigma'$
$\tau'_\pm=\Delta^{\pm}(\rho',\sigma')$ and $\varepsilon'$ correspondingly. It implies
\begin{equation}\label{omega-star+}
\frac{1}{1+\varepsilon'}\,\rho+\frac{\varepsilon'}{1+\varepsilon'}\,\Lambda(\tau'_-)=\Lambda(\omega_*)=
\frac{1}{1+\varepsilon'}\,\sigma+\frac{\varepsilon'}{1+\varepsilon'}\,\Lambda(\tau'_+).
\end{equation}
By applying inequalities (\ref{LAA-1}) and (\ref{LAA-2}) to both decompositions of $\Lambda(\omega_*)$ (as in the proof of Theorem \ref{AFW-1}) we conclude
that inequality (\ref{AFW-2+}) holds with $r(\varepsilon)$ replaced by $\varepsilon'$. Since the function $\,x\mapsto (1+x)(a_f+b_f)(x/(1+x))\,$ is non-decreasing on $[0,1]$ due to the assumption (\ref{a-b-assump}), simple continuity arguments complete the proof.

To prove the last assertion it suffices to show that the function $f$ is continuous on $\S_0$, since this set is compact by  the continuity of $\Lambda$ (which follows from its affinity by the
remark after Corollary \ref{AFW-1-c+}). This can be done easily by using  the compactness of $\S'_0$ and the continuity of $\Lambda$, since Theorem \ref{AFW-1} implies that the LAA function $f\circ\Lambda$ is continuous on $\S'_0$. $\square$\smallskip

\begin{example}\label{H-set} Let $H$ be a positive densely defined operator on $\H$ and
$\S_0$  the convex set of all states $\rho$ in $\S(\H)$ such that $\Tr H\rho$ defined in (\ref{H-fun}) is finite.
Theorem \ref{AFW-2} allows us to show that any LAA bounded function on $\S_0$ is uniformly
continuous on $\S_0$ and hence it has a unique  uniformly
continuous extension to the set $\S(\H)$. To show this assume that $\H'=\H\otimes \H_R$, where $\H_R$ is a separable Hilbert space,
$\S_0'$ is the set of all states $\rho'$ in $\S(\H')$ such that $\,\Tr (H\otimes I_{R})\rho'<+\infty\,$ and $\S_*'$ is the subset of $\S_0'$
consisting of pure states. Then all the conditions in Theorem \ref{AFW-2} are satisfied with the affine map $\Lambda(\rho')=\Tr_R\rho'$ from $\S_0'$ onto $\S_0$.
Indeed, the observation from the proof of Theorem 1 in \cite{AFM} shows that $\Delta^{\pm}(\rho',\sigma')\in\S'_0$ for any different states  $\rho'$ and $\sigma'$ in $\S'_*$,
while equality (\ref{F-Tn-eq}) and  inequality (\ref{F-Tn-ineq}) imply that $r(\varepsilon)\leq\sqrt{\varepsilon(2-\varepsilon})$.
\end{example}\smallskip

Theorem \ref{AFW-2} allows to prove the following\smallskip
\begin{corollary}\label{AFW-2-c} \emph{If a LAA function $f$ takes finite values
at a finite set of states then it is uniformly continuous on the convex hull of these states.}
\end{corollary}\smallskip

\emph{Proof.} Note first that any LAA function $f$ taking finite values at states $\rho_1$,...,$\rho_n$ is bounded on the set $\,\S_0=\mathrm{conv}\{\rho_1,...,\rho_n\}$. Indeed,
by using inequalities (\ref{LAA-1}) and (\ref{LAA-2}) sequentially it is easy to show that
$$
\min_i f(\rho_i) -(n-1)\max_{x\in[0,1]}a_f(x)\leq f(\rho) \leq\max_i f(\rho_i)+(n-1)\max_{x\in[0,1]}b_f(x)\quad\; \forall \rho\in\S_0.
$$
Let $\H'=\H\otimes \H_R$, where $\H_R$ is a $n$-dimensional Hilbert space,
$\S_0'$  the convex hull of the states $\rho_1\otimes|1\rangle\langle1|,...,\rho_n\otimes|n\rangle\langle n|$, where
$\{|k\rangle\}_{k=1}^n$ is a basis in $\H_R$, $\S_*'=\S_0'$ and $\Lambda(\cdot)=\Tr_{R}(\cdot)$  is an affine map from $\S_0'$ onto $\S_0$.
Since the convex set $\S'_0$ is $\Delta$-invariant and compact, Theorem \ref{AFW-2} implies that $f$ is uniformly continuous on $\S_0$. $\square$

\subsubsection{Application to characteristics of composite quantum systems.} Among characteristics of a $n$-partite
finite-dimensional quantum system $A_{1}...A_{n}$ there are many that satisfy inequalities (\ref{LAA-1}) and (\ref{LAA-2})
with the functions $a_f$ and $b_f$ proportional to the binary entropy (defined after (\ref{S-LAA-2})) and the inequality
\begin{equation}\label{Cm}
-c^-_f C_m(\rho)\leq f(\rho)\leq c^+_f C_m(\rho),\quad C_m(\rho)=\sum_{k=1}^m S(\rho_{A_k}),\;\; m\leq n,
\end{equation}
for any state $\rho$ in $\S(\H_{A_{1}..A_{n}})$, where $c^-_f$ and $c^+_f$ are nonnegative numbers.

Following \cite{CBM} introduce the class $L_n^m(C,D)$ of functions on $\S(\H_{A_{1}..A_{n}})$
satisfying inequalities  (\ref{LAA-1}) and (\ref{LAA-2})
with $a_f(p)=d_f^-h_2(p)$ and $b_f=d_f^+h_2(p)$ and inequality (\ref{Cm}) with the parameters $c^{\pm}_f$ and  $d^{\pm}_f$ such that $c^-_f+c^+_f=C$ and $d^-_f+d^+_f=D$.
We also consider the class $\widehat{L}^{m}_n(C,D)$ obtained by adding to the class $L^{m}_n(C,D)$ all functions of the form
$$
f(\rho)=\inf_{\lambda}f_{\lambda}(\rho)\quad \textrm{and} \quad f(\rho)=\sup_{\lambda}f_{\lambda}(\rho),
$$
where $\{f_{\lambda}\}$ is some family of functions in $L^{m}_n(C,D)$.

To generalize the above notation assume that $\S_0$ is a convex subset of $\S(\H_{A_{1}..A_{n}})$. Denote by $L_n^m(C,D|\S_0)$
the class of functions on the set $\S_0$ that satisfy \emph{on this set} inequalities  (\ref{LAA-1}) and (\ref{LAA-2})
with $a_f(p)=d_f^-h_2(p)$ and $b_f=d_f^+h_2(p)$ and inequality (\ref{Cm}) with the parameters $c^{\pm}_f$ and  $d^{\pm}_f$ such that $c^-_f+c^+_f=C$ and $d^-_f+d^+_f=D$.
The class $\widehat{L}^{m}_n(C,D|\S_0)$ is defined in the same way as $\widehat{L}^{m}_n(C,D)$.

Assume  now that $f$ is an arbitrary  function satisfying inequality (\ref{Cm}) on some set $\S_0\subseteq\S(\H_{A_{1}...A_{n}})$. If $A_1$,...,$A_m$ are finite-dimensional systems then
$$
-c^-_f \sum_{k=1}^m \ln \dim\H_{A_k}\leq f(\rho)\leq c^+_f \sum_{k=1}^m \ln \dim\H_{A_k}
$$
for any state $\rho$ in $\S_0$. So, by applying Theorem \ref{AFW-1} in Section 3.1.1 and by noting that this theorem gives one and the same continuity bound for any function from a given class $L_n^m(C,D|\S_0)$ one can  strengthen the observation in Section III-A in \cite{CBM} as follows.\smallskip

\begin{theorem}\label{CB-L} \emph{Let $\S_0$ be a convex subset of $\S(\H_{A_{1}..A_{n}})$ possessing the $\Delta$-invariance property (\ref{Inv-P}) and
$f$ a function on $\S_0$ belonging to the class $\widehat{L}^{m}_{n}(C,D|\S_0)$. If the  subsystems $A_1$,...,$A_m$ are finite-dimensional then
\begin{equation*}
 |f(\rho)-f(\sigma)|\leq C\varepsilon \sum_{k=1}^m \ln \dim\H_{A_k}+Dg(\varepsilon)
\end{equation*}
for any states $\rho$ and $\sigma$ in $\S(\H_{A_1...A_n})$ such that $\;\frac{1}{2}\|\shs\rho-\sigma\|_1\leq\varepsilon$, where $g(x)$
is the function defined in (\ref{g-def}).}
\end{theorem}\smallskip

We will see in Section 5 that Theorem \ref{CB-L} gives quite accurate continuity bounds for many important characteristics of multipartite quantum
states.

Following \cite{CBM} introduce the class $N^{m}_{n,1}(C,D)$ of functions
$f$ on the set $\S(\H_{A_1..A_n})$ constructed via some function $h$ on the set $\S(\H_{A_1..A_nA_{n+1}..A_{n+l}})$
belonging to the class $\widehat{L}^{m}_{n+l}(C,D)$ (where $l>0$) by one of the expressions
\begin{equation}\label{N-class-f}
f(\rho)=\inf_{\hat{\rho}\in \mathfrak{M}_1(\rho)}h(\hat{\rho}),\qquad
f(\rho)=\sup_{\hat{\rho}\in \mathfrak{M}_1(\rho)}h(\hat{\rho}),
\end{equation}
where $\,\mathfrak{M}_1(\rho)=\{\hat{\rho}\in\mathfrak{S}(\mathcal{H}_{A_{1}..A_{n+l}})\,|\,\hat{\rho}_{A_{1}..A_{n}}=\rho\,\}\,$ is the set of all extensions of $\rho$
  to a state of $\S(\H_{A_1..A_{n+l}})$.

Introduce also the classes $N^{m}_{n,s}(C,D)$, $s=2,3$ of functions
$f$ on the set $\S(\H_{A_1..A_n})$ constructed via some function $h$ on $\S(\H_{A_1..A_{n+1}})$
belonging to the class $\widehat{L}^{m}_{n+1}(C,D|\S_{\rm qc})$ by one of the expressions in (\ref{N-class-f}) with
$\mathfrak{M}_1(\rho)$ replaced by $\mathfrak{M}_s(\rho)=\mathfrak{M}_1(\rho)\cap\S^s_{\rm qc}$, where
\begin{itemize}
  \item $\S^2_{\rm qc}=\S_{\rm qc}$ -- the convex subset of $\S(\H_{A_1..A_{n+1}})$ consisting of all states of the form
  \begin{equation}\label{m-2}
       \hat{\rho}=\sum_k p_k \rho_k\otimes |k\rangle\langle k|,
  \end{equation}
  where $\{p_k\}$ is a probability distribution, $\{\rho_k\}$ is a finite or countable set of states in $\S(\H_{A_{1}..A_n})$ and $\{|k\rangle\}$ is a basis in $\H_{A_{n+1}}$
  \item $\S^3_{\rm qc}$ is the subset of $\S(\H_{A_1..A_{n+1}})$ consisting of all states of the form
(\ref{m-2}) in which all the states $\rho_k$ are pure.
\end{itemize}

By using Theorem \ref{CB-L} and the arguments from the proof of Proposition 3 in \cite{CBM} one can obtain the following  continuity bound for functions in $N^{m}_{n,s}(C,D)$.
\smallskip

\begin{corollary}\label{CB-N} \emph{Let $f$ be a function on the set $\S(\H_{A_1..A_n})$ belonging to the class $N^{m}_{n,s}(C,D)$ and $\delta\in(0,1]$. If
the subsystems $A_1$,...,$A_m$ are finite-dimensional then
\begin{equation*}
 |f(\rho)-f(\sigma)|\leq C\delta \sum_{k=1}^m \ln \dim\H_{A_k}+Dg(\delta),
\end{equation*}
for any states $\rho$ and $\sigma$ in $\S(\H_{A_1..A_n})$ such that either $F(\rho,\sigma)\geq 1-\delta^2$ or $\,\varepsilon(2-\varepsilon)\leq\delta^2$, where $\,\varepsilon=\textstyle\frac{1}{2}\|\rho-\sigma\|_1$, $F(\rho,\sigma)=\|\sqrt{\rho}\sqrt{\sigma}\|_1^2\,$
and $g(x)$ is the function defined in (\ref{g-def}).}
\end{corollary}\smallskip

The classes $N^{m}_{n,s}(C,D)$ contain important entanglement measures in bipartite and multipartite quantum systems and characteristics related to the classical
capacity of a quantum channel (see Sections 5.4-5.6).

\subsection{Modifications of the AFW-method for functions on sets of states with bounded energy}

Many characteristics of infinite-dimensional quantum systems are functions well defined only on subsets of
$\S(\H)$ consisting of states with bounded values of a given energy-type functional. In this subsection we describe
modifications of the  AFW-method designed to analyze uniform continuity of such functions.

Let $H$ be a positive (semidefinite)  operator on a Hilbert space $\H$.
Then
\begin{equation}\label{mainset}
\C_{H,E}=\{\rho\in\S(\H)\,|\,\Tr H\rho\leq E\shs\},\quad E>E_0,
\end{equation}
($E_0\doteq\inf\limits_{\|\varphi\|=1}\langle\varphi|H|\varphi\rangle$) is a closed convex  subset of $\S(\H)$. If $H$ is the Hamiltonian (energy observable) of the quantum system described by the space $\H$ then $\C_{H,E}$ is the set of states of this system with the mean energy not exceeding $E$.

The main problem of direct application of the AFW-method  to LAA functions defined on $\C_{H,\infty}\doteq\bigcup_{E>0}\C_{H,E}$ consists in
difficulty to estimate the value of $\Tr H\tau_{\pm} $, where $\tau_-=\Delta^-(\rho,\sigma)$ and $\tau_+=\Delta^+(\rho,\sigma)$ are the auxiliary  states
constructed via some  states $\rho$ and $\sigma$ in $\C_{H,E}$ in the first step of this method (see the proof of Theorem \ref{AFW-1}). Moreover, it is not clear how to guarantee that these states belong to the set $\C_{H,\infty}$.

Below we consider several approaches that allow us to construct continuity bounds for LAA functions on the set $\C_{H,E}$.

\subsubsection{Winter's approach based on approximation.} In this subsection we describe in a general form the two-step approach based
on approximation of states with bounded energy followed by the AFW-method, which was used by Winter in \cite{W-CB} to obtain continuity bounds
for the entropy and conditional entropy under the energy constraint. This approach is also used in \cite{SE} to obtain continuity bound for the quantum conditional mutual information and in the proofs of  Theorems \ref{L-1-ca} and \ref{L-m-ca} below.

Let $f$ be a function defined on the convex set $\C_{H,\infty}\doteq\bigcup_{E\geq 0}\C_{H,E}$ that satisfies inequalities (\ref{LAA-1}) and
(\ref{LAA-2}) on this set with some functions $a_f$ and $b_f$.

Let $\{\Lambda_m\}_{m\in\mathbb{N}}$ be a family of maps from
$\C_{H,E}$ to $\S(\H)$ such that
\begin{itemize}
  \item $\lim_{m\to+\infty}\Lambda_m(\rho)=\rho\,$ for any state $\rho$ in $\C_{H,E}$;
  \item $\frac{1}{2}\|\Lambda_m(\rho)-\Lambda_m(\sigma)\|_1\leq r_m(\frac{1}{2}\|\rho-\sigma\|_1)$ for any states $\rho$ and $\sigma$ in $\C_{H,E}$, where $r_m(x)$ is a
  particular function such that $r_m(x)\leq1$ for all $x\leq \varepsilon_0\leq1$, $m\geq m_0>0$;
  \item the states $\Lambda_m(\rho)$, $\Delta^-(\Lambda_m(\rho),\Lambda_m(\sigma))$ and $\Delta^+(\Lambda_m(\rho),\Lambda_m(\sigma))$ belong to the domain of $f$ for any states $\rho$ and $\sigma$ in $\C_{H,E}$ provided that $\Lambda_m(\rho)\neq\Lambda_m(\sigma)$;
  \item   $-A^-_m\leq f(\rho)-f(\Lambda_m(\rho))\leq A^+_m$ for any state $\rho\in\C_{H,E}$,  $A^\pm_m\in\mathbb{R}_+$;
  \item   $-B^-_m\leq f(\Delta^\pm(\Lambda_m(\rho),\Lambda_m(\sigma)))\leq B^+_m$ for any states $\rho,\sigma\in\C_{H,E}$,  $B^\pm_m\in\mathbb{R}_+$,
\end{itemize}
where $\Delta^-(\cdot,\cdot)$ and $\Delta^+(\cdot,\cdot)$ are the operations introduced in Section 3.1.1.\smallskip\pagebreak

\begin{theorem}\label{W-2-step}  \emph{Let $f$ be a function on $\C_{H,E}$ satisfying inequalities (\ref{LAA-1}) and (\ref{LAA-2}).
If there exists a family $\{\Lambda_m\}_{m\in\mathbb{N}}$ of maps with the above properties then
\begin{equation}\label{W-2-step+}
|f(\rho)-f(\sigma)|\leq \inf_{m\geq m_0} \left\{A_m+r_m(\varepsilon)B_m+D_f(r_m(\varepsilon))\shs\right\}
\end{equation}
for any states $\rho$ and $\sigma$ in $\C_{H,E}$ s.t. $\frac{1}{2}\|\shs\rho-\sigma\shs\|_1\leq\varepsilon\leq\varepsilon_0$, where $A_m=A^{-}_m+A^{+}_m$, $B_m=B^{-}_m+B^{+}_m\,$ and
$\,D_f(x)=\displaystyle(1+x)(a_f+b_f)\!\left(\frac{x}{1+x}\right)$.} \emph{Continuity bound (\ref{W-2-step+}) is faithful (i.e. the r.h.s. of (\ref{W-2-step+}) tends to zero as $\,\varepsilon\to 0$) if and only if
there is a function  $\,\varepsilon \mapsto m_{\varepsilon}$ such that}
\begin{equation}\label{faith-cond}
\lim_{\varepsilon \to 0}A_{m_{\varepsilon}}=\lim_{\varepsilon \to 0} r_{m_\varepsilon}(\varepsilon)B_{m_\varepsilon}=\lim_{\varepsilon \to 0} r_{m_\varepsilon}(\varepsilon)=0.
\end{equation}
\end{theorem}\smallskip

\emph{Proof.} W.l.o.g. we may assume that $f(\rho)\geq f(\sigma)$. For each $m\geq m_0$ we have
$$
\begin{array}{c}
f(\rho)-f(\sigma)=(f(\rho)-f(\Lambda_m(\rho)))+(f(\Lambda_m(\sigma))-f(\sigma))+(f(\Lambda_m(\rho))-f(\Lambda_m(\sigma)))\\\\
\leq A_m^++A_m^-+|f(\Lambda_m(\sigma))-f(\Lambda_m(\rho))|.
\end{array}
$$
By applying the AFW-technique presented in the proof of Theorem \ref{AFW-1}  we obtain
$$
|f(\Lambda_m(\sigma))-f(\Lambda_m(\rho))|\leq r_m(\varepsilon)B_m+(1+r_m(\varepsilon))\shs d_f\!\left(\frac{r_m(\varepsilon)}{1+r_m(\varepsilon)}\right).
$$
The above two inequalities imply (\ref{W-2-step+}). The last assertion is obvious. $\square$\smallskip

\begin{example}\label{W-2-step-ex} Let $f=S$ be the von Neumann entropy and $H$ be a positive operator on $\H$
having the form (\ref{H-form}) with $E_0=0$. Let $P_m$ be the spectral projector of $H$ corresponding to eigenvalues
$E_0,...,E_{m-1}$. Consider the family o maps
$$
\Lambda_m(\rho)=(\Tr P_m\rho)^{-1} P_m\rho P_m
$$
from $\C_{H,E}$ into $\S(\H)$. The compactness of $\C_{H,E}$  implies that $\Tr P_m\rho>0$ for all $\rho\in\C_{H,E}$  and all $m$ large enough.
In the proof of Lemma 16 in \cite{W-CB} it is shown (in implicit form) that the family $\Lambda_m$ satisfies all the conditions
stated before Theorem \ref{W-2-step} with
$$
A^-_m=\delta_m F_{H}\!\left(\frac{E}{\;\delta_m}\right),\; A^+_m=A^-_m+h_2(\delta_m),\; B_m^-=0,\; B_m^+=F_{H}\!\left(\frac{E}{\;\delta_m}\right),\; r_m(x)=\frac{x+\delta_m}{1-\delta_m},
$$
where $\delta_m=E/E_m$ and $F_H$ is the function defined in (\ref{F-def}). So, since the function $f=S$ is concave and satisfies inequality (\ref{LAA-2}) with $b_f=h_2$, Theorem \ref{W-2-step} implies that
$$
|S(\rho)-S(\sigma)|\leq \inf_{m\geq m_0} \left\{(2\delta_m+r_m(\varepsilon))F_{H}\!\left(\frac{E}{\;\delta_m}\right)+g(r_m(\varepsilon))+h_2(\delta_m)\right\}
$$
for any states $\rho$ and $\sigma$ in $\C_{H,E}$ such that $\frac{1}{2}\|\shs\rho-\sigma\shs\|_1\leq\varepsilon\leq\frac{1}{2}$, where $g(x)$ is the function defined in (\ref{g-def})
and $m_0$ is the minimal solution of the inequality $\delta_m\leq\frac{1}{4}$. This is Winter's continuity
bound for the von Neumann entropy from Lemma 16 in \cite{W-CB} presented in slightly different form (in the continuity
bound from Lemma 16 the role of "free parameter" is plaid by $\varepsilon'=r_m(\varepsilon)$). The faithfulness condition
(\ref{faith-cond}) is valid if and only if
\begin{equation*}
\lim_{x\to 0^+} x F_{H}\!\left(\frac{E}{x}\right)=0.
\end{equation*}
By Proposition 1 in \cite{EC}  this condition  holds if and only if  the operator $H$ satisfies the Gibbs condition (\ref{H-cond}).
Since the Gibbs condition is a criterion of continuity of the entropy on the \emph{compact} set $\C_{H,E}$ \cite{W,EC},
we see that in this case the faithfulness condition (\ref{faith-cond}) in Theorem \ref{W-2-step} \emph{is also necessary}
for the uniform continuity of the function $f=S$.
\end{example}

Note that the continuity bounds obtained by the above approximation technique for the entropy, the quantum conditional entropy and for the quantum conditional mutual information
(the first two in \cite{W-CB}, the last one in \cite{SE}) are quite accurate: they are asymptotically tight for large energy bound $E$ in the sense of the following
\smallskip

\begin{definition}\label{AT}  A continuity bound $\;\displaystyle\sup_{\rho,\sigma\in \S_p}|f(\rho)-f(\sigma)|\leq C_p(\|\rho-\sigma\|_1)\;$ depending on a parameter $\,p\,$ is \emph{asymptotically tight} for large $\,p\,$ if $\;\displaystyle\limsup_{p\rightarrow+\infty}\sup_{\rho,\sigma\in \S_p}\frac{|f(\rho)-f(\sigma)|}{C_p(\|\rho-\sigma\|_1)}=1$.
\end{definition}

\subsubsection{The approach based on initial purification.} Applicability of the approximation method described in the previous subsection is restricted
by necessity to find uniform estimates of the difference $f(\rho)-f(\Lambda_m(\rho))$ for any $\rho\in\C_{H,E}$, i.e. the sequences $\{A_m^-\}$ and $\{A_m^+\}$ tending to zero as $m\to+\infty$. Anyway, this method does not allow us to obtain an universal continuity bound, that can be applied to a wide class of functions on the set of energy-constrained states.

A general way to construct an universal continuity bound for LAA functions on the set of states with bounded value of some energy-type functional
proposed  in \cite{AFM} is based on initial purification of states followed by the standard AFW-technique.

Assume that $H$ is a positive semidefinite operator on a Hilbert space $\H$,  $\C_{H,E}$ is the set defined in (\ref{mainset}) and $\C_{H,\infty}=\bigcup_{E}\C_{H,E}$.

As mentioned before, the main problem of direct application of the AFW-method  to LAA functions defined on the set $\C_{H,\infty}$ consists in
difficulty to estimate the value of the energy functional at the auxiliary  states  $\tau_-=\Delta^-(\rho,\sigma)$ and $\tau_+=\Delta^+(\rho,\sigma)$ (introduced in the proof of Theorem \ref{AFW-1}). This problem however does not occur if the states $\rho$ and $\sigma$ are pure. Indeed, simple estimates based on using the Schwarz inequality show that
\begin{equation}\label{H-est}
\Tr H\rho,\Tr H\sigma\leq E\quad \Rightarrow\quad  \Tr H\tau_-,\Tr H\tau_+\leq 2E/\varepsilon^2
\end{equation}
in this case, where $\varepsilon=\frac{1}{2}\|\rho-\sigma\|_1$ (these  estimates are made in the proof of Theorem 1 in \cite{AFM}). So, the AFW-technique is applied and shows that
\begin{equation}\label{U-CB}
|f(\rho)-f(\sigma)|\leq \varepsilon C_f\!\left(\frac{2E}{\varepsilon^2}\right)+D_f(\varepsilon)
\end{equation}
for any function $f$ on the set $\C_{H,\infty}$ satisfying inequalities (\ref{LAA-1}) and (\ref{LAA-2}) and any
pure states $\rho$ and $\sigma$ in $\C_{H,E}$, where $\,D_f(\varepsilon)=\displaystyle(1+\varepsilon)(a_f+b_f)\!\left(\frac{\varepsilon}{1+\varepsilon}\right)$ and
\begin{equation}\label{Bf-def}
C_f(E)\doteq\sup_{\varrho,\varsigma\in\C_{H,E}}|f(\varrho)-f(\varsigma)|.
\end{equation}

This observation suggests a simple way to overcome the problem mentioned before by applying the first step of the AFW-method to purifications $\bar{\rho}$ and  $\bar{\sigma}$ of the states $\rho$ and $\sigma$ chosen as close to each other as possible. The following theorem is a modified version of Theorem 1 in \cite{AFM}.

\begin{theorem}\label{AFW-ip} \emph{Let $\,f$ be a function on the set $\,\C_{H,\infty}$  satisfying inequalities (\ref{LAA-1}) and (\ref{LAA-2}). Let $\varepsilon\in(0,1]$.   Then
\begin{itemize}
  \item inequality (\ref{U-CB}) holds for any pure states $\rho$ and $\sigma$ in $\,\C_{H,E}$ such that $\,\frac{1}{2}\|\rho-\sigma\|_1\leq\varepsilon$;
  \item inequality (\ref{U-CB}) holds for any states $\rho$ and $\sigma$ in $\,\C_{H,E}$ such that
$\,F(\rho,\sigma)\geq 1-\varepsilon^2$, where $\,F(\rho,\sigma)=\|\sqrt{\rho}\sqrt{\sigma}\|^2_1\,$ is the fidelity between the states $\rho$ and $\sigma$.
\end{itemize}}

\emph{Continuity bound (\ref{U-CB}) is faithful (its r.h.s. tends to zero as $\shs\varepsilon\to0$) if and only if}
\begin{equation}\label{faithful}
C_f(E)=o\shs(\sqrt{E})\;\textit{ as }\;E\rightarrow+\infty.
\end{equation}
\end{theorem}

\emph{Sketch of the proof.} The first claim is proved before the theorem. Since for any states $\rho$ and $\sigma$ such that
$\,F(\rho,\sigma)\geq 1-\varepsilon^2$ there exist purifications $\bar{\rho}$ and $\bar{\sigma}$ such that $\,\frac{1}{2}\|\bar{\rho}-\bar{\sigma}\|_1\leq\varepsilon$ (this follows from inequality (\ref{F-Tn-ineq})), the second claim is proved by applying the AFW-technique to these purifications with the help of (\ref{H-est}) (see details in \cite{AFM}). $\Box$\smallskip

\begin{corollary}\label{AFW-ip-c} \emph{Any LAA function $f$ on the set $\,\C_{H,\infty}$ satisfying condition (\ref{faithful}) is uniformly continuous on the set
$\,\C_{H,E}$ for any $E>0$.}
\end{corollary}\smallskip

\begin{remark}\label{AFW-r}  In Theorem \ref{AFW-ip} it is assumed that the  function $f$ is defined and satisfies inequalities (\ref{LAA-1}) and (\ref{LAA-2}) on
the set $\,\C_{H,\infty}\doteq\bigcup_E \C_{H,E}$. In fact, all the arguments from the proof of this theorem remain valid if we replace the set $\C_{H,E}$
by any its convex subset $\C^0_{H,E}$  provided that the following invariance property holds:
\begin{equation*}
 \Tr_R \Delta^{\pm}(\bar{\rho},\bar{\sigma})\in \C^0_{H,\infty}\doteq\bigcup_{E>0} \C^0_{H,E},
\end{equation*}
for any purifications $\bar{\rho}$ and $\bar{\sigma}$ in $\S(\H\otimes\H_R)$ of arbitrary states $\rho$ and $\sigma$ in $\C^0_{H,E}$. In this case the set $\C_{H,E}$ in definition (\ref{Bf-def}) of  the function $C_f$ can be replaced by $\C^0_{H,E}$.
\end{remark}

\subsubsection{Applications to characteristics of composite quantum systems.}
Let $A_1$,...,$A_n$ be arbitrary infinite-dimensional quantum systems. Generalizing the notation introduced in Section 3.1.2 denote by $L^{m}_n(C,D)$, $m\leq n$, the class of
all functions $f$ on the set
\begin{equation}\label{Set-m}
\mathfrak{S}_m(\mathcal{H}_{A_1..A_n})\doteq\left\{\rho\in\mathfrak{S}(\mathcal{H}_{A_1..A_n})\,|\,S(\rho_{A_1}),..., S(\rho_{A_m})<+\infty\shs\right\}
\end{equation}
satisfying inequalities  (\ref{LAA-1}) and (\ref{LAA-2})
with $a_f(p)=d_f^-h_2(p)$ and $b_f=d_f^+h_2(p)$ and inequality (\ref{Cm}) with the parameters $c^{\pm}_f$ and  $d^{\pm}_f$ such that $c^-_f+c^+_f=C$ and $d^-_f+d^+_f=D$.
We define the classes $\widehat{L}^{m}_n(C,D)$ and $N^{m}_{n,s}(C,D)$, $s=1,2,3$, of functions on the set $\S_m(\mathcal{H}_{A_1..A_n})$ in the same way as in Section 3.1.2 (everywhere replacing $\S(\mathcal{H}_{A_1..A_{n+l}})$ with
$\S_m(\mathcal{H}_{A_1..A_{n+l}})$, $l\geq 0$).

In this section we apply the approach based on initial purification to obtain universal continuity bounds for functions from the classes
$\widehat{L}_n^{m}(C,D)$ and $N_{n,s}^{m}(C,D)$ under the constraint  corresponding to the positive operator $H_m\otimes I_{A_{m+1}}\otimes...\otimes I_{A_n}$, where
\begin{equation*}
H_{m}=H_{A_1}\otimes I_{A_2}\otimes...\otimes I_{A_m}+\cdots+I_{A_1}\otimes... \otimes I_{A_{m-1}}\otimes H_{A_m}
\end{equation*}
is a positive operator on the space $\H_{A_1...A_m}$ determined by positive operators $H_{A_1}$,....,$H_{A_m}$ on the spaces $\H_{A_1}$,....,$\H_{A_m}$ (it is assumed that $H_1=H_{A_1}$). It is essential that the operator $H_{m}$ satisfies condition (\ref{H-cond+}) if and only if all the operators $H_{A_1}$,....,$H_{A_m}$  satisfy this condition \cite[Lemma 4]{CBM}. Note that $\Tr H_{m}\rho=\sum_{k=1}^{m}\Tr H_{A_k}\rho_{A_k}$ for any $\rho$ in $\S(\H_{A_1..A_m})$.
We will use the function
\begin{equation}\label{F-H-m}
F_{H_{m}}(E)=\sup\shs\{ S(\rho)\,|\,\rho\in\S(\H_{A_1..A_m}),\,\Tr H_{m}\rho\leq E \}.
\end{equation}
If all the operators $H_{A_1}$,....,$H_{A_m}$  are unitarily equivalent to each other then it is easy to see that $F_{H_{m}}(E)=mF_{H_{\!A_1}}(E/m)$, where $F_{H_{\!A_1}}$ is the function defined in (\ref{F-def}).\smallskip

The following theorem is a strengthened version of Theorem 5 in \cite{CBM}.\smallskip

\begin{theorem}\label{L-1-ip} \emph{Let $H_{\!A_1}$,...,$H_{\!A_m}$ be positive operators on the Hilbert spaces $\H_{A_1}$,....,$\H_{A_m}$ satisfying condition (\ref{H-cond+})
and $F_{H_{m}}$ the function defined in (\ref{F-H-m}). Let $\delta\in(0,1]$ and $\,g(x)$ be the function defined in (\ref{g-def}). Then
\begin{equation}\label{L-1-ip+}
    |f(\rho)-f(\sigma)|\leq C\delta F_{H_{m}}\!\!\left[\frac{2mE}{\delta^2}\right]+Dg(\delta)
\end{equation}
for any function $f$ in $\widehat{L}_n^{m}(C,D)\cup N_{n,s}^{m}(C,D)$, $s=1,2,3$, and any states $\rho$ and $\sigma$ in $\S(\H_{A_1...A_n})$ such that $\,\sum_{k=1}^{m}\Tr H_{A_k}\rho_{A_k},\,\sum_{k=1}^{m}\Tr H_{A_k}\sigma_{A_k}\leq mE\,$ and
$$
   \textit{either}\quad F(\rho,\sigma)\doteq\|\sqrt{\rho}\sqrt{\sigma}\|^2_1\geq 1-\delta^2\quad \textit{or} \quad \varepsilon(2-\varepsilon)\leq\delta^2,\quad \textit{where} \quad  \varepsilon=\textstyle\frac{1}{2}\|\rho-\sigma\|_1.
$$
For pure states $\rho$ and $\sigma$
inequality (\ref{L-1-ip+}) holds with $\shs\delta$ replaced by $\,\varepsilon$.}
\emph{The right hand sides of  (\ref{L-1-ip+})  tends to zero as $\,\delta\to 0$.}
\end{theorem}\smallskip

\emph{Proof.} It suffices to assume that $F(\rho,\sigma)\geq 1-\delta^2$, since this inequality follows
from the condition $\varepsilon(2-\varepsilon)\leq\delta^2$ due to the left inequality in (\ref{F-Tn-ineq}).

Assume that $f$ is a function from the class $L_n^{m}(C,D)$. It follows from the equality
$\mathrm{Tr} H_{A^m}(\rho_{A_1}\otimes...\otimes\rho_{A_m})=\sum_{k=1}^m \mathrm{Tr} H_{A_k}\rho_{A_k}$ that
\begin{equation*}
\sum_{k=1}^m S(\rho_{A_k})=S(\rho_{A_1}\otimes...\otimes\rho_{A_m})\leq F_{H_{\!A^m}}(mE)
\end{equation*}
for any state $\rho\in \S(\H_{A_1...A_n})$ such that $\Tr H_{A^m}\rho_{A^m}=\sum_{k=1}^m\mathrm{Tr} H_{A_k}\rho_{A_k}\leq mE$. Thus, for any such state $\rho$  inequality (\ref{Cm}) implies that
\begin{equation*}
-c_f^-F_{H_{\!A^m}}(mE)\leq f(\rho)\leq c_f^+F_{H_{\!A^m}}(mE).
\end{equation*}
Hence,  inequality (\ref{L-1-ip+})
directly follows from Theorem \ref{AFW-ip} in Section 3.1.2. Since the r.h.s. of
(\ref{L-1-ip+}) does not depend on $f$, it remains valid for any function
from the class $\widehat{L}_n^{m}(C,D)$.

Let $f$ be a function in $N^{m}_{n,1}(C,D)$ defined  by the first expression in (\ref{N-class-f}) by means of some function $h$ in $\widehat{L}^{m}_{n+l}(C,D)$. By using the isometrical
equivalence of all purifications of a given state one can show (see Section IV in \cite{C&W}) that
\begin{equation}\label{Lambda}
f(\rho)=\inf_{\Lambda}\shs h(\id_{A_1...A_n}\otimes \Lambda(\bar{\rho})),
\end{equation}
where $\bar{\rho}$ is any pure state in $\,\S(\H_{A_1..A_nR})$ such that $\Tr_R\shs\bar{\rho}=\rho$ and the infimum is over
all channels  $\Lambda:\T(\H_R)\rightarrow\T(\H_{A_{n+1}..A_{n+l}})$.

By equality (\ref{F-Tn-eq})  there are pure states $\bar{\rho}$ and $\bar{\sigma}$ in $\,\S(\H_{A_1..A_nR})$ such that
$\,\frac{1}{2}\|\shs\bar{\rho}-\bar{\sigma}\|_1=\delta$, $\Tr_R\shs\bar{\rho}=\rho$ and $\Tr_R\shs\bar{\sigma}=\sigma$.
Since the function $\varrho\mapsto h(\id_{A_1...A_n}\otimes \Lambda(\varrho))$ on $\,\S(\H_{A_1..A_nR})$ belongs to the class $\widehat{L}^{m}_{n+1}(C,D)$,
by using the proved claim of Theorem \ref{L-1-ip} and taken the purity of the states $\bar{\rho}$ and $\bar{\sigma}$ into account  we obtain
\begin{equation}\label{vu-in}
|h(\Lambda^{\rm e}(\bar{\rho}))-h(\Lambda^{\rm e}(\bar{\sigma}))| \leq C\delta F_{H_m}\!\!\left[\frac{2mE}{\delta^2}\right]+Dg(\delta),
\end{equation}
where $\Lambda^{\rm e}=\id_{A_1...A_n}\otimes \Lambda$.  As the right hand side of this inequality is independent on $\Lambda$, representation (\ref{Lambda}) implies (\ref{L-1-ip+}).

Let $f$ be a function in $N^{m}_{n,2}(C,D)$ defined  by the first expression in (\ref{N-class-f}) by means of some function $h$ in $\widehat{L}^{m}_{n+1}(C,D|\S_{\rm qc})$, where $\S_{\rm qc}$ is the set of states in $\S_m(\H_{A_1...A_{n+1}})$ having the form (\ref{m-2}). By generalizing Nielsen's construction of a continuity bound for the entanglement of formation (cf.\cite{Nielsen}) assume that $f(\rho)\leq f(\sigma)$. Let $\epsilon>0$ be arbitrary and let
$\hat{\rho}$ be an extension of the state $\rho$ belonging to the set $\S^2_{\rm qc}=\S_{\rm qc}$ such that
\begin{equation}\label{tmp-in}
h(\hat{\rho})\leq f(\rho)+\epsilon.
\end{equation}
The Schrodinger-Gisin–Hughston–Jozsa–Wootters theorem (cf.\cite{Schr,Gisin,HJW}) implies that $\rho=\Tr_R\shs\bar{\rho}$ and $p_k\rho_k=\Tr_R [I_{A_1...A_n}\otimes M_k]\shs\bar{\rho}\,$ for all $k$, where $\bar{\rho}$ is some pure state in $\S(\H_{A_1...A_nR})$ and $\{M_k\}$ is the corresponding POVM
on appropriate Hilbert space $\H_R$. So, we have $\hat{\rho}=\id_{A_1...A_n}\otimes \Lambda(\bar{\rho})$, where
$\Lambda(\varrho)=\sum_k [\Tr M_k\varrho]|k\rangle\langle k|$
is a q-c channel from $R$ to $A_{n+1}$. As $F(\rho,\sigma)\geq 1-\delta^2$, it follows from (\ref{F-Tn-eq}) that there is a pure state  $\bar{\sigma}$ in $\S_m(\H_{A_1...A_nR})$ such that $\sigma=\Tr_R\shs\bar{\sigma}$ and $\,\frac{1}{2}\|\shs\bar{\rho}-\bar{\sigma}\|_1=\delta$.

Since $\id_{A_1...A_n}\otimes \Lambda(\varrho)$ is a state in $\S_{\rm qc}$ for any state $\varrho$ in $\,\S_m(\H_{A_1..A_nR})$, the function
$\varrho\mapsto h(\id_{A_1...A_n}\otimes \Lambda(\varrho))$ on $\,\S_m(\H_{A_1..A_nR})$ belongs to the class $\widehat{L}^{m}_{n+1}(C,D)$. Thus,
by using the proved part of Theorem \ref{L-1-ip} and taken the purity of the states $\bar{\rho}$ and $\bar{\sigma}$ into account we see
that continuity bound (\ref{vu-in}) holds. By applying it we obtain
$$
h(\hat{\rho})=h(\Lambda^{\rm e}(\bar{\rho}))\geq h(\Lambda^{\rm e}(\bar{\sigma}))-C\delta F_{H_m}\!\!\left[\frac{2mE}{\delta^2}\right]-Dg(\delta),\quad \Lambda^{\rm e}=\id_{A_1...A_n}\otimes \Lambda.
$$
Since $\,\Lambda^{\rm e}(\bar{\sigma})\,$ is an extension of the state $\sigma$ belonging to the set $\S^2_{\rm qc}=\S_{\rm qc}$, by using this inequality and (\ref{tmp-in}) we obtain
$$
f(\rho)\geq f(\sigma)-C\delta F_{H_{m}}\!\!\left[\frac{2mE}{\delta^2}\right]-Dg(\delta)-\epsilon.
$$

If $f$ is a  function in $N^{m}_{n,3}(C,D)$ then we can repeat the above arguments by noting that in this case the POVM $\{M_k\}$ consists of 1-rank operators, and hence
$\Lambda^{\rm e}(\bar{\sigma})$ is an extension of the state $\sigma$ belonging to the set $\S^3_{\rm qc}$. $\square$

\subsubsection{Advanced continuity bound for the class $\widehat{L}_n^{1}(C,D)$.}
The continuity bound for functions from the class $\widehat{L}_n^{1}(C,D)$ given by Theorem \ref{L-1-ip}
with $m=1$ is universal but not too accurate, especially, if the trace norm is used as a measure of closeness of quantum states. More accurate continuity bounds for functions from this class can be obtained
by combining the approximation approach with the approach based on initial purification (described in Sections 3.2.1 and 3.2.2 correspondingly).

Let $H_{A_1}$ be a positive (semidefinite) operator on a Hilbert space $\H_{A_1}$ that satisfies condition (\ref{H-cond+}) and (\ref{star}). As mentioned in Section 2.2 in this case the
function $F_{H_{A_1}}$ defined in (\ref{F-def}) is concave on $[0,+\infty)$ and satisfies the asymptotic property (\ref{H-cond+a}). It follows from Proposition 1
in \cite{MCB} that we can always find a continuous function $G$ on $[0,+\infty)$  such that
\begin{equation}\label{G-c1}
G(E)\geq F_{H_{A_1}}\!(E)\quad \forall E>0,\quad G(E)=o\shs(\sqrt{E})\quad\textrm{as}\quad E\rightarrow+\infty
\end{equation}
and
\begin{equation}\label{G-c2}
G(E_1)\leq G(E_2),\quad G(E_1)/\sqrt{E_1}\geq G(E_2)/\sqrt{E_2}
\end{equation}
for any $E_2>E_1>0$. Moreover, if the operator $H_{A_1}$  satisfies condition (\ref{B-D-cond}) then
the above function $G$ can be chosen in such a way that
\begin{equation}\label{G-c3}
G(E)=F_{H_{A_1}}(E)(1+o(1))\quad\textrm{as}\quad E\to+\infty.
\end{equation}
A concrete example of a function $G$ satisfying all the conditions in (\ref{G-c1}), (\ref{G-c2}) and (\ref{G-c3}) is presented
after the following\smallskip

\begin{theorem}\label{L-1-ca} \cite{MCB} \emph{Let $H_{A_1}$ be a positive operator  on $\H_{A_1}$ satisfying
conditions (\ref{H-cond+}) and (\ref{star}),  $G$  any function satisfying conditions (\ref{G-c1}) and  (\ref{G-c2}) and $d_0$  the minimal natural number such that  $\,\ln d_0>G(0)\,$. Let $\,E>0$, $\varepsilon>0$, $\,T=(1/\varepsilon)\min\{1, \sqrt{E/G^{-1}(\ln d_0)}\}$ and $\Delta=1/d_0+\ln2$. If $f$ is any function from the class $\widehat{L}_n^{1}(C,D)$ then
\begin{equation}\label{L-1-ca+}
|f(\rho)-f(\sigma)|\leq\min_{t\in(0,T]}\mathbb{CB}_{\shs t}(E,\varepsilon\,|\,C,D),
\end{equation}
where
$$
\mathbb{CB}_{\shs t}(E,\varepsilon\,|\,C,D)=C\varepsilon(1+4t)\!\left(G\!\left[\!\frac{E}{(\varepsilon t)^2}\!\right]+\mathrm{\Delta}\right)+D(2g(\varepsilon t)+g(\varepsilon(1+2t))),
$$
for any states $\rho$ and $\sigma$ in $\S(\H_{A_1..A_n})$ s.t. $\,\Tr H_{A_1}\rho_{A_1},\Tr H_{A_1}\sigma_{A_1}\leq E\,$ and $\,\frac{1}{2}\|\rho-\sigma\|_1\leq \varepsilon$.}

\emph{If the functions $F_{H_{A_1}}$ and $G$ satisfy, respectively,  conditions (\ref{B-D-cond-a}) and (\ref{G-c3})
and the function $f$ satisfies inequality (\ref{Cm}) with the parameters
$c^-_f$ and $c^+_f$ such that
\begin{equation}\label{as-t-c-1}
\lim_{E\rightarrow+\infty} \left(c^-_f+\frac{\inf_{\rho\in\C_{E}}  f(\rho)}{F_{H_{A_1}}(E)}\right)=\lim_{E\rightarrow+\infty}\left(c^+_f-\frac{\sup_{\rho\in\C_{E}}f(\rho)}{F_{H_{A_1}}(E)}\right)=0,
\end{equation}
where $\C_{E}=\{\shs\rho\in\S(\H_{A_1..A_n})\shs|\,\Tr H_{A_1}\rho_{A_1}\leq E\shs\}$, then continuity bound (\ref{L-1-ca+}) is asymptotically tight for large $E$ (Definition \ref{AT} in Section 3.2.1).}
\end{theorem}\smallskip

The necessity of condition (\ref{as-t-c-1}) for the asymptotical tightness of continuity bound (\ref{L-1-ca+}) is obvious, since it means  "tightness" of each of the inequalities
\begin{equation*}
-c^-_fF_{H_{A_1}}(E)\leq  f(\rho)\leq c^+_fF_{H_{A_1}}(E)
\end{equation*}
valid for any state $\rho$ in $\C_{E}$ which are deduced from (\ref{Cm}) and used in the proof of (\ref{L-1-ca+}).

If $H_{A_1}$ is the grounded Hamiltonian of the $\,\ell$-mode quantum oscillator with the frequencies $\,\omega_1,...,\omega_{\ell}\,$,
i.e.
\begin{equation}\label{H-osc}
H_{A_1}=\sum_{i=1}^{\ell}\hbar \omega_i a_i^*a_i,
\end{equation}
where $a_i$ and $a^*_i$ are the annihilation and creation operators of the $i$-th mode \cite{H-SCI}, then
the function
\begin{equation}\label{F-osc}
G_{\ell,\omega}(E)\doteq \ell\ln \frac{E+2E_0}{\ell E_*}+\ell,\quad E_0= \frac{1}{2}\sum_{i=1}^{\ell}\hbar \omega_i, \;\; E_*=\left[\prod_{i=1}^{\ell}\hbar\omega_i\right]^{1/\ell}\!\!,\vspace{-5pt}
\end{equation}
satisfies conditions (\ref{G-c1}),(\ref{G-c2}) and (\ref{G-c3}) \cite{CID}. If follows that the function $F_{H_{A_1}}$ satisfies  condition (\ref{B-D-cond-a}).
Theorem \ref{L-1-ca} with the function $G=G_{\ell,\omega}$  can be formulated as\smallskip

\begin{corollary}\label{L-1-ca-c} \cite{MCB}
\emph{Let $H_{A_1}$ be the operator defined in (\ref{H-osc}). Let $E>0$, $\varepsilon>0$, $\,T_*=(1/\varepsilon)\min\{1, \sqrt{E/E_0}\}$ and $\mathbb{CB}^{*}_{\shs t}(E,\varepsilon\,|\,C,D)$ denotes the function $\mathbb{CB}_{\shs t}(E,\varepsilon\,|\,C,D)$
defined in Theorem \ref{L-1-ca} with $G=G_{\ell,\omega}$ and $\,\Delta=e^{-\ell}+\ln2$. If $f$ is any function from the class $\widehat{L}_n^{1}(C,D)$ then
\begin{equation}\label{L-1-ca-c+}
|f(\rho)-f(\sigma)|\leq \min_{t\in(0,T_*]}\mathbb{CB}^{*}_{\shs t}(E,\varepsilon\,|\,C,D)
\end{equation}
for any states $\rho$ and $\sigma$ in $\S(\H_{A_1..A_n})$ s.t. $\Tr H_{A_1}\rho_{A_1},\Tr H_{A_1}\sigma_{A_1}\leq E$ and $\frac{1}{2}\|\shs\rho-\sigma\|_1\leq\varepsilon$.}

\emph{If the function $f$ satisfies condition (\ref{as-t-c-1}) then continuity bound (\ref{L-1-ca-c+}) is asymptotically tight for large $E$ (Definition \ref{AT} in Section 3.2.1).}
\end{corollary}

\subsubsection{Advanced continuity bound for the class $\widehat{L}_n^{m}(C,D)$, $m>1$.}

Continuity bounds for functions from the classes $\widehat{L}_n^{m}(C,D)$  given by  Theorem \ref{L-1-ip}
in the case $m>1$ can be also improved assuming that all the subsystems on which  the energy-type constraint is imposed are isomorphic.\smallskip

\begin{theorem}\label{L-m-ca} \cite{CBM} \emph{Let $H_{\!A_1}$,..,$H_{\!A_m}$ be positive operators on the Hilbert spaces $\H_{A_1}$,..,$\H_{A_m}$ that are unitary equivalent to each other
and satisfy conditions (\ref{H-cond+}) and (\ref{star}). Let $G$ be any function satisfying conditions (\ref{G-c1}) and  (\ref{G-c2}).
If $f$ is a function from the class $\widehat{L}^{m}_n(C,D)$ then
\begin{equation}\label{L-m-ca+}
|f(\rho)-f(\sigma)|\leq\min_{t\in(0,1/\varepsilon)}\mathbb{VB}^m_{\shs t}(E,\varepsilon\,|\,C,D),
\end{equation}
where
$$
\begin{array}{rl}
\mathbb{VB}^m_{\shs t}(E,\varepsilon\,|\,C,D)\!& =\,\displaystyle
Cm\!\left((\varepsilon+\varepsilon^2t^2)G\!\left[\frac{mE}{\varepsilon^2t^2}\right]+2\sqrt{2\varepsilon t}G\!\left[\frac{E}{\varepsilon t}\right]\right)\\\\
& +\,D\!\left(g\!\left(\varepsilon+\varepsilon^2t^2\right)+2g(\sqrt{2\varepsilon t})\right),
\end{array}
$$
for arbitrary quantum states $\rho$ and $\sigma$ in $\S(\H_{A_1..A_n})$ such that $\;\frac{1}{2}\|\shs\rho-\sigma\|_1\leq\varepsilon$ and $\,\sum_{k=1}^{m}\mathrm{Tr} H_{A_k}\rho_{A_k},\,\sum_{k=1}^{m}\mathrm{Tr} H_{A_k}\sigma_{A_k}\leq mE$.}

\emph{If the functions $F_{H_{A_1}}$ and $G$ satisfy, respectively,  conditions (\ref{B-D-cond-a}) and (\ref{G-c3}) and the function $f$ satisfies inequality (\ref{Cm}) with the parameters $c^-_f$ and $c^+_f$ such that
\begin{equation}\label{as-t-c-2}
\!\lim_{E\rightarrow+\infty} \left(c_f^-+\frac{\inf_{\rho\in\C_{E}} f(\rho)}{mF_{H_{A_1}}(E)}\right)=\lim_{E\rightarrow+\infty}\left(c^+_f-\frac{\sup_{\rho\in\C_{E}}f(\rho)}{mF_{H_{A_1}}(E)}\right)=0,
\end{equation}
where $\C_{E}=\{\shs\rho\in\S(\H_{A_1..A_n})\shs|\,\sum_{k=1}^{m}\mathrm{Tr} H_{A_k}\rho_{A_k}\leq mE\shs\}$ and $F_{H_{A_1}}$ is the function defined in (\ref{F-def}), then   continuity bound (\ref{L-m-ca+}) is asymptotically tight for large $E$ (Definition \ref{AT} in Section 3.2.1).}
\end{theorem}\smallskip

Condition (\ref{as-t-c-2}) is generalization of the "tightness condition" (\ref{as-t-c-1}) to the case $m>1$.

If $H_{A_1}$ is the grounded Hamiltonian of the $\,\ell$-mode quantum oscillator with the frequencies $\,\omega_1,...,\omega_{\ell}\,$
defined in (\ref{H-osc}) then the function $F_{H_{A_1}}$ satisfies  condition (\ref{B-D-cond-a}) and the role of function $G$ in Theorem \ref{L-m-ca} can be played by the function $G_{\ell,\omega}$ defined in (\ref{F-osc}) that satisfies conditions (\ref{G-c1}),(\ref{G-c2}) and (\ref{G-c3}).
So, in this case continuity bound (\ref{L-m-ca+}) is asymptotically tight for large $E$
for any function $f$  satisfying  condition (\ref{as-t-c-2}).

\section{General methods of local continuity analysis}

\subsection{The use of lower semicontinuity}

A very simple but effective way   to obtain sufficient conditions of local continuity of characteristics of quantum systems and channels is based on the fact that many of these characteristics are lower semicontinuous functions on the set of quantum set. The basic characteristics possessing this property are the following:
\begin{itemize}
   \item the von Neumann entropy $S(\rho)$, the output entropy $S(\Phi(\rho))$ and  the entropy exchange $S(\Phi,\rho)$ of a quantum channel $\Phi$ \cite{H-SCI,L-2,W};
   \item the quantum relative entropy $D(\rho\shs\|\shs\sigma)$ as a function of a pair $(\rho,\sigma)$ \cite{H-SCI,L-2,W};
   \item the quantum conditional mutual information $I(A\!:\!B|C)_{\rho}$ \cite{CMI};
   \item the relative entropy distance to a set of quantum states satisfying the particular regularity condition (see \cite{L&Sh}),
   in particular, the relative entropy of entanglement.
   \item the constrained Holevo capacity and the mutual information of a quantum channel (as functions of input state) \cite{cmp-2,H-SCI}.
\end{itemize}

The method we what to describe in this subsection is based on the following very simple observation.\smallskip
\begin{lemma}\label{L-S} \emph{If $f(x)$ and $g(x)$ are lower semicontinuous nonnegative functions on a metric space $X$ and $\{x_n\}_{n}\subset X$ a sequence converging to $x_0\in X$ such that there exists
$$
\lim_{n\to+\infty}(f+g)(x_n)=(f+g)(x_0)<+\infty
$$
then}
$$
\lim_{n\to+\infty}f(x_n)=f(x_0)<+\infty\quad \textit{and}\quad \lim_{n\to+\infty}g(x_n)=g(x_0)<+\infty.
$$
\end{lemma}\smallskip

Thus, if $f$ is a lower semicontinuous function on some subset $\S_0$ of $\S(\H)$ that is majorized on this set
by some continuous function $h$ then to prove that $f$ is continuous on $\S_0$
it suffices to show that $h-f$ is a lower semicontinuous function on $\S_0$.\smallskip

\begin{example}\label{L-S-e} Let $f(\rho)=S(\rho)$ be the von Neumann entropy of a bipartite state $\rho\in\S(\H_{AB})$
and $h(\rho)=S(\rho_A)+S(\rho_B)$. Since $(h-f)(\rho)=D(\rho\shs\|\shs\rho_A\otimes\rho_B)$ for any state $\rho$ with finite entropy,
the lower semicontinuity of the entropy and of the relative entropy imply, by the above lemma, that
$$
\lim_{n\to+\infty}S(\rho_n)=S(\rho_0)<+\infty
$$
for any sequence $\{\rho_n\}\subset\S(\H_{AB})$ converging to a state $\rho_0$ such that
$$
\lim_{n\to+\infty}S([\rho_n]_A)=S([\rho_0]_A)<+\infty\quad \textup{and}\quad\lim_{n\to+\infty}S([\rho_n]_B)=S([\rho_0]_B)<+\infty.
$$
\end{example}

Nontrivial examples of using Lemma \ref{L-S} are considered in Section 5.

\subsection{Dini-type lemma and its corollaries}

In  this section we describe an effective  method for qualitative continuity analysis of characteristic
of composite quantum systems based on uniform approximation of these characteristics. Since the main arguments of the proof of basic technical result of the method  are close to those used in the proof of the classical Dini's lemma,  we call it Dini-type lemma.

\subsubsection{Dini-type lemma.} Let $\A$ be a subset of $\S(\H)$ and $\{\Psi_m\}_{m\in\mathbb{N}}$ a family of maps from
$\A$ to $\T_{+}(\H)$ such that
\begin{itemize}
  \item $\Psi_m(\rho)\leq\Psi_{m+1}(\rho)\leq\rho\,$ for any $\,\rho\in\A$ and all $\,m$;
  \item $\lim_{m\to+\infty}\Psi_m(\rho)=\rho\,$ for any $\rho\in\A$;
  \item for any state $\rho\in\A$ there is a countable subset  $M_{\rho}$ of $\mathbb{N}$ such that
  $\,\lim_{n\to+\infty}\Psi_m(\rho_n)=\Psi_m(\rho)$ for any $\,m\in M_{\rho}$ and any sequence $\{\rho_n\}\subset\A$ converging to a state $\rho\in\A$.
\end{itemize}

By using the classical Dini lemma and the properties of the family $\{\Psi_m\}$ it is easy to prove the following\smallskip

\begin{lemma}\label{DTL-L} \emph{If $\,\C$ is a compact subset of $\,\A$ then}
$
\lim\limits_{m\to+\infty}\inf\limits_{\rho\in\C}\Tr\Psi_m(\rho)=1.
$
\end{lemma}

For any compact subset $\C$ of $\,\A$  let
\begin{equation}\label{N-def}
  \mathbb{N}_{\C}=\left\{m\in\mathbb{N}\,|\,\Tr\Psi_m(\rho)\geq1/2\;\;\forall \rho\in\C\,\right\}.
\end{equation}

The examples of families $\{\Psi_m\}_{m\in\mathbb{N}}$  are described at the end of this subsection.

Throughout this section we will use \textbf{the following notation}
\begin{equation}\label{BN}
  [\sigma]\doteq \sigma/\Tr\sigma,\;\; \sigma\in\T_+(\H)\setminus\{0\},\quad\textrm{and}\quad[0]=0.
\end{equation}

We will consider the state $[\Psi_m(\rho)]$ as an approximation
for $\rho$. So, for a given characteristic $f$ it is natural to explore conditions that imply
\begin{equation}\label{lr-1}
\lim_{m\to+\infty} f([\Psi_m(\rho)])=f(\rho)\quad \forall\rho\in\A
\end{equation}
and conditions that guarantee that this convergence is uniform on a given subset of $\A$.

Note that (\ref{lr-1}) is not valid  in general even for an affine function $f$ on $\,\A=\S(\H)$
if it is not bounded. Indeed, for any such function (\ref{lr-1}) holds if (and in the case $f(\rho)<+\infty$ only if)
\begin{equation}\label{lr-2}
\lim_{m\to+\infty} (1-\Tr\Psi_m(\rho))f([\rho-\Psi_m(\rho)])=0,
\end{equation}
where the notation (\ref{BN}) is used. But if the function $f$ is not bounded then we can not
prove (\ref{lr-2}) despite the convergence of $\Tr\Psi_m(\rho)$ to $1$ as $m\to+\infty$. The simplest example is given by the affine function
$f$ equal to $0$ on the set of finite rank states and to $+\infty$ on its complement and the map $\Psi_m$ defined by formula (\ref{Psi-m-1}) below.

The above relations between (\ref{lr-1}) and (\ref{lr-2}) also hold for any LAA function $f$, i.e.
a function $f$ defined on a convex subset  of $\S(\H)$ that satisfies the inequalities (\ref{LAA-1}) and (\ref{LAA-2}) on this set.

Since many entropic and information characteristic used in quantum information theory
are functions that satisfy either one of inequalities (\ref{LAA-1}) and (\ref{LAA-2}) or both of these inequalities at once (cf.\cite{AFM}),
it is natural to explore conditions implying the validity of (\ref{lr-1}) for such functions.

It follows from the above remarks that some additional properties of a function $f$ satisfying
(\ref{LAA-1}) and (\ref{LAA-2}) are necessary  to ensure the validity of (\ref{lr-1}). The following proposition implies, in particular, that
(\ref{lr-1}) holds in the following two cases:
\begin{itemize}
  \item $f$ is a nonnegative lower semicontinuous function satisfying inequality (\ref{LAA-1});
  \item $f$ satisfies inequalities (\ref{LAA-1}) and (\ref{LAA-2}) and $|f|$  is majorized
  by some lower semicontinuous function  satisfying inequality (\ref{LAA-1}) that is finite at the state $\rho$.
\end{itemize}

\begin{proposition}\label{DTL} \emph{Let $\A$ be a subset of $\S(\H)$ and $\{\Psi_m\}_{m\in\mathbb{N}}$ a family of maps from
$\A$ to $\T_+(\H)$ with the above stated properties. Let $\S_0$ be a convex subset of $\S(\H)$ that
contains $\A$ and all states proportional to the nonzero operators $\Psi_m(\rho)$, $\rho\in\A$, $m\in\mathbb{N}$.}

\emph{Let $f$ be a lower semicontinuous nonnegative function on $\,\S_0$ satisfying inequality (\ref{LAA-1}).
Then
\begin{equation}\label{p-w-c}
\lim_{m\to+\infty}f([\Psi_m(\rho)])=f(\rho)\leq+\infty\quad \forall \rho\in\A,
\end{equation}
where the notation (\ref{BN}) is used.}

\emph{Let $\C$ be a compact subset of $\,\A$ on which the function $\,f$ is continuous. Then
\begin{equation}\label{A-1}
\lim_{m\to+\infty}\sup_{\rho\in\C}\left|f(\rho)-f([\Psi_m(\rho)])\right|=0
\end{equation}
and
\begin{equation}\label{B-1}
\lim_{m\to+\infty}\sup_{\rho\in\C} \Tr\Delta_m(\rho)f([\Delta_m(\rho)])=0,
\end{equation}
where $\Delta_m=\id_{\H}-\Psi_m$ and it is assumed that $f(0)=0$.}

\emph{Let $\,g$ be any function on $\S_0$ such that $|g(\rho)|\leq f(\rho)$ for any $\rho$ in $\S_0$. If
$g$ satisfies inequality (\ref{LAA-1}) then
\begin{equation}\label{C-1}
\lim_{m\to+\infty}\sup_{\rho\in\C}\left(g([\Psi_m(\rho)])-g(\rho)\right)\leq 0.
\end{equation}
If $\,g$ satisfies inequality (\ref{LAA-2}) then
\begin{equation}\label{C-2}
\lim_{m\to+\infty}\sup_{\rho\in\C}\left(g(\rho)-g([\Psi_m(\rho)])\right)\leq 0.
\end{equation}
If $\,g$ satisfies both inequalities (\ref{LAA-1}) and (\ref{LAA-2}) then}
\begin{equation*}
\lim_{m\to+\infty}\sup_{\rho\in\C}\left|g(\rho)-g([\Psi_m(\rho)])\right|=0.
\end{equation*}

\end{proposition}

\emph{Proof.} Note first that (\ref{p-w-c}) follows from the lower semicontinuity of $f$, since inequality (\ref{LAA-1}) and the nonnegativity of $f$ imply
\begin{equation}\label{u-r}
f(\rho)\geq \mu_m f([\Psi_m(\rho)])-a_f(1-\mu_m),\quad \forall \rho\in\A,
\end{equation}
where $\,\mu_m=\Tr\Psi_m(\rho)=1-o(1)\,$ as $\,m\to+\infty$ (since $\Psi_m(\rho)$ tends to $\rho$).

Assume that (\ref{A-1}) is not valid. Then there exist $\,\varepsilon>0$, a sequence $\{m_n\}\subset \mathbb{N}$ tending to $+\infty$ and a sequence $\{\rho_n\}\subset\C$ such that
$$
|f(\rho_n)-f([\Psi_{m_n}(\rho_n)])|\geq\varepsilon.
$$
Since the set $\C$ is compact, we may assume that $\{\rho_n\}$ is a converging sequence. So, to prove (\ref{A-1}) it suffices to show that
\begin{equation}\label{d-rel}
\lim_{m\to+\infty}\sup_{n\geq 0}\left|f(\rho_n)-f([\Psi_m(\rho_n)])\right|=0
\end{equation}
for any sequence $\{\rho_n\}\subset \C$ converging to a state $\rho_0$.

Let $\mu_n^m=\Tr\Psi_m(\rho_n)$. It follows from Lemma \ref{DTL-L} that
\begin{equation}\label{mu-u-c}
\lim_{m\to+\infty}\inf_{n\geq 0}\mu_n^m=1.
\end{equation}

Let $\varepsilon\in(0,1/2)$ be arbitrary. It follows from (\ref{p-w-c}) and (\ref{mu-u-c}) that
there is $m_\varepsilon\in M_{\rho_0}$ such that
\begin{equation}\label{m-e}
|f(\rho_0)-f([\Psi_{m_\varepsilon}(\rho_0)])|<\varepsilon,\quad \mu_n^{m_\varepsilon}>1-\varepsilon\quad  \textrm{and}\quad  a_f(1-\mu_n^{m_\varepsilon})<\varepsilon
\end{equation}
for all $n\geq0$. It follows from inequality (\ref{LAA-1}) and the nonnegativity of $f$ that
$$
f(\rho_n)\geq \mu_n^m f([\Psi_{m}(\rho_n)])-a_f(1-\mu_n^m)\geq (1-\varepsilon)f([\Psi_{m}(\rho_n)])-\varepsilon\quad \forall n\geq0\;\, \forall  m\geq m_\varepsilon,
$$
where we used the assumption (\ref{a-b-assump}). So, the continuity of $f$ on $\C$ implies
\begin{equation}\label{u-b}
f(\rho_n)\leq C<+\infty,\quad f([\Psi_{m}(\rho_n)])\leq 2C+1 \quad \forall n\geq0\;\, \forall  m\geq m_\varepsilon.
\end{equation}
It follows from the above inequalities that
\begin{equation}\label{e-way}
f([\Psi_{m}(\rho_n)])-f(\rho_n)\leq 2\varepsilon(C+1)\quad \forall n\geq0\;\,\forall m\geq m_\varepsilon.
\end{equation}

Since
$$
\mu_n^{m+k}[\Psi_{m+k}(\rho_n)]=\mu_n^{m}[\Psi_{m}(\rho_n)]+(\mu_n^{m+k}-\mu_n^{m})[\sigma_n^m]
$$
for any natural numbers $m\geq m_{\varepsilon}$ and $k>0$, where $\sigma_n^m=\Psi_{m+k}(\rho_n)-\Psi_{m}(\rho_n)$ is a positive operator, inequality (\ref{LAA-1}) and the nonnegativity of $f$ along with assumption (\ref{a-b-assump}) imply
\begin{equation}\label{m-k}
\begin{array}{l}
 f([\Psi_{m+k}(\rho_n)])\geq \frac{\mu_n^{m}}{\mu_n^{m+k}}f([\Psi_{m}(\rho_n)])-a_f\!\left(1-\frac{\mu_n^{m}}{\mu_n^{m+k}}\right)\\\\\geq \mu_n^{m}f([\Psi_{m}(\rho_n)])-a_f(1-\mu_n^{m})\geq (1-\varepsilon)f([\Psi_{m}(\rho_n)])-\varepsilon.
\end{array}
\end{equation}

Note that $[\Psi_{m_\varepsilon}(\rho_n)]$ tends to $[\Psi_{m_\varepsilon}(\rho_0)]$ as $n\to+\infty$ because $m_\varepsilon\in M_{\rho_0}$.
Since the function $f$ is lower semicontinuous on $\S_0$ and continuous on $\C$, there is $n_\varepsilon$ such that
$$
f(\rho_n)-f(\rho_0)\leq \varepsilon \quad \textrm{and} \quad f([\Psi_{m_\varepsilon}(\rho_0)])-f([\Psi_{m_\varepsilon}(\rho_n)])\leq \varepsilon \quad \forall n\geq n_\varepsilon.
$$
So, it follows from (\ref{m-e}), (\ref{u-b}) and (\ref{m-k}) with $m=m_\varepsilon$ and $k\geq0$ that
$$
f(\rho_n)-f([\Psi_{m_\varepsilon+k}(\rho_n)])\leq f(\rho_n)-f([\Psi_{m_\varepsilon}(\rho_n)])+\varepsilon(2C+2)\leq\varepsilon(2C+5)\quad \forall n\geq n_\varepsilon.
$$
This along with (\ref{p-w-c}) and (\ref{e-way}) imply (\ref{d-rel}).

To prove relation (\ref{B-1}) note that inequality (\ref{LAA-1}) implies
$$
\Tr\Delta_m(\rho)f([\Delta_m(\rho)])\leq f(\rho)-\Tr\Psi_{m}(\rho)f([\Psi_{m}(\rho)])+a_f(1-\Tr\Psi_{m}(\rho))\qquad \forall\rho\in\A.
$$
Since boundedness of the function $f$ on $\C$ follows from its continuity, relation (\ref{u-r}) implies  boundedness of the function $(\rho,m)\mapsto f([\Psi_{m}(\rho)])$ on $\C\times \mathbb{N}_{\C}$,
where $\mathbb{N}_{\C}$ is defined in (\ref{N-def}). So, the above inequality allows to derive (\ref{B-1}) from (\ref{A-1}) by using Lemma \ref{DTL-L}.

Assume that $g$ is a function on $\S_0$ such that $|g(\rho)|\leq f(\rho)$ for any $\rho$ in $\S_0$. If $g$ satisfies inequality (\ref{LAA-1})
then
$$
g([\Psi_m(\rho)])-g(\rho)\leq (1-\Tr\Psi_{m}(\rho))(f([\Psi_m(\rho)])+f([\Delta_m(\rho)]))+a_g(1-\Tr\Psi_{m}(\rho))
$$
for any $\rho$ in $\C$ and $m\in\mathbb{N}_{\C}$. Thus, Lemma \ref{DTL-L} and the boundedness of the function $(\rho,m)\mapsto f([\Psi_{m}(\rho)])$ on $\C\times \mathbb{N}_\C$ mentioned before allows to derive (\ref{C-1}) from (\ref{B-1}). If $g$ satisfies inequality (\ref{LAA-2}) with some function $b_g$ then $-g$ satisfies inequality (\ref{LAA-1}) with the function $a_{-g}=b_{g}$. So, (\ref{C-2}) follows from (\ref{C-1}). The last assertion is a corollary of the previous ones. $\square$
\medskip

\begin{corollary}\label{DTL-c} \emph{Let $\{\rho_n\}$ be a sequence in $\S(\H)$ converging to a state $\rho_0$. Let $\{\Psi_m\}$
be a family of maps from $\,\A=\{\rho_n\}\cup\{\rho_0\}$ to $\T_+(\H)$ possessing the properties described before Lemma \ref{DTL-L} and $M_{\rho_0}$ the corresponding countable subset of $\,\mathbb{N}$. Let $\S_0$ be a convex subset of $\S(\H)$ that
contains the states $\rho_n$ and $[\Psi_m(\rho_n)]$, $n\geq0$, $m\in\mathbb{N}_\A$  (see (\ref{N-def})).}

\emph{Let $f$ be a nonnegative lower semicontinuous function on $\S_0$  satisfying inequality (\ref{LAA-1}) such that
$$
\lim_{n\to+\infty} f(\rho_n)=f(\rho_0)<+\infty
$$
and $g$ a function on $\S_0$ such that $|g(\rho)|\leq f(\rho)$ for any $\rho$ in $\S_0$. Then
\begin{itemize}
  \item if the function $g$ satisfies inequality (\ref{LAA-1}),
\begin{equation}\label{1-st-cond-1}
\limsup_{m\to+\infty,\; m\in M_{\rho_0}}\! g([\Psi_m(\rho_0)])\geq g(\rho_0)
\end{equation}
and
\begin{equation}\label{1-st-cond-2}
\liminf_{n\to+\infty} g([\Psi_m(\rho_n)])\geq g([\Psi_m(\rho_0)])\quad \forall m\in M_{\rho_0}
\end{equation}
then
$$
\liminf_{n\to+\infty} g(\rho_n)\geq g(\rho_0);
$$
\item if the function $g$ satisfies inequality  (\ref{LAA-2}),
$$
\liminf_{m\to+\infty,\; m\in M_{\rho_0}} g([\Psi_m(\rho_0)])\leq g(\rho_0)\quad \textit{and} \quad \limsup_{n\to+\infty} g([\Psi_m(\rho_n)])\leq g([\Psi_m(\rho_0)])
$$
for any $m\in M_{\rho_0}$ then
$$
\limsup_{n\to+\infty} g(\rho_n)\leq g(\rho_0);
$$
\item if the function $g$ satisfies both inequalities (\ref{LAA-1}) and (\ref{LAA-2}) and
for any $m\in M_{\rho_0}$ there exists
$$
\lim\limits_{n\to+\infty} g([\Psi_m(\rho_n)])=g([\Psi_m(\rho_0)])\quad \textit{then} \quad
\lim_{n\to+\infty} g(\rho_n)=g(\rho_0).
$$
\end{itemize}}
\end{corollary}

\emph{Proof.}  Suppose that the first statement is not valid. By passing to a subsequence we may assume that there exists
\begin{equation}\label{as-rel}
\lim_{n\to+\infty} g(\rho_n)<g(\rho_0)-\Delta,\quad \Delta>0.
\end{equation}
Proposition \ref{DTL} and  relation (\ref{1-st-cond-1}) imply the existence of $m$ in $M_{\rho_0}$
such that
$$
g([\Psi_m(\rho_n)])\leq g(\rho_n)+\Delta/4\quad\forall n \quad \textrm{and} \quad  g([\Psi_m(\rho_0)])\geq g(\rho_0)-\Delta/4.
$$
Hence, by  relation (\ref{1-st-cond-2}) we have
$$
g(\rho_n)+\Delta/4\geq g([\Psi_m(\rho_n)])\geq g([\Psi_m(\rho_0)])-\Delta/4\geq g(\rho_0)-\Delta/2
$$
for all sufficiently large $n$. This contradicts to (\ref{as-rel}).

The second and third statements are proved similarly by using the
corresponding statements of Proposition \ref{DTL}. $\square$

Corollary \ref{DTL-c} is used essentially in the proofs of Proposition \ref{CB-DB}B in Section 5.3.3 and
Proposition \ref{EF-2}B in Section 5.4.3 (continuity conditions for $C_B$, $D_B$, $E_F$ and $E^{\infty}_F$).

We will use the following proposition that is proved by  natural generalization of the arguments from the proof of Proposition \ref{DTL}.\smallskip

\begin{proposition}\label{DTL+} \emph{Let $\{\rho_n\}$ be a sequence in $\S(\H)$ converging to a state $\rho_0$. Let $\{\Psi_m\}$
be a family of maps from $
\,\A=\{\rho_n\}\cup\{\rho_0\}$ to $\T_+(\H)$ possessing the properties described before Lemma \ref{DTL-L} and $M_{\rho_0}$ the corresponding countable subset of $\,\mathbb{N}$. Let $\S_0$ be a convex subset of $\S(\H)$ that
contains  the states $\rho_n$ and $[\Psi_m(\rho_n)]$, $n\geq0$, $m\in\mathbb{N}_{\A}$ (see (\ref{N-def})).}

\emph{Let $f$ be a  nonnegative lower semicontinuous function on $\S_0$ satisfying inequality (\ref{LAA-1}) such that  $f(\rho_n)$ is finite for all $n\geq0$ and
$A\doteq\limsup_{n\to+\infty} f(\rho_n)-f(\rho_0)$. Then
\begin{equation*}
0\leq\liminf_{m\to+\infty}\inf_{n\geq0}\left(f(\rho_n)-f([\Psi_m(\rho_n)])\right)\leq\limsup_{m\to+\infty}\sup_{n\geq0}\left(f(\rho_n)-f([\Psi_m(\rho_n)])\right)\leq A
\end{equation*}
and
\begin{equation*}
\limsup_{m\to+\infty}\sup_{n\geq0} \Tr\Delta_m(\rho_n)f([\Delta_m(\rho_n)])\leq A,\quad \Delta_m=\id_{\H}-\Psi_m.
\end{equation*}}

\emph{Let $g$ be a function on $\S_0$ such that $|g(\rho)|\leq f(\rho)$ for any $\rho$ in $\S_0$. If
the function $g$  satisfies inequality (\ref{LAA-1}) then
\begin{equation*}
\limsup_{m\to+\infty}\sup_{n\geq0}\left(g([\Psi_m(\rho_n)])-g(\rho_n)\right)\leq A.
\end{equation*}
If the function $g$  satisfies inequality (\ref{LAA-2}) then
\begin{equation*}
\limsup_{m\to+\infty}\sup_{n\geq0}\left(g(\rho_n)-g([\Psi_m(\rho_n)])\right)\leq A.
\end{equation*}
If the function $g$  satisfies both inequalities (\ref{LAA-1}) and  (\ref{LAA-2}) then}
\begin{equation*}
\limsup_{m\to+\infty}\sup_{n\geq0}\left|g(\rho_n)-g([\Psi_m(\rho_n)])\right|\leq A.
\end{equation*}
\end{proposition}
\medskip

In this review  we will mainly use the families $\{\Psi_m\}$ of the following two  types.

1) Let
\begin{equation}\label{Psi-m-1}
  \Psi_m(\rho)=P_m^{\rho}\rho
\end{equation}
for any state $\rho$ in $\,\A=\S(\H)$, where $P_m^{\rho}$ is the spectral projector of
$\rho$ corresponding to its $m$ maximal eigenvalues (taken the multiplicity into account). To avoid the ambiguity of the definition of $P_m^{\rho}$  associated with multiple eigenvalues we fix some basic $\{\varphi_i\}$ in $\H$ and will construct an ordered basic $\{\psi_i\}$ in any given finite-dimensional subspace $\H_0$ of $\H$ by \textbf{the following rule:}
let $\psi_1$ be the vector $P_0\varphi_i/\|P_0\varphi_i\|$, where $P_0$ is the projector on $\H_0$ and $\varphi_i$ is the first vector in the basic  $\{\varphi_i\}$ such that
$P_0\varphi_i\neq0$, let $\psi_2$ be the vector obtained by the same construction with $\H_0$  replaced by  $\H_0\ominus\{c\psi_1\}$, etc.

Let $\rho$ be an arbitrary state in $\S(\H)$.  It is clear that $\Psi_m(\rho)\leq \Psi_{m+1}(\rho)\leq\rho$ for all $m$. It is also clear $\Psi_m(\rho)$ tends to $\rho$ as $m\to+\infty$. Denote by $\{\lambda^{\rho}_i\}$ the sequence
of eigenvalues of $\rho$ taken in the non-increasing order. Let
\begin{equation}\label{M-s}
M^{s}_{\rho}\doteq\left\{m\in \mathbb{N}\,|\,\{\lambda^{\rho}_{m+1}< \lambda^{\rho}_m\}\vee\{\lambda^{\rho}_{m}=0\}\right\}.
\end{equation}
Then  $P_m^{\rho_n}\rho_n$ tends to $P_m^{\rho_0}\rho_0$ as $n\to+\infty$ in the trace norm for any $m$ in $M^s_{\rho_0}$ and
any sequence $\{\rho_n\}$ in $\S(\H)$ converging to a state $\rho_0$. Indeed, by the Mirsky inequality
we have
\begin{equation}\label{W-th}
  \sum_i|\lambda^{\rho_n}_i-\lambda^{\rho_0}_i|\leq \|\rho_n-\rho_0\|_1.
\end{equation}
Assume first that $\lambda^{\rho_0}_{m+1}< \lambda^{\rho_0}_m$. If $\,a=(\lambda^{\rho_0}_{m}+\lambda^{\rho_0}_{m+1})/2\,$ and $b=2$ then it follows from (\ref{W-th}) that
the spectral projector of $\rho_n$ corresponding to the interval $(a,b)$ coincides with $P_m^{\rho_n}$ for
all sufficiently large $n$ and for $n=0$. Thus, it follows from Theorem VIII.23 in \cite{R&S} that
$P_m^{\rho_n}$ tends to $P_m^{\rho_0}$ as $n\to+\infty$ in the norm topology. Since
$$
\|P_m^{\rho_n}\rho_n-P_m^{\rho_0}\rho_0\|_1\leq \|P_m^{\rho_n}(\rho_n-\rho_0)\|_1+\|(P_m^{\rho_n}-P_m^{\rho_0})\rho_0\|_1\leq \|\rho_n-\rho_0\|_1+\|P_m^{\rho_n}-P_m^{\rho_0}\|,
$$
this implies that $P_m^{\rho_n}\rho_n$ tends to $P_m^{\rho_0}\rho_0$ as $n\to+\infty$ in the trace norm.

If $\lambda^{\rho_0}_{m}=0$ then it follows from (\ref{W-th}) that $\sum_{i=m+1}^{+\infty}\lambda^{\rho_n}_{i}$
tends to zero and hence $\|(I_{\H}-P_m^{\rho_n})\rho_n\|_1$ vanishes  as $\,n\to+\infty$. This implies that
$P_m^{\rho_n}\rho_n$ tends to $\rho_0=P_m^{\rho_0}\rho_0$ in the trace norm.

Note that in general $P_m^{\rho_n}\rho_n$ does not tend to $P_m^{\rho_0}\rho_0$ as $\,n\to+\infty\,$ for arbitrary $m$.

If $\A=\{\rho_n\}\cup\{\rho_0\}$, where $\{\rho_n\}$ is a sequence converging to a state $\rho_0$, then
we may consider a simple modification of the above family.
Let
\begin{equation}\label{Psi-m-1+}
  \Psi_m(\rho)=P_{\widehat{m}(\rho_0)}^{\rho}\rho
\end{equation}
be a map from $\A$ to $\T_+(\H)$, where $\widehat{m}(\rho_0)$ is the maximal
number in $M^s_{\rho_0}$ not exceeding $m$ ($\widehat{m}(\rho_0)$ is well defined for all $m$ large enough). The above observations show that
the family of maps defined in (\ref{Psi-m-1+}) possesses all the properties described before Lemma \ref{DTL-L} with $M_{\rho_0}=\mathbb{N}\cap[m_0,+\infty)$ for some $m_0>0$.

2) Let $\{\rho_n\}$ and $\{\tau_n\}$ be sequences converging, respectively, to states $\rho_0$ and $\tau_0$  such that
$c\rho_n\leq\tau_n$ for some $c>0$. Assume that $\A=\{\tau_n\}\cup\{\tau_0\}$ and
\begin{equation}\label{Psi-m-2}
  \Psi_m(\tau_n)=c P_{\widehat{m}(\rho_0)}^{\rho_n}\rho_n+P_{\widehat{m}(\sigma_0)}^{\sigma_n}\sigma_n, \quad n\geq0,
\end{equation}
where $\sigma_n=\tau_n-c\rho_n$, $P_{\widehat{m}(\rho_0)}^{\rho_n}$ and $P_{\widehat{m}(\sigma_0)}^{\sigma_n}$
are the spectral
projectors of $\rho_n$ and $\sigma_n$ defined according to the rule described after (\ref{Psi-m-1+}).
It is clear that $\Psi_m(\tau_n)\leq \Psi_{m+1}(\tau_n)\leq\tau_n$ for all $m$ and $n\geq0$. It is also clear that $\Psi_m(\tau_n)$ tends to $\tau_n$ as $m\to+\infty$ for all $n\geq0$.
The arguments used in analysis of the family (\ref{Psi-m-1}) show that
$\Psi_m(\tau_n)$ tends to $\Psi_m(\tau_0)$ as $n\to+\infty$ for any $m\in \mathbb{N}\cap[m_0,+\infty)$, where $m_0$ is some natural number.

The list of families $\{\Psi_m\}$ that can be used in the study  of characteristics of quantum systems is not limited to the two families described above.
Look, for example, at the family $\Psi^{\vartheta}_m$ used in the proof of Theorem \ref{convex-m} in Section 5.2.3. To obtain results concerning uniform approximation of the output Holevo quantity of a quantum channel one should use the family of maps
$$
 \Psi_m:\; \sum_{k=1}^{+\infty} p_k\rho_k\otimes |k\rangle\langle k|\;\mapsto\;  \sum_{k=1}^{m} p_k\rho_k\otimes |k\rangle\langle k|
$$
defined on the set of quantum-classical states (see details in \cite{DTL}).

\subsubsection{Quantifying discontinuity of nonnegative LAA functions.}
The results of Section 4.2.1 allow to obtain useful expression for discontinuity jumps
of a nonnegative lower semicontinuous function $f$ on $\S(\H)$ satisfying inequalities (\ref{LAA-1}) and (\ref{LAA-2}). This expression
can be used, in particular, in proving conditions of local continuity of such functions.\smallskip

\begin{proposition}\label{DJ} \emph{Let $\{\rho_n\}$ be a sequence in $\S(\H)$ converging to a state $\rho_0$. Let $\{\Psi_m\}$
be a family of maps from $
\,\A=\{\rho_n\}\cup\{\rho_0\}$ to $\T_+(\H)$ possessing the properties described before Lemma \ref{DTL-L} and $M_{\rho_0}$ the corresponding countable subset of $\,\mathbb{N}_\A$ (see (\ref{N-def})).}

\emph{Let $f$ be a nonnegative lower semicontinuous function on $\S(\H)$ satisfying inequalities (\ref{LAA-1}) and (\ref{LAA-2}) such that $f(\rho_0)$ is finite and
one of the following conditions holds:
\begin{itemize}
  \item $\,\sup_{n\geq0}\rank\Psi_m(\rho_n)<+\infty\;$ for all $m\quad$ and $\quad\sup_{\varphi\in\H, \|\varphi\|=1}f(|\varphi\rangle\langle\varphi|)<+\infty$;
  \item for each $\,m\in M_{\rho_0}\cap\mathbb{N}_\A$ there exists
\begin{equation}\label{DJ-cond}
\lim_{n\to+\infty} f([\Psi_m(\rho_n)])=f([\Psi_m(\rho_0)])<+\infty,
\end{equation}
where $[\sigma]$ denotes the state proportional to a positive operator $\sigma$ (see (\ref{BN})).
\end{itemize}
Then
$$
\limsup_{n\to+\infty} f(\rho_n)-f(\rho_0)=\lim_{m\to+\infty,\, m\in M_{\rho_0}}\limsup_{n\to+\infty}\shs(1-\Tr\Psi_m(\rho_n))f([\rho_n-\Psi_m(\rho_n)]),
$$
where both sides may take the value $+\infty$ and it is assumed that $f([0])=0$. The above relation  remains valid with
$\,\lim\limits_{m\to+\infty,\, m\in M_{\rho_0}}\,$ replaced by $\;\limsup\limits_{m\to+\infty}$.}
\end{proposition}

\begin{remark}\label{DJ-cond-r}
To prove that
$$
\limsup_{n\to+\infty} f(\rho_n)-f(\rho_0)\geq\limsup_{m\to+\infty}\limsup_{n\to+\infty}\shs(1-\Tr\Psi_m(\rho_n))f([\rho_n-\Psi_m(\rho_n)])
$$
it suffices to assume that $f$ is a nonnegative lower semicontinuous function on $\S(\H)$
satisfying inequality (\ref{LAA-1}) such that $f(\rho_0)$ is finite (no additional conditions are necessary). If $\,f(\rho_n)<+\infty\,$ for all sufficiently large $n$ then this inequality directly follows from Proposition \ref{DTL+}. If $\,f(\rho_n)=+\infty\,$ for infinitely many $n$ then the l.h.s. of the above inequality is equal to $\,+\infty$.
\end{remark}\smallskip

\emph{Proof.} Denote the quantity $\,\limsup_{n\to+\infty} f(\rho_n)-f(\rho_0)$ by $A$.  By Remark \ref{DJ-cond-r} it suffices to prove that
\begin{equation}\label{DJ-1++}
A\leq\liminf_{m\to+\infty,\, m\in M_{\rho_0}}\limsup_{n\to+\infty}\shs(1-\Tr\Psi_m(\rho_n))f([\rho_n-\Psi_m(\rho_n)]).
\end{equation}
If $\,\sup_{\varphi\in\H, \|\varphi\|=1}f(|\varphi\rangle\langle\varphi|)<+\infty$ then
the function $f$ is continuous on any  set of states with bounded rank by Lemma \ref{S-m} below. So, to prove
the proposition it suffices to assume that condition (\ref{DJ-cond}) holds. Denote by $\rho_n^m$ and $\mu_n^m$  the state $[\Psi_m(\rho_n)]$
and the positive number $\Tr\Psi_m(\rho_n)$ for all $n\geq0$ and $m\in\mathbb{N}_{\A}$.

If $\,f(\rho_n)=+\infty\,$ for infinitely many $n$ then the r.h.s. of (\ref{DJ-1++}) is equal to $\,+\infty$.
Indeed, the condition (\ref{DJ-cond}) implies that $f(\rho_n^m)<+\infty$ for all sufficiently large $n$ for each $m$ in $M_{\rho_0}\cap\mathbb{N}_{\A}$.
By inequality (\ref{LAA-2}) this shows that $(1-\Tr\Psi_m(\rho_n))f([\rho_n-\Psi_m(\rho_n)])=+\infty\,$ for infinitely many $n$ for each $m$ in $M_{\rho_0}\cap\mathbb{N}_{\A}$.

Thus, we may assume in what follows that $f(\rho_n)<+\infty$ for all sufficiently large $n$.

Assume first that $A<+\infty$. Let $\varepsilon\in(0,1/2)$ be arbitrary. There is a countable subset $N_{\varepsilon}$ of $\N$ such that
\begin{equation}\label{DJ-two-1}
f(\rho_n)\geq f(\rho_0)+A-\varepsilon \quad \forall n\in N_{\varepsilon}.
\end{equation}
By Lemma \ref{DTL-L} there is $m_{\varepsilon}$ such that $\mu_n^m>1-\varepsilon$ for all $m\geq m_{\varepsilon}$ and all $n\geq0$. It follows form
inequality (\ref{LAA-1}) and assumption (\ref{a-b-assump}) that
\begin{equation}\label{DJ-one}
f(\rho_0^m)\leq (f(\rho_0)+a_f(\varepsilon))/(1-\varepsilon)\quad m\geq m_{\varepsilon}.
\end{equation}
By the condition (\ref{DJ-cond}) for each $m\in M_{\rho_0}$ there is $n_{\varepsilon,m}$ such that
\begin{equation}\label{DJ-two-2}
f(\rho^m_n)\leq f(\rho^m_0)+\varepsilon  \quad \forall n\geq n_{\varepsilon,m}.
\end{equation}
Inequality (\ref{LAA-2}) and inequalities (\ref{DJ-two-1}), (\ref{DJ-one}) and (\ref{DJ-two-2}) imply that
$$
(1-\mu_n^m)f([\rho_n-\mu_n^m\rho^m_n])\geq f(\rho_n)-f(\rho_n^m)-b_f(\varepsilon)\geq A-(\varepsilon f(\rho_0)+a_f(\varepsilon))/(1-\varepsilon)-b_f(\varepsilon)-2\varepsilon
$$
for all $m\in M_{\rho_0} \cap [m_{\varepsilon},+\infty)$ and all $n\in N_{\varepsilon} \cap [n_{\varepsilon,m},+\infty)$. Since the r.h.s. of this inequality can be made arbitrarily close to $A$ by choosing sufficiently small $\varepsilon$, it shows the validity of (\ref{DJ-1++}).

If $A=+\infty$ then by replacing (\ref{DJ-two-1}) with the inequality $f(\rho_n)\geq 1/\varepsilon$ and by repeating the above arguments
it is easy to show that the r.h.s. of (\ref{DJ-1++}) is equal to $\,+\infty$.
$\square$ \medskip

\begin{example}\label{DTL-e0} Let $f(\rho)=S(\rho)$ be the von Neumann entropy of a state $\rho$
and $\{\Psi_m\}$ the family of maps from $\S(\H)$ to $\T_+(\H)$ defined by formula (\ref{Psi-m-1}), i.e.
$\Psi_m(\rho)=P_m^{\rho}\rho$, where $P_m^{\rho}$ is the spectral projector of $\rho$ corresponding to its $m$ maximal eigenvalues.
In this case $M_{\rho}$ coincides with the set $M^s_{\rho}$ defined in (\ref{M-s}) and $\,\rank\Psi_m(\rho)\leq m<+\infty\,$ for all $m$ and any state $\rho$. So,
since the entropy is a nonnegative lower semicontinuous concave function on $\S(\H)$ that
satisfies inequality (\ref{S-LAA-2}) and equals to zero on the set of pure states, Proposition \ref{DJ} implies that
\begin{equation}\label{H-dj}
\limsup_{n\to+\infty}S(\rho_n)-S(\rho_0)=\lim_{m\to+\infty,\, m\in M^s_{\rho_0}}\limsup_{n\to+\infty}\shs r^m_n S(\sigma^m_n),
\end{equation}
for any sequence
$\{\rho_n\}$ converging to a state $\rho_0$ such that  $S(\rho_0)<+\infty$, where $\,r^m_n=1-\Tr P_m^{\rho_n}\rho_n$, $\,\sigma^m_n=(\rho_n-P_m^{\rho_n}\rho_n)/r^m_n\,$,
and $\,\lim\limits_{m\to+\infty,\, m\in M^s_{\rho_0}}$ can be replaced by $\,\limsup\limits_{m\to+\infty}$.

By using expression (\ref{H-dj}) one can easily prove many of the conditions of local continuity of the entropy
obtained in \cite[the Appendix]{Ruskai}, \cite{SSP}, \cite{W}. For example, it gives a simple proof of continuity
of the entropy on the set $\C_{H,E}$ consisting of all states $\rho$ in $\S(\H)$ such that $\Tr H\rho\leq E$
provided that the positive operator $H$ satisfies the Gibbs condition (\ref{H-cond}). Indeed, if $\{\rho_n\}$
is a sequence converging to a state $\rho_0$ such that $\Tr H\rho_n\leq E$ for all $n$ then
$\Tr H\sigma^m_n\leq E/r^m_n$ for all $n$ and $m$, where $\sigma^m_n$ and
$r^m_n$ are the state and the parameter defined after (\ref{H-dj}). Hence, assuming for simplicity that condition (\ref{star}) holds we obtain
$$
\limsup_{n\to+\infty}\shs r^m_n S(\sigma^m_n)\leq \limsup_{n\to+\infty}\shs r^m_n F_H(E/r^m_n)\leq \delta_m F_H(E/\delta_m),\quad \textrm{where}\;\; \delta_m=\sup_n r^m_n,
$$
$F_H(E)=S(\gamma_H(E))$ is the entropy of the Gibbs state (\ref{Gibbs}) and the second inequality follows from (\ref{W-L}).
By Lemma \ref{DTL-L} $\delta_m$ vanishes as $m\to+\infty$. So, since the Gibbs condition (\ref{H-cond}) is equivalent to (\ref{H-cond-a}), it implies that
the r.h.s. of the last inequality tends to zero as $m\to+\infty$. Thus, the r.h.s. of (\ref{H-dj})
is equal to zero. As $S(\rho_n)\leq F_H(E)$ for all $n$, this and the lower semicontinuity of the entropy implies that $\,\lim_{n\to+\infty}S(\rho_n)=S(\rho_0)<+\infty$.
\end{example}
\smallskip

By using the above arguments and a simple generalization of inequality (\ref{W-L}) to nonnegative functions satisfying inequality (\ref{LAA-1}) it is easy to obtain the following\smallskip

\begin{corollary}\label{DJ-c+} \emph{Let $f$ be a nonnegative lower semicontinuous function on $\S(\H)$
satisfying inequalities (\ref{LAA-1}) and (\ref{LAA-2}) that is bounded on the set of pure states. The function
$f$ is continuous on the set $\C_{H,E}\doteq\{\rho\in\S(\H)\,|\,\Tr H\rho\leq E\}$ for any $E>E_0$ if}
\begin{equation}\label{sp-c-cond}
\sup_{\rho\in\C_{H,E}}f(\rho)=o(E)\quad \textit{as} \;\; E\to+\infty.
\end{equation}
\end{corollary}
Since the entropy is not continuous on the set $\C_{H,E}$ if the operator $H$ does not satisfy the Gibbs condition (\ref{H-cond}) \cite[Proposition 1]{EC},
the equivalence of this condition and (\ref{H-cond-a}) shows the necessity of condition (\ref{sp-c-cond}) in Corollary \ref{DJ-c+}.\smallskip

\begin{lemma}\label{S-m} \emph{Let $f$ be a function on $\S(\H)$ satisfying inequalities (\ref{LAA-1}) and (\ref{LAA-2}) that is bounded on the set of pure states
in $\S(\H)$. Then $f$ is uniformly continuous on the set}
$$
\S_m(\H)=\{\rho\in\S(\H)\,|\,\rank\rho\leq m\}\quad \forall m\in\mathbb{N}.
$$
\end{lemma}

\emph{Proof.} By using inequality (\ref{LAA-2}) it is easy to show
that $|f(\rho)|\leq C_m$ for all $\rho$ in $\S_m(\H)$ and some $C_m<+\infty$. By this observation the uniform continuity of $f$ on $\S_m(\H)$ can be proved by the   Alicki-Fannes-Winter technique presented in the proof of Theorem \ref{AFW-1} (it suffices to note that for any different states $\rho$ and $\sigma$ in $\S_m(\H)$, the rank of the states
$\tau_+=\Delta^+(\rho,\sigma)$ and $\tau_-=\Delta^-(\rho,\sigma)$ does not exceed $2m-1$). $\square$
\smallskip

Proposition \ref{DJ} and Remark \ref{DJ-cond-r} imply the following\smallskip

\begin{corollary}\label{DJ-c} \emph{Let $\{\rho_n\}$ be a sequence in $\S(\H)$ converging to a state $\rho_0$ and $\{\Psi_m\}$
be a family of maps from $\,\A=\{\rho_n\}\cup\{\rho_0\}$ to $\T_+(\H)$ possessing the properties described before Lemma \ref{DTL-L}.}

\emph{Let $f$ be a function satisfying the condition of Proposition \ref{DJ} and $h$  a nonnegative lower semicontinuous function on $\S(\H)$
satisfying inequality (\ref{LAA-1}) such that $f(\rho)\leq h(\rho)$ for all $\rho$ in $\S(\H)$ and $h(\rho_0)<+\infty$. Then}
\begin{equation*}
\limsup_{n\to+\infty} f(\rho_n)-f(\rho_0)\leq\limsup_{n\to+\infty} h(\rho_n)-h(\rho_0).
\end{equation*}
\end{corollary}

\begin{example}\label{DJ-c-e1} The mutual information $I(\Phi,\rho)$ of a  quantum channel $\Phi:A\to B$ at a state $\rho$ in $\S(\H_A)$
defined in (\ref{MI-Ch-def}) is a lower semicontinuous nonnegative concave function of $\rho$. It is equal to zero on the set of pure states and satisfies inequality
(\ref{LAA-2}) with $b_f=2h_2$, where $h_2$ is the binary entropy \cite[Section 4.3]{MCB}. Since $I(\Phi,\rho)\leq2S(\rho)$ for any $\rho$ in $\S(\H_A)$, by using Corollary \ref{DJ-c} with the family $\{\Psi_m\}$ of maps defined in (\ref{Psi-m-1})
one can show that
\begin{equation*}
\limsup_{n\to+\infty} I(\Phi,\rho_n)-I(\Phi,\rho_0)\leq2\left(\limsup_{n\to+\infty} S(\rho_n)-S(\rho_0)\right)
\end{equation*}
for any sequence  $\{\rho_n\}$ in $\S(\H_A)$ converging to a state $\rho_0$ with finite entropy and arbitrary channel $\Phi$. It gives another way to show that continuity of
the function $\rho\mapsto S(\rho)$ on some set of states implies continuity of function $\rho\mapsto I(\Phi,\rho)$ on this set.
\end{example}

\subsubsection{Simon-type dominated convergence theorem and beyond.}
One of Simon's convergence theorems for the von Neumann entropy presented in \cite[the Appendix]{Ruskai} can be formulated as follows: \emph{if $\{\rho_n\}$
is a sequence of states converging to a state $\rho_0$ such that $\,c\rho_n\leq \tau$ for all $\,n\geq 0$, where $\tau$ is a state such that $\,S(\tau)<+\infty$ and $c>0$, then}
\begin{equation}\label{H-l-r}
\lim_{n\to+\infty }S(\rho_n)=S(\rho_0)<+\infty.
\end{equation}

This result called Simon's dominated convergence theorem for the entropy is strengthened in \cite{SSP}, where it is proved that relation (\ref{H-l-r})
holds for a sequence $\{\rho_n\}$ of states converging to a state $\rho_0$  provided that there exists a sequence $\{\tau_n\}$ of states converging to a state $\tau_0$
such that
$$
c\rho_n\leq \tau_n\;\; \textrm{for some} \;\,c>0\;\, \textrm{and all} \,\; n\geq 0\quad\textrm{ and }\quad  \lim_{n\to+\infty }S(\tau_n)=S(\tau_0)<+\infty.
$$

In fact, the above dominated convergence theorem for the entropy is a partial case of part A of the following
general theorem  proved by using Propositions \ref{DTL} and \ref{DJ}.\smallskip

\begin{theorem}\label{DCT-II} \emph{ Let $\S_0$ be a convex subset of $\,\S(\H)$ that
contains all finite rank states. Let $\{\rho_n\}$ and $\{\tau_n\}$  be sequences in $\S_0$ converging to states $\rho_0$ and $\tau_0$
in $\S_0$ such that $c\rho_n\leq \tau_n$ for some $\,c>0$ and all $\,n\geq0$. Let $f$ be a lower semicontinuous nonnegative function on $\S_0$ satisfying
inequalities (\ref{LAA-1}) and (\ref{LAA-2}) such that $f(\tau_n)$ and hence $f(\rho_n)$ are finite for all $n\geq0$.}

A) \emph{If the function  $f$ is bounded on the set of pure states
in $\,\S(\H)$ then
\begin{equation}\label{tau-rho-DJ}
\limsup_{n\to+\infty}f(\rho_n)-f(\rho_0)\leq \frac{1}{c} \left(\limsup_{n\to+\infty}f(\tau_n)-f(\tau_0)\right).
\end{equation}
In particular, if
\begin{equation}\label{DCT-II-tau}
\exists\lim_{n\to+\infty}f(\tau_n)=f(\tau_0)<+\infty
\end{equation}
then}
\begin{equation}\label{DCT-II-rho}
\exists\lim_{n\to+\infty}f(\rho_n)=f(\rho_0)<+\infty.
\end{equation}

B) \emph{Assume that  relation (\ref{DCT-II-tau})  holds and there is a function $G(x)$ on $\mathbb{R}_+$ tending to $G(0)=0$ as $x\to0^+$ such that
\begin{equation}\label{gen-con-II}
\limsup_{n\to+\infty} f(\rho'_n)-f(\rho'_0)\leq G\!\left(\limsup_{n\to+\infty} f(\tau'_n)-f(\tau'_0)\right)
\end{equation}
for any sequences $\{\rho'_n\}$ and $\{\tau'_n\}$ converging to states $\rho'_0$ and $\tau'_0$
such that $\rho'_n\leq 2\rho_n$, $\,c\rho'_n\leq 2\tau'_n\leq 4\tau_n$ for all $\,n\geq0$ and $\,\sup_n\rank\tau'_n<+\infty$. Then relation (\ref{DCT-II-rho}) is valid.}
\end{theorem}\smallskip

\begin{remark}\label{DCT-II-r}
Condition (\ref{gen-con-II}) is well defined, since relation (\ref{DCT-II-tau}) and the assumed properties of the sequences
$\{\rho'_n\}$ and $\{\tau'_n\}$ imply, due to the lower semicontinuity of $f$ and inequality (\ref{LAA-1}), that $f(\sigma'_0)\leq\limsup_{n\to+\infty} f(\sigma'_n)<+\infty$, $\sigma=\rho,\tau$. By the proof of Theorem \ref{DCT-II}B presented below  it suffices to require that condition (\ref{gen-con-II}) holds
for the sequences  $\{\rho'_n=\rho_n^m\}$ and $\{\tau'_n=\tau_n^m\}$ introduced in that proof.
\end{remark}\smallskip

\emph{Proof of Theorem \ref{DCT-II}.} Let $\{\Psi_m\}$ be the family of maps from $\A=\{\tau_n\}\cup\{\tau_0\}$ to $\T_+(\H)$
defined by formula (\ref{Psi-m-2}). In this case $M_{\tau_0}=\mathbb{N}\cap[m_*,+\infty)$ for some $m_*>0$.

 Let $\sigma_n=\tau_n-c\rho_n$ for all $n\geq 0$. Let $\mu_n^m=\Tr P^{\rho_n}_{\widehat{m}(\rho_0)}\rho_n$ and $\nu_n^m=\Tr\Psi_m(\tau_n)$ for all $n$ and $m$ (the notation is explained after (\ref{Psi-m-2})). By Lemma \ref{DTL-L} we have
\begin{equation}\label{mu-nu}
\lim_{m\to+\infty}\inf_{n\geq0}\mu^m_n=\lim_{m\to+\infty}\inf_{n\geq0}\nu^m_n=1.
\end{equation}

By relation (\ref{mu-nu}) there is $m_0\geq m_*$ such that $\mu_n^m>1/2$ for all $n\geq0$ and all $m\geq m_0$.  Denote by $\rho_n^m$ and $\tau_n^m$ the states proportional to the operators $P^{\rho_n}_{\widehat{m}(\rho_0)}\rho_n$ and
$\Psi_m(\tau_n)=cP^{\rho_n}_{\widehat{m}(\rho_0)}\rho_n+\sigma^m_n$ for all $n\geq0$ and $m\geq m_0$,
where  $\sigma^m_n=P_{\widehat{m}(\sigma_0)}^{\sigma_n}\sigma_n$.

To prove part A note first that the finiteness of $f(\tau_0)$, the equality $\tau_0=c\rho_0+\sigma_0$, where $\sigma_0\geq0$,
and inequality (\ref{LAA-1}) imply the finiteness of $f(\rho_0)$.

Since $\sup_n\rank\rho_n^m$ and $\sup_n\rank\tau_n^m$ are finite for each $m$,
Proposition \ref{DJ}  implies that
\begin{equation}\label{tau-DJ}
\limsup_{n\to+\infty}f(\tau_n)-f(\tau_0)=\lim_{m\to+\infty}\limsup_{n\to+\infty}\shs(1-\nu^m_n)f([\tau_n-\nu_n^m\tau_n^m]),
\end{equation}
\begin{equation}\label{rho-DJ}
\limsup_{n\to+\infty}f(\rho_n)-f(\rho_0)=\lim_{m\to+\infty}\limsup_{n\to+\infty}\shs(1-\mu^m_n)f([\rho_n-\mu_n^m\rho_n^m]).
\end{equation}

Since $\tau_n=c\rho_n+\sigma_n$ and $\nu_n^m\tau_n^m=c\mu_n^m\rho_n^m+\sigma_n^m$, we have
\begin{equation}\label{imp-dec}
(1-\nu_n^m)[\tau_n-\nu_n^m\tau_n^m]=c(1-\mu_n^m)[\rho_n-\mu_n^m\rho_n^m]+(\sigma_n-\sigma_n^m).
\end{equation}
So, it follows from inequality (\ref{LAA-1}) and the nonnegativity
of $f$ that
\begin{equation}\label{tau-n}
c(1-\mu_n^m)f([\rho_n-\mu_n^m\rho_n^m])\leq(1-\nu_n^m)(f([\tau_n-\nu_n^m\tau_n^m])+a^*_f).
\end{equation}
where $a^*_f=\max_{x\in[0,1]}a_f(x)$. Note that (\ref{tau-n}) holds trivially if $\nu_n^m=1$, since in this case $\mu_n^m=1$. This inequality, relation (\ref{mu-nu}) and representations (\ref{tau-DJ}) and (\ref{rho-DJ}) imply (\ref{tau-rho-DJ}).
The last claim  follows from (\ref{tau-rho-DJ}) by the lower semicontinuity of $f$.

To prove part B assume that (\ref{DCT-II-tau}) holds. By Proposition \ref{DTL} this implies existence of vanishing sequences $\{\alpha_m\}$ and $\{\beta_m\}$ such that
\begin{equation}\label{two-rel}
|f(\tau_n)-f(\tau_n^m)|\leq\alpha_m\quad \textrm{and} \quad (1-\nu_n^m)|f([\tau_n-\nu_n^m\tau_n^m])|\leq\beta_m
\end{equation}
for all $n\geq 0$ and $m>m_0$.  Since $\tau_n=c\rho_n+\sigma_n=c\mu_n^m\rho^m_n+c(\rho_n-\mu_n^m\rho^m_n)+\sigma_n$, by using the boundedness of the sequence $\{f(\tau_n)\}_{n\geq0}$ and inequality (\ref{LAA-1}) it is easy to show the boundedness of the sequence $\{f(\rho_n)\}_{n\geq0}$ and of the double sequence $\{f(\rho_n^m)\}_{n\geq0,\shs m\geq m_0}$. By using inequalities (\ref{LAA-1}) and (\ref{LAA-2}) it is easy to obtain
$$
|f(\rho_n)-f(\rho_n^m)|\leq (1-\mu_n^m)(f([\rho_n-\mu_n^m\rho_n^m])+f(\rho_n^m))+(a_f+b_f)(1-\mu_n^m)\},
$$
for all $n\geq0,\shs m\geq m_0$. This inequality, inequality (\ref{tau-n}), the second inequality in (\ref{two-rel}) and relation (\ref{mu-nu}) imply, by the boundedness of the sequence $\{f(\rho_n^m)\}_{n\geq0,\shs m\geq m_0}$
mentioned before, the existence of a vanishing sequence $\{\gamma_m\}$ such that
\begin{equation}\label{gamma}
|f(\rho_n)-f(\rho_n^m)|\leq\gamma_m\quad\forall n\geq 0\;\, \forall m\geq m_0.
\end{equation}

Since the sequence $\{\tau^m_n\}_n$ tends to the state
$\tau^m_0$ for all $m\geq m_0$, limit relation (\ref{DCT-II-tau}) and the first inequality in (\ref{two-rel}) show that
\begin{equation}\label{tau-gap}
A_m\doteq\limsup_{n\to+\infty}f(\tau^m_n)-f(\tau^m_0)\leq 2\alpha_m.
\end{equation}
By  relation (\ref{mu-nu}) we have  $\,2\tau_n^m\geq 2c(\mu_n^m/\nu_n^m)\rho_n^m\geq c\rho_n^m\,$ for all $n\geq0$ and all sufficiently large $m$. Hence, the condition (\ref{gen-con-II}) implies that
$$
\limsup_{n\to+\infty}f(\rho^m_n)-f(\rho^m_0)\leq G(A_m)
$$
for all such $m$. Thus, by using (\ref{gamma}) we obtain
$$
\limsup_{n\to+\infty}f(\rho_n)-f(\rho_0)\leq 2\gamma_m+G(A_m).
$$
It follows from (\ref{tau-gap}) that $G(A_m)=o(1)$ as $m\to+\infty$. So, the above inequality and the lower semicontinuity of $f$ imply (\ref{DCT-II-rho}).  $\square$\smallskip

\begin{example}\label{DCT-e1} The mutual information $I(\Phi,\rho)$ of a  quantum channel $\Phi:A\to B$ at a state $\rho$ in $\S(\H_A)$
defined in (\ref{MI-Ch-def}) is a lower semicontinuous nonnegative concave function of $\rho$. It is equal to zero on the set of pure states and satisfies inequality
(\ref{LAA-2}) with $b_f=2h_2$, where $h_2$ is the binary entropy \cite[Section 4.3]{MCB}.
So, relation (\ref{tau-rho-DJ}) and the implication (\ref{DCT-II-tau})$\Rightarrow$(\ref{DCT-II-rho}) is valid for the
function $f(\rho)=I(\Phi,\rho)$ by Theorem \ref{DCT-II}A.

This observation generalizes the dominated
convergence theorem for the von Neumann entropy mentioned at the begin of this subsection,
since $I(\Phi,\rho)=2S(\rho)$ if $\Phi=\id_{\H}$ is the identity channel.
\end{example}\medskip

Theorem \ref{DCT-II}B  is used essentially in the proof of dominated convergence theorem for the quantum mutual information (Proposition \ref{DCT-MI}A
in Section 5.2.3).\medskip

Let $f$ be a nonnegative lower semicontinuous function satisfying the conditions of Theorem \ref{DCT-II}B. If
$\,\{\rho_n\}$ and $\,\{\sigma_n\}$ are sequences of states converging, respectively,
to states  $\rho_0$ and $\sigma_0$ such that
\begin{equation}\label{23-one}
\lim_{n\to+\infty}f(p_n\rho_n+\bar{p}_n\sigma_n)=f(p_0\rho_0+\bar{p}_0\sigma_0)<+\infty,\quad  \bar{p}_n=1-p_n,
\end{equation}
for some sequence $\,\{p_n\}$ of numbers in $[0,1]$  converging to  $p_0\in(0,1)$ then Theorem \ref{DCT-II}B implies that
\begin{equation}\label{23-two}
\lim_{n\to+\infty} f(\rho_n)=f(\rho_0)<+\infty\quad\textrm{and}\quad\lim_{n\to+\infty} f(\sigma_n)=f(\sigma_0)<+\infty.
\end{equation}
The following theorem gives conditions under which the validity of both relations in (\ref{23-two})  implies the validity of (\ref{23-one})
for any sequence $\,\{p_n\}\subset[0,1]$  converging to  $p_0\in[0,1]$.\smallskip

\begin{theorem}\label{convex-m}
 \emph{ Let $\S_0$ be a convex subset of $\,\S(\H)$ and $\widetilde{\S}_0$ the cone generated by $\S_0$. Let $f$ be a nonnegative lower semicontinuous function on $\S_0$ satisfying inequalities (\ref{LAA-1}) and (\ref{LAA-2}). Let $\,\{\rho_n\}$ and $\,\{\sigma_n\}$ be sequences of states in $\S_0$ converging, respectively,
to states  $\rho_0$ and $\sigma_0$ in $\S_0$ such that  both limit relations in (\ref{23-two}) hold.
If there exist  families $\{\Psi_m^{\rho}\}$ and $\{\Psi_m^{\sigma}\}$ of maps from $\,\A_{\rho}\doteq\{\rho_n\}_{n\geq0}$ and $\,\A_{\sigma}\doteq\{\sigma_n\}_{n\geq0}$ to $\widetilde{\S}_0$
satisfying the conditions stated at the begin of Section 4.2.1 with the sets $M_{\rho_0}$ and $M_{\sigma_0}$  such that $\,\mathrm {card}(M_{\rho_0}\cap M_{\sigma_0})=+\infty\,$ and
\begin{equation}\label{sum-rel}
\lim_{n\to+\infty}f(\tau^m_n)=f(\tau^m_0)<+\infty,\quad \tau^m_n=p_n[\Psi_m^{\rho}(\rho_n)]+\bar{p}_n[\Psi_m^{\sigma}(\sigma_n)],
\end{equation}
for any sequence $\,\{p_n\}\subset[0,1]$  converging to $p_0\in[0,1]$ and all sufficiently large $m$ in $M_{\rho_0}\cap M_{\sigma_0}$, where $\,\bar{p}_n=1-p_n$, then  (\ref{23-one})  holds
for any sequence $\,\{p_n\}\subset[0,1]$  converging to $p_0\in[0,1]$.}
\end{theorem}
\medskip

\emph{Proof.} Let  $\mu_n^m=\Tr \Psi_m^{\rho}(\rho_n)$ and $\nu_n^m=\Tr \Psi_m^{\sigma}(\sigma_n)$.  By Lemma \ref{DTL-L} we have
\begin{equation}\label{mu-nu+gc}
\lim_{m\to+\infty}\inf_{n\geq0}\mu^m_n=\lim_{m\to+\infty}\inf_{n\geq0}\nu^m_n=1.
\end{equation}
Denote by $\rho_n^m$ and $\sigma_n^m$ the states $(\mu_n^m)^{-1}\Psi_m^{\rho}(\rho_n)$ and $(\nu_n^m)^{-1}\Psi_m^{\sigma}(\sigma_n)$ for all sufficiently large $m$.

By applying Proposition \ref{DTL}
to the function $f$, we conclude that the limit relations in (\ref{23-two})
imply that
\begin{equation}\label{2-r-e-gc}
\begin{array}{l}
 \displaystyle\lim_{m\to+\infty}\sup_{n\geq0}(1-\mu_n^m)f([\rho_n-\mu_n^m\rho_n^m])=0,\\\displaystyle\lim_{m\to+\infty}\sup_{n\geq0}(1-\nu_n^m)f([\sigma_n-\nu_n^m\sigma_n^m])=0
\end{array}
\end{equation}
and that
\begin{equation}\label{2-rel+gc}
\sup_{n\geq0}f(\rho^m_n),\; \sup_{n\geq0}f(\sigma^m_n)\leq C<+\infty
\end{equation}
for all sufficiently large $m$.

Let $\vartheta_n=p_n\rho_n+\bar{p}_n\sigma_n$ for a given arbitrary sequence $\,\{p_n\}\subset[0,1]$  converging to $p_0\in[0,1]$. Consider the family $\Psi^{\vartheta}_m$ of maps from $\,\A_{\vartheta}\doteq\{\vartheta_n\}_{n\geq0}$ to $\widetilde{\S}_0$ defined as
$$
\Psi^{\vartheta}_m(\vartheta_n)=p_n\Psi^{\rho}_m(\rho_n)+\bar{p}_n\Psi^{\sigma}_m(\sigma_n)=p_n\mu_n^m\rho^m_n+\bar{p}_n\nu_n^m\sigma^m_n
$$
It satisfies all the conditions stated at the begin of Section 4.2.1 with
$M_{\vartheta_0}\doteq M_{\rho_0}\cap M_{\sigma_0}$.

It follows from (\ref{mu-nu+gc}) that condition (\ref{sum-rel}) implies
\begin{equation}\label{2-rel++gc}
\lim_{n\to+\infty}f([\Psi^{\vartheta}_m(\vartheta_n)])=f([\Psi^{\vartheta}_m(\vartheta_0)])<+\infty
\end{equation}
for all sufficiently large $m$ in $M_{\vartheta_0}$.

Since the function $f$ satisfies  inequality (\ref{LAA-1}) and inequality (\ref{LAA-2}), we have
$$
\begin{array}{ccc}
  |f(\vartheta_n)-f([\Psi^{\vartheta}_m(\vartheta_n)])|\leq \delta^m_n (f([\vartheta_n-\Psi^{\vartheta}_m(\vartheta_n)])+f([\Psi^{\vartheta}_m(\vartheta_n)]))+(a_f+b_f)(\delta^m_n)\\\\
  \leq p_n(1-\mu_n^m)f([\rho_n-\mu_n^m\rho_n^m])+\bar{p}_n(1-\nu_n^m)f([\sigma_n-\nu_n^m\sigma_n^m])+\delta^m_nb^*_f\\\\
  +p_n\mu_n^m \delta^m_n(1-\delta^m_n)^{-1}f(\rho_n^m)+\bar{p}_n\nu_n^m\delta^m_n(1-\delta^m_n)^{-1}f(\sigma_n^m)+\delta^m_nb^*_f+(a_f+b_f)(\delta^m_n),
\end{array}
$$
where $\delta^m_n=1-p_n\mu^m_n-\bar{p}_n\nu^m_n$ and $b^*_f=\max_{x\in[0,1]}b_f(x)$, for all $n\geq0$. Thus, it follows from (\ref{mu-nu+gc}), (\ref{2-r-e-gc})  and (\ref{2-rel+gc}) that
$$
 \lim_{m\to+\infty,\, m\in M_{\vartheta_0}}\sup_{n\geq0}|f(\vartheta_n)-f([\Psi^{\vartheta}_m(\vartheta_n)])|=0.
$$
This and (\ref{2-rel++gc}) imply (\ref{23-one}). $\square$ \medskip

\begin{example}\label{convex-m-ex}
By using Theorem \ref{convex-m} it is easy to show that
$$
\lim_{n\to+\infty}S(p_n\rho_n+\bar{p}_n\sigma_n)=S(p_0\rho_0+\bar{p}_0\sigma_0)<+\infty,\quad \bar{p}_n=1-p_n,
$$
for any sequences $\,\{\rho_n\}$ and $\,\{\sigma_n\}$  of states in $\S(\H)$ converging, respectively,
to states  $\rho_0$ and $\sigma_0$ and any sequence $\,\{p_n\}\subset[0,1]$  converging to $p_0\in[0,1]$ provided that
$$
\lim_{n\to+\infty}S(\rho_n)=S(\rho_0)<+\infty\quad \textup{and}
\quad \lim_{n\to+\infty}S(\sigma_n)=S(\sigma_0)<+\infty.
$$
Indeed, let $\Psi_m^{\rho}(\rho_n)=P_{\widehat{m}(\rho_0)}^{\rho_n}\rho_n$ and $\Psi_m^{\sigma}(\sigma_n)=P^{\sigma_n}_{\widehat{m}(\sigma_0)}\sigma_n$,
where $P_{\widehat{m}(\rho_0)}^{\rho_n}$ and $P_{\widehat{m}(\sigma_0)}^{\sigma_n}$
are the spectral projectors of $\rho_n$ and $\sigma_n$ defined according to the rule described after (\ref{Psi-m-1+}).
Then all the conditions of Theorem \ref{convex-m} are satisfied with $f=S$, since the von Neumann entropy is a concave
lower semicontionuous nonnegative function on $\S(\H)$ that is continuous on any set of states with bounded rank and satisfies inequality (\ref{LAA-2}) with $b_f=h_2$.

This result was proved previously in \cite{SSP} by using rather sophisticated technique.
\end{example}

Theorem \ref{convex-m}  is used essentially in the proof of Proposition \ref{DCT-MI}
in Section 5.2.3.

\section{Concrete characteristics of composite quantum systems}

\subsection{Quantum conditional entropy}

\subsubsection{Definitions and the Alicki-Fannes-Winter continuity bound.} The quantum conditional entropy (QCE) of a state $\omega$ of a finite-dimensional bipartite system $AB$ defined as
\begin{equation}\label{ce-def}
S(A|B)_{\omega}=S(\omega)-S(\omega_{B})
\end{equation}
plays essential role in analysis of quantum systems \cite{H-SCI,Wilde}.
In contrast to the classical conditional entropy the QCE
may take negative values but its absolute value is bounded above by the entropy of $\omega_A$, i.e.
\begin{equation}\label{ce-ub}
|S(A|B)_{\omega}|\leq S(\omega_A).
\end{equation}
The QCE $S(A|B)_{\omega}$ is a concave function of a state $\omega$ \cite{H-SCI,Wilde}. By using concavity of the von Neumann entropy and inequality (\ref{S-LAA-2}) it is easy to show that
\begin{equation}\label{ce-LAA-2}
S(A|B)_{p\rho+(1-p)\sigma}\leq p S(A|B)_{\rho}+(1-p)S(A|B)_{\sigma}+h_2(p)
\end{equation}
for any states  $\rho$ and $\sigma$ in $\S(\H_{AB})$ and $p\in[0,1]$, where $\,h_2\,$ is the binary entropy.

Definition (\ref{ce-def}) remains valid for a state $\omega$ of an infinite-dimensional bipartite system $AB$
with finite marginal entropies
$S(\omega_A)$ and $S(\omega_B)$  (since the finiteness of $S(\omega_A)$ and $S(\omega_B)$ are equivalent to the finiteness of $S(\omega)$ and $S(\omega_B)$).
For a state $\omega$ with finite $S(\omega_A)$ and arbitrary $S(\omega_B)$ one can define the QCE
by the formula
\begin{equation}\label{ce-ext}
S(A|B)_{\omega}=S(\omega_{A})-D(\omega\shs\Vert\shs\omega_{A}\otimes\omega_{B})
\end{equation}
proposed and analysed by Kuznetsova in \cite{Kuz} (the finiteness of $S(\omega_{A})$ implies the finiteness of $D(\omega\shs\Vert\shs\omega_{A}\otimes\omega_{B})$ by inequality (\ref{MI-UB}) in Section 5.2.1). The QCE  extented by the above formula to the convex set $\,\{\omega\in\S(\H_{AB})\,|\,S(\omega_A)<+\infty\}\,$ possesses all basic properties of the QCE valid in finite dimensions \cite{Kuz,CMI}. In particular, it is concave and satisfies inequalities (\ref{ce-ub}) and (\ref{ce-LAA-2}). So, it belongs to the class $L_2^1(2,1)$ in terms of Section 3.1.2 (in the settings $A_1=A$, $A_2=B$).

An important step in quantitative continuity analysis of characteristics of quantum systems was made by
Alicki and Fannes in 2004, who obtained a continuity  bound for the QCE  $S(A|B)$ depending only on the dimension of system $A$
and the trace norm distance between quantum states \cite{A&F}. The method used by Alicki and Fannes was then modified by different authors \cite{SR&H,M&H}.
A  final (optimal, in a sense) version of this method (described in Section 3.1) was proposed by Winter in \cite{W-CB} and applied to optimize
the Alicki-Fannes  continuity bound for the QCE. The resulting continuity bound for the extended QCE defined in (\ref{ce-ext}) and its refinements for special cases are presented in the following\smallskip

\begin{proposition}\label{AFW-CB} \cite{W-CB} \emph{Let $A$ be a finite-dimensional system and $B$ an arbitrary system. Then
\begin{equation}\label{AFW-CB+}
|S(A|B)_{\rho}-S(A|B)_{\sigma}|\leq 2\varepsilon \ln \dim\H_A+g(\varepsilon)
\end{equation}
for any states $\rho$ and $\sigma$ in $\S(\H_{AB})$ such that $\frac{1}{2}\|\rho-\sigma\|_1\leq\varepsilon$, where $S(A|B)$ is the extended QCE defined in (\ref{ce-ext}) and $\,g(x)$ is the function defined in (\ref{g-def}).}

\emph{Factor $2$ in the right hand side of (\ref{AFW-CB+}) can be omitted if $\rho$ and $\sigma$ are either quantum-classical (q-c) states, i.e.
\begin{equation}\label{qc-states}
\rho=\sum_{i} p_i\, \rho_i\otimes |i\rangle\langle i|\quad \textrm{and}\quad \sigma=\sum_{i} q_i\, \sigma_i\otimes |i\rangle\langle i|,
\end{equation}
where $\{p_i,\rho_i\}$ and $\{q_i,\sigma_i\}$ are ensembles of states in $\S(\H_A)$ and $\{|i\rangle\}$ a fixed orthonormal basis in $\H_B$,
or classical-quantum (c-q) states, i.e.
\begin{equation}\label{cq-states}
\rho=\sum_{i} p_i\, |i\rangle\langle i| \otimes \rho_i\quad \textrm{and}\quad \sigma=\sum_{i} q_i\, |i\rangle\langle i|\otimes \sigma_i,
\end{equation}
where $\{p_i,\rho_i\}$ and $\{q_i,\sigma_i\}$ are ensembles of states in $\S(\H_B)$ and $\{|i\rangle\}$ a fixed orthonormal basis in $\H_A$.}

\emph{Continuity bound (\ref{AFW-CB+}) and its  refinements  for q-c and c-q  states are asymptotically tight for large $\dim\H_A$ (Definition \ref{AT} in Section 3.2.1).}
\end{proposition}\smallskip

Continuity bound (\ref{AFW-CB+}) and its versions for q-c and c-q states were  proved in \cite{W-CB} for the QCE defined by formula (\ref{ce-def}), i.e. assuming that
this formula is well defined for the states $\rho$ and $\sigma$ and for the auxiliary states $\Delta^{\pm}(\rho,\sigma)$ involved in the AFW-technique.
The generalization of this result to the extended QCE  presented in Proposition \ref{AFW-CB} directly follows from
Theorem \ref{CB-L}. It suffices to note that the extended QCE belongs  to the classes $L^{1}_{2}(2,1)$, $L^{1}_{2}(1,1|\S_{\rm qc})$ and $L^{1}_{2}(1,1|\S_{\rm cq})$,
where $\S_{\rm qc}$ and $\S_{\rm cq}$ are, respectively, the sets of q-c states and  c-q states in $\S(\H_{AB})$. This follows from the nonnegativity of QCE
on the sets $\S_{\rm qc}$ and $\S_{\rm cq}$, which are convex and possess the $\Delta$-invariance property (\ref{Inv-P}).

The refinements of continuity bound (\ref{AFW-CB+}) for q-c and c-q  states presented in Proposition \ref{AFW-CB} are important, since states of these types are often used in analysis of information abilities of quantum systems and channels. By using definition (\ref{ce-ext}) of QCE one can show (see the proof of Corollary 3 in \cite{Wilde-CB}) that
$$
S(A|B)_{\rho}=\sum_ip_iS([\rho_i]_A)\quad \textrm{if}\quad\rho=\sum_{i} p_i \rho_i\otimes |i\rangle\langle i|
$$
and
$$
S(A|B)_{\rho}=H(\{p_i\})-\chi(\{p_i,\rho_i\})\quad \textrm{if}\quad\rho=\sum_{i} p_i |i\rangle\langle i| \otimes \rho_i,
$$
where $H(\{p_i\})$ is the Shannon entropy of the probability distribution $\{p_i\}$ and $\chi(\{p_i,\rho_i\})$ is the Holevo quantity of the ensemble $\{p_i,\rho_i\}$ (see Section 2).

\subsubsection{Wilde's optimal continuity bound for q-c states.}  In 2019  Alhejji and Smith obtained the optimal uniform continuity bound for the classical conditional entropy
\begin{equation}\label{AG-CB}
  |H(X|Y)_{p}-H(X|Y)_{q}|\leq \varepsilon \ln(|X|-1)+h_2(\varepsilon),
\end{equation}
where $h_2$ is the binary entropy and $\varepsilon$ is the total variation distance between probability distributions $p_{XY}$ and $q_{XY}$ \cite{A&G}. This continuity bound is valid if $\,\varepsilon\leq 1-1/|X|$.

Starting from this result and applying the special construction based on using a conditional dephasing channel Wilde obtained the optimal continuity bound
for the QCE on the set of quantum-classical states presented in the following\smallskip

\begin{proposition}\label{Wilde-CB} \cite{Wilde-CB} \emph{Let $A$ be a finite-dimensional system and $B$ an arbitrary system. Let
$\rho$ and $\sigma$ be q-c states in $\S(\H_{AB})$ defined in (\ref{qc-states}) such that  $\varepsilon=\frac{1}{2}\|\rho-\sigma\|_1\leq 1-1/\dim\H_A$.
Then
\begin{equation}\label{Wilde-CB+}
|S(A|B)_{\rho}-S(A|B)_{\sigma}|\leq \varepsilon \ln(\dim\H_A-1)+h_2(\varepsilon),
\end{equation}
where $S(A|B)$ in the extended QCE defined in (\ref{ce-ext}).}

\emph{The continuity bound (\ref{Wilde-CB+}) is optimal: for each $n$ and $\varepsilon\in[0,1-1/n]$ there exist q-c states $\rho$ and $\sigma$
in $\S(\H_{AB})$, where $\dim\H_A=n$, satisfying the above conditions such that $"="$ holds in (\ref{Wilde-CB+}).}
\end{proposition}\smallskip

The optimality of continuity bound (\ref{Wilde-CB+}) can be shown by considering the q-c states
$\rho$ and $\sigma$ defined in (\ref{qc-states}), in which $p_i=q_i=1/n$, $\{\rho_i\}$ is a collection  of $n$ mutually orthogonal pure states,  $\sigma_i=(1-p)\rho_i+(p/n)I_A$, $p=\varepsilon/(1-1/n)$, $i=\overline{1,n}$.  Then it is easy to see that $\,\frac{1}{2}\sum_{i=1}^n p_i \|\rho_i-\sigma_i\|_1=\varepsilon\,$ and that
$$
S(A|B)_{\sigma}-S(A|B)_{\rho}=S((1-p)\rho_i+(p/n)I_A)=\varepsilon\ln (n-1)+h_2(\varepsilon)
$$

Continuity bound (\ref{Wilde-CB+}) improves the continuity bound for q-c states given by Proposition \ref{AFW-CB}. It allows to
refine the Nielsen-Winter continuity bound for the entanglement of formation (see Proposition \ref{EF-CB-1} in Section 5.4.1). Since
$$
S(A|B)_{\rho}-S(A|B)_{\sigma}=\chi(\{q_i,\sigma_i\})-\chi(\{p_i,\rho_i\})
$$
for any  q-c states $\rho$ and $\sigma$ in $\S(\H_{AB})$ defined in (\ref{qc-states}) provided that
$S(\sum_ip_i\rho_i)=S(\sum_iq_i\sigma_i)$,  Proposition \ref{Wilde-CB} gives optimal continuity bound
for the Holevo quantity of discrete ensembles with the same average states, which is used in
the proof of Proposition \ref{QD-CB-1} in Section 5.3.1 (the optimality of the last  continuity bound also follows from the above example, since the ensembles considered therein have the same average state).

\subsubsection{One-side optimal continuity bound for c-q states.}  By using the Alhejji-Smith continuity bound (\ref{AG-CB})
and the corresponding modification of the arguments from \cite{Wilde-CB} one can prove the following\smallskip

\begin{proposition}\label{hl-ub} \emph{Let $A$ be a finite-dimensional system and $B$ an arbitrary system. Let
$\rho$ and $\sigma$ be c-q states in $\S(\H_{AB})$ defined in (\ref{cq-states}) such that  $\varepsilon=\frac{1}{2}\|\rho-\sigma\|_1\leq 1-1/\dim\H_A$.
If all the states $\{\rho_i\}$ are mutually commuting (in particular, orthogonal) then
\begin{equation}\label{UB-1}
S(A|B)_\sigma-S(A|B)_\rho\leq \varepsilon\ln(\dim\H_A-1)+h_2(\varepsilon).
\end{equation}
The one-side continuity bound (\ref{UB-1}) is optimal: for each $n$ and $\varepsilon\in[0,1-1/n]$ there exist c-q states $\rho$ and $\sigma$
in $\S(\H_{AB})$, where $\dim\H_A=n$, satisfying the above conditions such that $"="$ holds in (\ref{UB-1}).}
\end{proposition}\medskip

\begin{remark}\label{hl-up-r2} Proposition \ref{hl-ub} \emph{does not assert} that
\begin{equation}\label{UB-1+}
|S(A|B)_\sigma-S(A|B)_\rho|\leq \varepsilon\ln(\dim\H_A-1)+h_2(\varepsilon)
\end{equation}
for any c-q states $\rho$ and $\sigma$ with the stated properties. But if we assume that all the states $\rho_i$ are \emph{mutually orthogonal} then
$S(A|B)_\rho=0$, so (\ref{UB-1+}) trivially follows from (\ref{UB-1}) and gives an upper bound on the nonnegative value of $S(A|B)_\sigma$. Note also
that Proposition \ref{hl-ub} implies the validity of (\ref{UB-1+}) in the case when both families $\{\rho_i\}$ and $\{\sigma_i\}$ consist of mutually commuting
states (states from different families may not commute).
\end{remark}\smallskip

\begin{remark}\label{hl-up+r} In terms of the Holevo quantity Proposition \ref{hl-ub} can be reformulated as follows: \emph{If $\,\{p_i,\rho_i\}$ and  $\,\{q_i,\sigma_i\}$ are ensembles of consisting of $\,n$  states on a separable Hilbert space such that all the states $\{\rho_i\}$ are mutually commuting (in particular, orthogonal) and $\,\varepsilon=\frac{1}{2}\sum_{i=1}^n  \|p_i\rho_i-q_i\sigma_i\|_1\leq 1-1/n\,$ then}
\begin{equation}\label{UB-1+++}
\chi(\{p_i,\rho_i\})-\chi(\{q_i,\sigma_i\})\leq \varepsilon\ln(n-1)+h_2(\varepsilon)+[H(\{p_i\})-H(\{q_i\})].\qquad
\end{equation}
\emph{The upper bound (\ref{UB-1+++}) is optimal: for each $n$ and $\varepsilon\in[0,1-1/n]$ there exist an ensemble
$\{p_i,\rho_i\}$ of $\,n$ orthogonal states and an ensemble
$\{q_i,\sigma_i\}$ of $\,n$ states  such that $\frac{1}{2}\sum_{i=1}^n \|p_i\rho_i-q_i\sigma_i\|_1=\varepsilon\,$ and $"="$ holds in (\ref{UB-1+++}).}
\smallskip

Inequality (\ref{UB-1+++}) can be interpreted  as follows. If we
consider $\{p_i,\rho_i\}$ as an original "non-perturbed" ensemble and $\{q_i,\sigma_i\}$ as
its perturbation (under classical or quantum noise) then the l.h.s. of  (\ref{UB-1+++}) is the
loss of Holevo information under this perturbation. Inequality (\ref{UB-1+++}) means that this loss is bounded above
by the loss of the Shannon entropy of the probability distributions plus the term
$\,\varepsilon\ln (n-1)+h_2(\varepsilon)\,$ depending on the metric divergence of the ensembles.
\end{remark}\smallskip

\emph{Proof of Proposition \ref{hl-ub}.} Assume first that the space $\H_B$ is finite-dimensional. Consider the c-q states (\ref{cq-states}).
Since all the states $\{\rho_i\}$ are mutually commuting, there is an orthonormal basis $\{|\phi_j\rangle\}$ of $\H_B$ such that
$\rho_i=\sum_{j}\lambda_{ij}|\phi_j\rangle\langle \phi_j|$.
Following \cite{Wilde-CB} consider the channel
$$
\Delta(\omega)=\sum_{i,j}[|i\rangle\langle i| \otimes|\phi_j\rangle\langle \phi_j|]\cdot \omega\cdot[|i\rangle\langle i| \otimes|\phi_j\rangle\langle \phi_j|]
$$
from $\S(\H_{AB})$ to itself. It is easy to see that
$$
\Delta(\rho)=\rho\quad \textrm{and}\quad \Tr_B\Delta(\sigma)=\Tr_B \sigma=\sum_{i=1}^n q_i|i\rangle\langle i|.
$$
Since $\Delta$ is a tensor product of local channels, the monotonicity of the quantum mutual information implies that
$I(A\!:\!B)_{\sigma}\geq I(A\!:\!B)_{\Delta(\sigma)}$. Since $\Tr_B \Delta(\sigma)=\Tr_B \sigma$,
this shows that
\begin{equation}\label{CE-one}
 S(A|B)_{\sigma}\leq S(A|B)_{\Delta(\sigma)}=H(X|Y)_{s},
\end{equation}
where $s$ denotes the joint probability distribution $s(i,j)=q_i\langle\phi_j|\sigma_i|\phi_j\rangle$.
Since $S(A|B)_{\rho}=H(X|Y)_r$,
where $r$ denotes the joint probability distribution $r(i,j)=p_i \lambda_{ij}$, it follows from (\ref{CE-one}) that
\begin{equation}\label{CE-two}
S(A|B)_{\sigma}-S(A|B)_{\rho}\leq H(X|Y)_s-H(X|Y)_r.
\end{equation}
By using the data processing inequality it is easy to show (cf.\cite{Wilde-CB}) that
the total variance distance between the probability distributions $r$ and $s$ does not exceed
$\varepsilon=\frac{1}{2}\|\rho-\sigma\|_1=\frac{1}{2}\sum_i\|p_i\rho_i-q_i\sigma_i\|_1$.
Thus, by using continuity bound (\ref{AG-CB}) we see that the r.h.s. of  (\ref{CE-two})
is less than or equal to $\,\varepsilon \ln(|X|-1)+h_2(\varepsilon)=\varepsilon \ln(\dim\H_A-1)+h_2(\varepsilon)$.

If $\H_B$ is an infinite-dimensional Hilbert space then validity of (\ref{UB-1}) can be proved by approximation.
Assume that $\varepsilon<1-1/n$ (if $\varepsilon=1-1/n$ then the r.h.s. of (\ref{UB-1}) is equal to $\ln n$ and inequality
(\ref{UB-1}) is obvious). For each natural $m$ let $P_m=\sum_{j=1}^m|\phi_j\rangle\langle \phi_j|$, where $\{|\phi_j\rangle\}$ is the orthonormal basis of $\H_B$ in
which all the states $\rho_i$ are diagonisable. We will assume that this basis is ordered in such a way that $P_m\rho_i\neq 0$ and $P_m\sigma_i\neq 0$ for all $i$ and $m>2n$. For each $m>2n$ consider the ensembles $\{p_i^m,\rho_i^m\}$ and $\{q_i^m,\sigma_i^m\}$ of $n$ states, where
$$
\begin{array}{c}
\rho^m_i=(1/r_i^m)P_m\rho_iP_m,\quad p^m_i=p_i r^m_i /r_m,\quad r^m_i=\Tr P_m\rho_i,\quad r_m=\sum_ip_ir^m_i ,\\\\
\sigma^m_i=(1/s_i^m)P_m\sigma_iP_m,\quad q^m_i=q_i s^m_i /s_m,\quad s^m_i=\Tr P_m\sigma_i,\quad s_m=\sum_iq_is^m_i.
\end{array}
$$
By monotonicity of the trace norm under the quantum operation $\rho\mapsto P_m \rho P_m$ we have
$$
\begin{array}{rl}
\displaystyle 2\varepsilon_m \doteq &\!\!\sum_i \|p^m_i\rho^m_i-q_i^m\sigma_i^m\|_1=\sum_i \|p_i P_{m}\rho_iP_{m}-(r_m/s_m) q_i P_{m}\sigma_iP_{m}\|_1/r_m
\\\\\displaystyle \leq & \!\!\sum_i \|p_i P_{m}\rho_iP_{m}-q_i P_{m}\sigma_iP_{m}\|_1/r_m+|1-s_m/r_m|\leq 2\varepsilon/r_m+|1-r_m/s_m|.
\end{array}
$$
Since $r_m$ and $s_m$ tends to $1$ as $m\to+\infty$, it follows that  $\varepsilon_m$ tends to $\varepsilon$ as $m\to+\infty$.
So, we may assume that $\varepsilon_m\leq 1-1/n$ for all sufficiently large $m$. Since  all the states $\rho_i^m$ are mutually commuting, the above part of the proof implies that
\begin{equation}\label{B-n}
\begin{array}{c}
  H(\{q^m_i\})-\chi(\{q^m_i,\sigma^m_i\})-H(\{p^m_i\})+\chi(\{p^m_i,\rho^m_i\})\\\\\leq \varepsilon_m\ln(n-1)+h_2(\varepsilon_m)
\end{array}
\end{equation}
for all sufficiently large $m$. It is clear that
$$
\lim_{m\to+\infty}H(\{p^m_i\})=H(\{p_i\})\quad \textrm{and} \quad \lim_{m\to+\infty}H(\{q^m_i\})=H(\{q_i\}).
$$
By using Corollary 17 in \cite{CHI} it is easy to show that
$$
\lim_{m\to+\infty}\chi(\{p^m_i,\rho^m_i\})=\chi(\{p_i,\rho_i\})\quad \textrm{and} \quad \lim_{m\to+\infty}\chi(\{q^m_i,\sigma^m_i\})=\chi(\{q_i,\sigma_i\}).
$$
Since $\,\varepsilon_m\to\varepsilon\,$ as $\,m\to+\infty$, the above limit relations and (\ref{B-n}) imply (\ref{UB-1}).

The optimality of one-side continuity bound (\ref{UB-1}) can be shown by considering the c-q states
$\rho$ and $\sigma$ defined in (\ref{cq-states}), where $p_i=q_i=1/n$, $\{\rho_i\}$ is a collection  of $n$ mutually orthogonal pure states,  $\sigma_i=(1-p)\rho_i+(p/n)\sum_i\rho_i$, $p=\varepsilon/(1-1/n)$, $i=\overline{1,n}$.  Then it is easy to see that
$\,\frac{1}{2}\sum_{i=1}^n p_i \|\rho_i-\sigma_i\|_1=\varepsilon\,$ and that
$$
S(A|B)_{\sigma}-S(A|B)_{\rho}=S\!\left((1-p)\rho_i+(p/n)\textstyle\sum_i\rho_i\right)=\varepsilon\ln (n-1)+h_2(\varepsilon).\quad\square
$$

\subsubsection{Continuity bound under the energy-type constraint on system $A$.}
Continuity bound for the function $\omega\mapsto S(A|B)_{\omega}$ under the energy-type constraint on
system $A$ determined  by a positive operator $H$ on $\H_A$ satisfying the Gibbs condition (\ref{H-cond}) can be obtained by using Theorem \ref{W-2-step}
in Section 3.2.1 with the family of maps
$$
\Lambda_m(\omega)=(\Tr P_m\omega_A)^{-1} P_m\otimes I_B\cdot\omega\cdot P_m\otimes I_B,
$$
where $P_m$ is the spectral projector of $H$ corresponding to the eigenvalues
$E_0,...,E_{m-1}$. In the following proposition we present this continuity bound in the original form  obtained by Winter in \cite{W-CB}.
\smallskip

\begin{proposition}\label{Winter-CB} \cite{W-CB} \emph{Let $H$ be a positive operator on the space $\H_A$ satisfying conditions (\ref{H-cond}) and (\ref{star}). Let $\rho$ and $\sigma$ be states in $\S(\H_{AB})$ such that $\Tr H\rho_A,\Tr H\sigma_A\leq E$ and
$\,\frac{1}{2}\|\rho-\sigma\|_1\leq \varepsilon<1$. Then
\begin{equation}\label{Winter-CB+}
\left|S(A|B)_{\rho}-S(A|B)_{\sigma}\right|\leq(2\varepsilon'+4\delta)F_H\left(E/\delta\right)
+g(\varepsilon')+2h_2(\delta),
\end{equation}
$\delta=(\varepsilon'-\varepsilon)/(1+\varepsilon')$, for any $\varepsilon'\in(\varepsilon,1]$, where $S(A|B)$ is the QCE defined in (\ref{ce-ext}), $F_H(E)$ is the function defined in (\ref{F-def}) and $g(x)$ is the function defined in (\ref{g-def}).}
\end{proposition}\smallskip

\begin{remark}\label{Winter-CB-r}
The quantity $\varepsilon'$ is a "free parameter" that can be used to optimize the continuity bound (\ref{Winter-CB+})
for given values of $E$ and $\varepsilon$. The faithfulness of this continuity bound follows from the equivalence of
(\ref{H-cond}) and (\ref{H-cond-a}), it can be shown by taking $\,\varepsilon'=\sqrt{\varepsilon}$.
\end{remark}

If $H$ is the grounded Hamiltonian of the $\ell$-mode quantum oscillator defined in (\ref{H-osc}) then
one can use the function $G_{\ell,\omega}$ defined in (\ref{F-osc}) as an upper bound for $F_H$. In this case continuity bound
(\ref{Winter-CB+}) with $F_H$ replaced by $G_{\ell,\omega}$ is asymptotically tight for large $E$ (Definition \ref{AT} in Section 3.2.1). This can be shown assuming that $\rho$ is a purification in $\S(\H_{AB})$ of the Gibbs state $\gamma_A(E)$ and $\sigma=(1-\varepsilon)\rho+\varepsilon \gamma_A(E)\otimes \tau$, where $\tau$ is any state in $\S(\H_B)$. Then $S(A|B)_{\sigma}-S(A|B)_{\rho}\geq 2\varepsilon F_H(E)=2\varepsilon G_{\ell,\omega}(E)+o(1)$ as $E\to+\infty$.

Since the function $\omega\mapsto S(A|B)_{\omega}$ belongs to the class $L_2^1(2,1)$ (in the settings $A_1=A$, $A_2=B$),
Theorem \ref{L-1-ip} with $m=1$ gives the following\smallskip

\begin{proposition}\label{Sh-CB} \emph{Let $H$ be a positive operator on $\H_A$ satisfying the condition (\ref{H-cond+}).   Let $\delta\in(0,1]$ and $\,g(x)$ be the function defined in (\ref{g-def}). Then
\begin{equation*}
    \left|S(A|B)_{\rho}-S(A|B)_{\sigma}\right|\leq 2\delta F_{H}\!\!\left[\frac{2E}{\delta^2}\right]+g(\delta)
\end{equation*}
for any states $\rho$ and $\sigma$ in $\S(\H_{AB})$ such that $\,\Tr H\rho_{A},\,\Tr H\sigma_{A}\leq E\,$ and}
$$
   \textit{either}\quad F(\rho,\sigma)\doteq\|\sqrt{\rho}\sqrt{\sigma}\|^2_1\geq 1-\delta^2\quad \textit{or} \quad \varepsilon(2-\varepsilon)\leq\delta^2,\quad \textit{where} \quad  \varepsilon=\textstyle\frac{1}{2}\|\rho-\sigma\|_1.
$$
\end{proposition}\smallskip

The continuity bound given by Proposition \ref{Sh-CB} is not asymptotically tight for large $E$ (in contrast to (\ref{Winter-CB+}))
in the case when $H$ is the  Hamiltonian of the $\ell$-mode quantum oscillator and
the function $G_{\ell,\omega}$ defined in (\ref{F-osc}) is used in the role of $F_H$. Nevertheless, it may be useful,
especially, if the fidelity $F(\rho,\sigma)$ is employed as a measure of closeness
of the states $\rho$ and $\sigma$ (see the proof of Proposition \ref{EF-CB-2} in Section 5.4.2).

\subsubsection{Condition of local continuity.}  If $\H_A$ is a finite-dimensional space then
the function $\omega\mapsto S(A|B)_{\omega}$  defined in (\ref{ce-ext}) is uniformly continuous on $\S(\H_{AB})$ by Proposition \ref{AFW-CB}.
But if the space $\H_A$ is infinite-dimensional then this function is not continuous on $\S(\H_{AB})$
regardless of the dimension of $\H_B$. In this case one can use the following simple sufficient condition of
its local continuity.\smallskip

\begin{proposition}\label{QCE-lc} \emph{Let $\{\omega_n\}\subset\S(\H_{AB})$ be a sequence converging to a state $\omega_0$. Then}
$$
\left\{\exists \lim_{n\to\infty}S([\omega_n]_A)=S([\omega_0]_A)<+\infty\right\}\;\;\Rightarrow\;\;\left\{\exists \lim_{n\to\infty}S(A|B)_{\omega_n}=S(A|B)_{\omega_0}<+\infty\right\}.
$$
\end{proposition}\smallskip
Proposition \ref{QCE-lc} directly follows from Proposition \ref{QCMI-lc} in Section 5.2.3.\pagebreak

\subsection{Quantum (conditional) mutual information}

\subsubsection{Definitions and the AFW-type continuity bounds.} The quantum mutual information (QMI) of a state $\omega$ of a finite-dimensional bipartite quantum system $AB$ is a quantity characterizing total correlation of this state, it is defined as
\begin{equation}\label{MI-e1}
I(A\!:\!B)_{\omega}=D(\omega\shs\Vert\shs\omega_{A}\otimes\omega_{B})=S(\omega_{A})+S(\omega_{B})-S(\omega),
\end{equation}
where the second formula is valid if $S(\omega)<+\infty$ \cite{L-mi}. This quantity is nonnegative and
\begin{equation}\label{MI-UB}
I(A\!:\!B)_{\omega}\leq 2\min\{S(\omega_A),S(\omega_B)\}
\end{equation}
for any state $\omega\in\S(\H_{AB})$. If $\omega$ is a separable state then factor 2 in (\ref{MI-UB}) can be omitted \cite{L-mi,MI-B,Wilde}.
These upper bounds and the below inequalities (\ref{MI-LAA-1}) and (\ref{MI-LAA-2}) with trivial system $C$ show that the function
$\omega\mapsto I(A\!:\!B)_{\omega}$ belongs to the classes $L_2^1(2,2)$ and $L_2^1(1,2|\S_{\rm qc})$, where
$\S_{\rm qc}$ is the set of q-c states in $\S(\H_{AB})$, in both settings $A_1=A,A_2=B$ and $A_1=B,A_2=A$ (we use the notation of Section 3.1.2).

The quantum conditional mutual information (QCMI) of a state $\omega$ of a finite-dimensional tripartite quantum system $ABC$  is defined by the formula
\begin{equation}\label{CMI-e1}
I(A\!:\!B|C)_{\rho}=S(\rho_{AC})+S(\rho_{BC})-S(\rho_{ABC})-S(\rho_{C}).
\end{equation}
This quantity is widely used in quantum information theory and has an operational interpretation as a
cost of a quantum state
redistribution protocol \cite{D&J}. It is nonnegative (cf.\cite{Ruskai}) and
\begin{equation}\label{CMI-UB}
I(A\!:\!B|C)_{\omega}\leq 2\min\{S(\omega_A),S(\omega_{AC}),S(\omega_B),S(\omega_{BC})\}
\end{equation}
for any state $\omega\in\S(\H_{ABC})$. This upper bound follows from upper bound (\ref{MI-UB})
and the representations
\begin{equation}\label{QCMI-rep}
\begin{array}{c}
  I(A\!:\!B|C)_{\omega}=I(AC\!:\!B)_{\omega}-I(B\!:\!C)_{\omega},\\
  I(A\!:\!B|C)_{\omega}=I(A\!:\!BC)_{\omega}-I(A\!:\!C)_{\omega}.
\end{array}
\end{equation}
If $\omega$ is a state of an infinite-dimensional tripartite quantum system $ABC$ then the r.h.s. of (\ref{CMI-e1}) and
of the above two representations  may contain the uncertainty $\infty-\infty$. In this case one can define the QCMI by one of the
following expressions (that are equivalent and coincide with the above formulae for any state $\omega$ at which these formulae are well defined)
\begin{equation}\label{QCMI-e1}
I(A\!:\!B|C)_{\rho}=\sup_{P_A}\left[\shs
I(A\!:\!BC)_{Q\rho Q}-I(A\!:\!C)_{Q\rho
Q}\shs\right]\!,\;\,Q=P_A\otimes I_{BC},
\end{equation}
\begin{equation}\label{QCMI-e2}
I(A\!:\!B|C)_{\rho}=\sup_{P_B}\left[\shs
I(AC\!:\!B)_{Q\rho Q}-I(B\!:\!C)_{Q\rho
Q}\shs\right]\!,\;\,Q=P_B\otimes I_{AC},
\end{equation}
where the suprema are taken over the sets of all finite rank projectors in
$\B(\H_A)$ and in $\B(\H_B)$ correspondingly and it is assumed that $I(X\!:\!Y)_{\sigma}=[\Tr
\sigma]I(X\!:\!Y)_{\sigma/\Tr\sigma}$ for any nonzero $\sigma$ in $\T_+(\H_{XY})$.

The QCMI defined by equivalent  expressions (\ref{QCMI-e1}) and (\ref{QCMI-e2}) is a nonnegative lower semicontinuous function on $\S(\H_{ABC})$ possessing all the basic properties of QMCI valid in the finite-dimensional case \cite[Theorem 2]{CMI}.

Since $I(A\!:\!B)$ is partial case of $I(A\!:\!B|C)$ corresponding to trivial system $C$, in what follows we will
formulate all the results valid for both these quantities in terms of $I(A\!:\!B|C)$ except for those that are proven only for $I(A\!:\!B)$.

The function $\omega\mapsto I(A\!:\!B|C)_{\omega}$ is not convex or concave, but it satisfies the inequalities
\begin{equation}\label{MI-LAA-1}
I(A\!:\!B|C)_{p\rho+(1-p)\sigma} \geq p I(A\!:\!B|C)_{\rho}+(1-p)I(A\!:\!B|C)_{\sigma}-h_2(p),
\end{equation}
\begin{equation}\label{MI-LAA-2}
I(A\!:\!B|C)_{p\rho+(1-p)\sigma} \leq p I(A\!:\!B|C)_{\rho}+(1-p)I(A\!:\!B|C)_{\sigma}+h_2(p)
\end{equation}
for any states $\rho$ and $\sigma$ in $\S(\H_{ABC})$ and  any $p\in(0,1)$  with possible values $+\infty$ in both sides.
If $\rho$ and $\sigma$ are states with finite marginal entropies then inequalities (\ref{MI-LAA-1}) and (\ref{MI-LAA-2}) can be proved by using
the concavity of the entropy and of the QCE along with inequalities (\ref{S-LAA-2}) and (\ref{ce-LAA-2}). Then one can use the approximation based on  Corollary 9 in \cite{CMI} to show the validity of these inequalities for arbitrary states $\rho$ and $\sigma$ in $\S(\H_{ABC})$.

Inequalities (\ref{MI-LAA-1}) and (\ref{MI-LAA-2}), the nonnegativity of QCMI  and
upper bound (\ref{CMI-UB}) show that the function $\omega\mapsto I(A\!:\!B|C)_{\omega}$  belongs to the class $L_3^1(2,2)$ in the notation of Section 3.1.2 (with $A_1=A$, $A_2=B$, $A_3=C$). Thus, Theorem \ref{CB-L} in Section 3.1.2
implies\smallskip
\begin{proposition}\label{CMI-CB} \cite{CHI} \emph{If $\,d=\min\{\dim\H_A,\dim\H_B\}<+\infty\,$ then
\begin{equation}\label{CMI-CB+}
 |I(A\!:\!B|C)_{\rho}-I(A\!:\!B|C)_{\sigma}|\leq 2 \ln d+2g(\varepsilon)
\end{equation}
for any states $\rho$ and $\sigma$ in $\S(\H_{ABC})$ such that $\;\frac{1}{2}\|\shs\rho-\sigma\|_1\leq\varepsilon$, where $g(x)$
is the function defined in (\ref{g-def}). Continuity bound (\ref{CMI-CB+}) is asymptotically tight for large $d$ (Definition \ref{AT} in Section 3.2.1).}
\end{proposition}\smallskip

The above result can be strengthened and generalized as follows. Let $A,B,C,D$ and $E$ be arbitrary quantum systems. Inequalities (\ref{CMI-UB}), (\ref{MI-LAA-1}), (\ref{MI-LAA-2}) and
\begin{equation}\label{I-monotone}
I(A\!:\!B|C)_{\omega}\leq I(A\!:\!BC)_{\omega}\leq I(AD\!:\!BCE)_{\omega},\quad \omega\in\S(\H_{ABCDE}),
\end{equation}
imply that the function $\omega\mapsto I(A\!:\!B|C)_{\omega}$ on $\S(\H_{ABCDE})$
belongs to the class $L^1_3(2,2)$ in the settings $A_1=[AD]$, $A_2=BE$, $A_3=C$, where $[AD]$
denotes the system corresponding to a given finite-dimensional subspace of $\H_{AD}$. Thus, by applying Theorem \ref{CB-L} in Section 3.1.2
one can obtain the first and the second claims of the following proposition which is used  in \cite{CID} to obtain continuity bounds for basic characteristics
of  quantum channels depending on their input dimension.
\smallskip

\begin{proposition}\label{CMI-CB-S} \cite{CID} \emph{If $\rho$ and $\sigma$ are states in $\S(\H_{ABCDE})$ such that $\;\frac{1}{2}\|\shs\rho-\sigma\|_1\leq\varepsilon$ and the supports of states  $\rho_{AD}$ and $\sigma_{AD}$ lie within some $d$-dimensional subspace of $\H_{AD}$ then $I(A\!:\!B|C)_{\rho}$ and $I(A\!:\!B|C)_{\sigma}$ are finite and
\begin{equation}\label{CMI-CB+S}
 |I(A\!:\!B|C)_{\rho}-I(A\!:\!B|C)_{\sigma}|\leq 2 \ln d+2g(\varepsilon).
\end{equation}
If $\rho$ and $\sigma$ are either q-c states or c-q states w.r.t. the decomposition (AD)(BCE) then the first factor $2$ in the r.h.s. of  (\ref{CMI-CB+S})
can be removed. If $\rho_{BC}=\sigma_{BC}$ then the second  factor $2$ in the r.h.s. of  (\ref{CMI-CB+S})
can be removed (independently of the first one).}
\end{proposition}\smallskip

The last claim of Proposition \ref{CMI-CB-S} is proved by using Corollary \ref{AFW-1-c} in Section 3.1.1 (see the proof of Lemma 2 in \cite{UFA}).

The QMI of a state $\omega$ of a $n$-partite system $A_1...A_n$ is defined as (cf. \cite{L-mi,Herbut,NQD})
\begin{equation}\label{MI-n-def}
     I(A_1\!:\!...\!:\!A_n)_{\omega}\doteq
    D(\omega\shs\|\shs\omega_{A_{1}}\otimes...\otimes\omega_{A_{n}})=\sum_{k=1}^n S(\omega_{A_{k}})-S(\omega),
\end{equation}
where the second formula is well defined if $S(\omega)<+\infty$. If  the  entropies $S(\omega_{A_1})$,..,$S(\omega_{A_{n}})$ are finite then the second formula in (\ref{MI-n-def}) implies that
\begin{equation}\label{MI-n-UB}
  I(A_1\!:\!...\!:\!A_n)_{\omega}\leq \sum_{k=1}^{n}S(\omega_{A_k}).
\end{equation}
The function $\,\omega\mapsto I(A_1\!:\!...\!:\!A_n)_{\omega}\,$ satisfies  inequalities (\ref{LAA-1}) and (\ref{LAA-2}) with  $a_f=1$ and $b_f=n-1$ \cite{CBM,CID}. So, the nonnegativity of the QMI and upper bound (\ref{MI-n-UB}) show that this function belongs to the class $L^{n}_{n}(1,n)$. Thus, Theorem \ref{CB-L}  in Section 3.1.2
implies the following\smallskip
\begin{proposition}\label{MI-n-CB} \cite{CBM} \emph{If $\,d_k\doteq\dim\H_{A_k}<+\infty\,$ for $\,k=\overline{1,n}\,$ then
\begin{equation}\label{MI-n-CB+}
 |I(A_1\!:\!...\!:\!A_n)_{\rho}-I(A_1\!:\!...\!:\!A_n)_{\sigma}|\leq \varepsilon \sum_{k=1}^n \ln d_k+ng(\varepsilon)
\end{equation}
for any states $\rho$ and $\sigma$ in $\S(\H_{A_1...A_n})$ such that $\;\frac{1}{2}\|\shs\rho-\sigma\|_1\leq\varepsilon$, where $g(x)$
is the function defined in (\ref{g-def}). If $d_1=...=d_n=d$ then continuity bound (\ref{MI-n-CB+}) is asymptotically tight for large $d$ (Definition \ref{AT} in Section 3.2.1).}
\end{proposition}\smallskip

The QCMI $I(A_1\!:\!...\!:\!A_n|C)_{\omega}$ of a state $\omega$ of a $(n+1)$-partite system $A_1...A_nC$ with finite marginal entropies can be defined by replacing all the entropies $S(\omega_{A_k})$
in the r.h.s. of (\ref{MI-n-def}) by the conditional entropies $S(A_k|C)_{\omega}$. To correctly define this quantity for any state $\omega$ in $\S(\H_{A_1...A_nC})$
one can use the representation (cf.\cite{Y&C})
\begin{equation}\label{n-QCMI-rep}
\begin{array}{rl}
     I(A_1\!:\!...\!:\!A_n|C)_{\omega}\!\!&=I(A_{n-1}\!:\!A_n|C)_{\omega} \\\\&+ \,I(A_{n-2}\!:\!A_{n-1}A_n|C)_{\omega}+...+
     I(A_1\!:\!A_2...A_{n}|C)_{\omega}
\end{array}
\end{equation}
and assume that all the tripartite QCMI in the r.h.s. of (\ref{n-QCMI-rep}) are defined by equivalent expressions (\ref{QCMI-e1}) and (\ref{QCMI-e2}) \cite[Proposition 5]{CMI}.

Representation (\ref{n-QCMI-rep}) and upper bound (\ref{CMI-UB}) imply that
\begin{equation}\label{CMI-n-UB-1}
  I(A_1\!:\!...\!:\!A_n|C)_{\omega}\leq 2\sum_{k=1}^{n-1}S(\omega_{A_k}).
\end{equation}
The function $\,\omega\mapsto I(A_1\!:\!...\!:\!A_n|C)_{\omega}\,$ satisfies  inequalities (\ref{LAA-1}) and (\ref{LAA-2}) with  $a_f=1$ and $b_f=n-1$ \cite{CBM,CID}. So, the nonnegativity of QCMI and upper bound (\ref{CMI-n-UB-1}) show that this function belongs to the class $L^{n-1}_{n+1}(2,n)$.  Thus, Theorem \ref{CB-L}  in Section 3.1.2
implies\smallskip
\begin{proposition}\label{CMI-n-CB} \cite{CBM} \emph{If $\,d_k\doteq\dim\H_{A_k}<+\infty\,$ for $\,k=\overline{1,n-1}\,$ then
\begin{equation*}
 |I(A_1\!:\!...\!:\!A_n|C)_{\rho}-I(A_1\!:\!...\!:\!A_n|C)_{\sigma}|\leq 2\varepsilon \sum_{k=1}^{n-1}\ln d_k+ng(\varepsilon)
\end{equation*}
for any states $\rho$ and $\sigma$ in $\S(\H_{A_1...A_nC})$ such that $\;\frac{1}{2}\|\shs\rho-\sigma\|_1\leq\varepsilon$, where $g(x)$
is the function defined in (\ref{g-def}).}
\end{proposition}

\subsubsection{Continuity bounds under the energy-type constraint.} Uniform continuity bound for the function $\omega\mapsto I(A\!:\!B|C)_{\omega}$ under the energy-type constraint on
system $A$ determined by a positive operator $H$ on $\H_A$ satisfying the Gibbs condition (\ref{H-cond}) can be obtained by using Theorem \ref{W-2-step}
in Section 3.2.1 with the family of maps
$$
\Lambda_m(\omega)=(\Tr P_m\omega_A)^{-1} P_m\otimes I_{BC}\cdot\omega\cdot P_m\otimes I_{BC},
$$
where $P_m$ is the spectral projector of $H$ corresponding to the eigenvalues
$E_0,...,E_{m-1}$. In the following proposition we present this continuity bound in the Winter form.
\smallskip

\begin{proposition}\label{CMI-W-CB} \cite{CHI} \emph{Let $H$ be a positive operator on $\H_A$ satisfying conditions (\ref{H-cond}) and (\ref{star}). Let $\rho$ and $\sigma$ be states in $\S(\H_{ABC})$ such that $\Tr H\rho_A,\Tr H\sigma_A\leq E$ and
$\frac{1}{2}\|\rho-\sigma\|_1\leq \varepsilon<1$. Then
\begin{equation}\label{CMI-W-CB+}
\left|I(A\!:\!B|C)_{\rho}\!-I(A\!:\!B|C)_{\sigma}\right|\leq(2\varepsilon'\!+\!4\delta)F_H\!\left(E/\delta\right)
+2g(\varepsilon')+4h_2(\delta),
\end{equation}
$\delta=(\varepsilon'-\varepsilon)/(1+\varepsilon')$, for any $\varepsilon'\in(\varepsilon,1]$, where $F_H$ is the function defined in (\ref{F-def}).}
\end{proposition}\smallskip

\begin{remark}\label{CMI-W-CB-r}
The quantity $\varepsilon'$ is a "free parameter" that can be used to optimize the continuity bound (\ref{CMI-W-CB+})
for given values of $E$ and $\varepsilon$. The faithfulness of this continuity bound follows from the equivalence of
(\ref{H-cond}) and (\ref{H-cond-a}), it can be shown by taking $\,\varepsilon'=\sqrt{\varepsilon}$.
\end{remark}\smallskip

\begin{remark}\label{CMI-W-CB-r+}
In Proposition 5 in \cite{CHI} continuity bound  (\ref{CMI-W-CB+}) is presented in the equivalent form in which the r.h.s. explicitly
depends on $\varepsilon$ and on the free parameter $t$.
\end{remark}

If $H$ is the grounded Hamiltonian of the $\ell$-mode quantum oscillator defined in (\ref{H-osc}) then
one can use the function $G_{\ell,\omega}$ defined in (\ref{F-osc}) as an upper bound for $F_H$. In this case continuity bound
(\ref{CMI-W-CB+}) with $F_H$ replaced by $G_{\ell,\omega}$ is asymptotically tight for large $E$ (Definition \ref{AT} in Section 3.1.2). This can be shown assuming that $\rho=\gamma_A(E)\otimes \tau_B\otimes \tau_C$, where $\gamma_A(E)$ is the Gibbs state of system $A$ and $\tau_X$ is a pure state in $\S(\H_X)$, $X=B,C$, and $\sigma=(1-\varepsilon)\rho+\varepsilon \omega_{AB}\otimes \tau'_C$, where $\omega_{AB}$
is a purification in $\S(\H_{AB})$ of $\gamma_A(E)$  and  $\tau'_C$ is a pure state in $\S(\H_C)$ orthogonal to $\tau_C$. Then
$\frac{1}{2}\|\rho-\sigma\|_1\leq\varepsilon$,  $I(A\!:\!B|C)_{\rho}=0$  and $I(A\!:\!B|C)_{\sigma}=2\varepsilon F_H(E)=2\varepsilon G_{\ell,\omega}(E)+o(1)$ as $E\to+\infty$.

The following energy-constrained version of Proposition \ref{CMI-CB-S} allows
to obtain continuity bounds for basic characteristics
of  quantum channels depending on their input energy bound (see details in \cite{CID}).\smallskip

\begin{proposition}\label{CMI-ECCB-S} \emph{Let $A$,$B$,$C$,$D$ and $E$ be arbitrary systems and $H$ a positive operator on $\H_{AD}$ satisfying  condition (\ref{H-cond+}). Let $\rho$ and $\sigma$ be states in $\S(\H_{ABCDE})$ such that $\Tr H\rho_{AD},\Tr H\sigma_{AD}\leq E$ and $\textstyle\frac{1}{2}\|\rho-\sigma\|_1\leq \varepsilon\leq 1$.}

\emph{Then $I(A\!:\!B|C)_{\rho}$ and $I(A\!:\!B|C)_{\sigma}$ are finite and
\begin{equation}\label{CMI-ECCB-S+}
    |I(A\!:\!B|C)_{\rho}-I(A\!:\!B|C)_{\sigma}|\leq 2\delta F_H\!\!\left[\frac{2E}{\delta^2}\right]+2g(\delta),
\end{equation}
for any $\delta\in(0,1]$ such that  either $F(\rho,\sigma)=\|\sqrt{\rho}\sqrt{\sigma}\|_1^2\geq 1-\delta^2$ or $\varepsilon(2-\varepsilon)\leq\delta^2$, where
$F_{H}$ is the function defined in (\ref{F-def}) and  $\,g(x)$ is the function defined in (\ref{g-def}). If $\rho$ and $\sigma$ are pure states then inequality (\ref{CMI-ECCB-S+}) holds with $\shs\delta$ replaced by $\,\varepsilon$. If $\rho_{BC}=\sigma_{BC}$ then  factor $2$ in the last term in the right hand side of (\ref{CMI-ECCB-S+}) and of its specification for pure states
can be removed.}

\emph{If $\,\inf\limits_{\|\varphi\|=1}\langle\varphi|H|\varphi\rangle=0\,$ and $G$ is any function  satisfying conditions (\ref{G-c1}) and  (\ref{G-c2}) then
\begin{equation}\label{CMI-ECCB-S++}
    |I(A\!:\!B|C)_{\rho}-I(A\!:\!B|C)_{\sigma}|\leq \min_{t\in(0,T]}\mathbb{CB}_{\shs t}(E,\varepsilon\,|\,2,2),
\end{equation}
where $T$ and $\mathbb{CB}_{\shs t}(E,\varepsilon\,|\,2,2)$ are defined in Theorem \ref{L-1-ca} in Section 3.2.4. If $\rho_{BC}=\sigma_{BC}$ then (\ref{CMI-ECCB-S++}) holds with $\mathbb{CB}_{\shs t}(E,\varepsilon\,|\,2,2)$ replaced by $\mathbb{CB}_{\shs t}(E,\varepsilon\,|\,2,1)$.}

\emph{If $\,H$ is the grounded Hamiltonian of the $\ell$-mode quantum oscillator defined in (\ref{H-osc})
then inequality (\ref{CMI-ECCB-S++}) holds with $\,T$ and $\,\mathbb{CB}_{\shs t}(E,\varepsilon\,|\,2,2)$ replaced  by
$\,T_*$ and $\,\mathbb{CB}^*_{\shs t}(E,\varepsilon\,|\,2,2)$ defined in Corollary \ref{L-1-ca-c} in Section 3.2.4. In this case  continuity bound
(\ref{CMI-ECCB-S++}) is asymptotically tight for large $E$ (Definition \ref{AT} in Section 3.1.2)}
\end{proposition}\smallskip

\emph{Proof.} By using inequalities (\ref{CMI-UB}), (\ref{MI-LAA-1}), (\ref{MI-LAA-2}) and (\ref{I-monotone}) it is easy to show that the function $\omega\mapsto I(A\!:\!B|C)_{\omega}$ on $\S(\H_{ABCDE})$
belongs to the class $L^1_3(2,2)$ in the settings $A_1=AD$, $A_2=BE$, $A_3=C$. Thus, continuity bound (\ref{CMI-ECCB-S+}), its specification for pure states  and continuity bound (\ref{CMI-ECCB-S++}) follow, respectively, from Theorems \ref{L-1-ip} and  \ref{L-1-ca} in Section 3.2.
The claims concerning the case $\rho_{BC}=\sigma_{BC}$ can be established by using the arguments from the proof of Lemma 2 in \cite{UFA}. $\Box$

Uniform continuity bounds for the multipartite QMI under the energy-type constraint can be obtained directly from Theorem \ref{L-1-ip} and Theorem \ref{L-m-ca} in Section 3.2 by noting
that the function $\,\omega\mapsto I(A_1\!:\!...\!:\!A_n)_{\omega}\,$ belongs to the class $L^{n}_{n}(1,n)$.
\smallskip

\begin{proposition}\label{mQMI-CB} \cite{CBM} \emph{Let $n\geq2$ be arbitrary and $H_{A_1},...,H_{A_n}$ be
positive operators on the spaces $\H_{A_1},...,\H_{A_n}$ satisfying  condition (\ref{H-cond+}).
Let $\rho$ and $\sigma$ be states in $\,\S(\H_{A_1..A_n})$ such that $\,\sum_{k=1}^{n}\Tr H_{A_k}\rho_{A_k},\,\sum_{k=1}^{n}\Tr H_{A_k}\sigma_{A_k}\leq nE$  and $\,\frac{1}{2}\|\shs\rho-\sigma\|_1\leq\varepsilon\leq 1$.}

\emph{If $\delta\in(0,1]$ is such that either $F(\rho,\sigma)\doteq\|\sqrt{\rho}\sqrt{\sigma}\|^2_1\geq 1-\delta^2$ or $\,\varepsilon(2-\varepsilon)\leq\delta^2$ then
\begin{equation}\label{mQMI-CB-1}
|I(A_1\!:\!...\!:\!A_n)_{\rho}-I(A_1\!:\!...\!:\!A_n)_{\sigma}|\leq \delta\shs F_{H_{n}}\!\!\left[\frac{2nE}{\delta^2}\right]+ng(\delta),
\end{equation}
where $F_{H_n}$ is the function defined in (\ref{F-H-m}) with $m=n$.}

\emph{If the operators $H_{\!A_1}$,...,$H_{\!A_n}$ are unitary equivalent to each other
and $G$ is  any function satisfying conditions (\ref{G-c1}) and  (\ref{G-c2}) then
\begin{equation}\label{mQMI-CB-2}
|I(A_1\!:\!...\!:\!A_n)_{\rho}-I(A_1\!:\!...\!:\!A_n)_{\sigma}|\leq \min_{t\in(0,1/\varepsilon)}\mathbb{VB}^{n}_{\shs t}(E,\varepsilon\,|\,1,n),
\end{equation}
where $\mathbb{VB}^{n}_{\shs t}(E,\varepsilon\,|\,1,n)$ is defined in Theorem \ref{L-m-ca} in Section 3.2.5.}

\emph{The right hand sides of (\ref{mQMI-CB-1}) and (\ref{mQMI-CB-2}) tend to zero as $\shs\delta,\varepsilon\to 0$ for given $E$.}

\emph{If the functions $F_{H_{A_1}}$ and $G$ satisfy, respectively,  conditions (\ref{B-D-cond-a}) and (\ref{G-c3})
then continuity bound (\ref{mQMI-CB-2}) is asymptotically tight for large $E$ (Definition \ref{AT} in Section 3.1.2).
This is true if $A_1$ is the $\ell$-mode quantum oscillator and $G=G_{\ell,\omega}$ is the function  defined in (\ref{F-osc}).}
\end{proposition}\smallskip

To obtain uniform continuity bounds for the multipartite QCMI under the energy-type constraint note that
 the function  $\,\omega\mapsto I(A_1\!:\!...\!:\!A_n|C)_{\omega}\,$ belong to the classes $L^{n-1}_{n+1}(2,n)$
and $L^{n}_{n+1}(2-2/n,n)$. This follows from the observations before Proposition \ref{CMI-n-CB} and from the inequality
\begin{equation*}
  I(A_1\!:\!...\!:\!A_n|C)_{\omega}\leq (2-2/n)\sum_{k=1}^{n}S(\omega_{A_k}),
\end{equation*}
which is obtained from (\ref{CMI-n-UB-1}) by simple symmetry arguments.

So, we may directly apply  Theorem \ref{L-1-ip} and Theorem \ref{L-m-ca}  in Section 3.2  to the  function $\,\omega\mapsto I(A_1\!:\!...\!:\!A_n|C)_{\omega}\,$ in both  cases $m=n-1$ and $m=n$. This allows as to obtain  continuity bounds for $I(A_1\!:\!...\!:\!A_n|C)_{\omega}$ under two forms of energy-type constraint:
\begin{itemize}
  \item the constraint on the marginal states $\,\omega_{A_1},...,\omega_{A_{n-1}}$ of a state $\shs\omega$;
  \item the constraint on the marginal states $\,\omega_{A_1},...,\omega_{A_{n}}$ of  a state $\shs\omega$.
\end{itemize}
These forms of constraints correspond to the cases $m=n-1$ and $m=n$ in the following
\smallskip

\begin{proposition}\label{mQCMI-CB} \cite{CBM} \emph{Let $n\geq2$ be arbitrary and $H_{A_1},...,H_{A_m}$ be
positive operators on the spaces $\H_{A_1},...,\H_{A_m}$ satisfying  condition (\ref{H-cond+}), where  either $\,m=n-1$ or $\,m=n$. Let $C_m=(n-1)/m$. Let $\rho$ and $\sigma$ be states in $\S(\H_{A_1..A_nC})$ such that $\,\sum_{k=1}^{m}\Tr H_{A_k}\rho_{A_k},\,\sum_{k=1}^{m}\Tr H_{A_k}\sigma_{A_k}\leq mE\,$ and $\,\textstyle\frac{1}{2}\|\rho-\sigma\|_1\leq\varepsilon\leq1$.}

\emph{If $\,\delta\in(0,1]$ is such that either $F(\rho,\sigma)\doteq\|\sqrt{\rho}\sqrt{\sigma}\|^2_1\geq 1-\delta^2$ or $\,\varepsilon(2-\varepsilon)\leq\delta^2$ then
\begin{equation}\label{mQCMI-CB-1}
|I(A_1\!:\!...\!:\!A_n|C)_{\rho}\!-\!I(A_1\!:\!...\!:\!A_n|C)_{\sigma}|\leq \displaystyle 2C_m\delta\shs F_{H_{m}}\!\!\left[\frac{2mE}{\delta^2}\right]\!+\!ng(\delta)
\end{equation}
where $F_{H_m}$ is the function defined in (\ref{F-H-m}).}

\emph{If the operators $H_{\!A_1}$,..,$H_{\!A_m}$ are unitary equivalent to each other
and $G$ is  any function satisfying conditions (\ref{G-c1}) and  (\ref{G-c2}) then
\begin{equation}\label{mQCMI-CB-2}
|I(A_1\!:\!...\!:\!A_n|C)_{\rho}-I(A_1\!:\!...\!:\!A_n|C)_{\sigma}|\leq \!\min_{t\in(0,1/\varepsilon)}\!\!\mathbb{VB}^{m}_{\shs t}(E,\varepsilon|\shs2C_m,n),
\end{equation}
where $\mathbb{VB}^{m}_{\shs t}(E,\varepsilon|\shs2C_m,n)$ is defined in Theorem \ref{L-m-ca} in Section 3.2.5.}

\emph{The right hand sides of (\ref{mQCMI-CB-1}) and (\ref{mQCMI-CB-2}) tend to zero  as $\shs\delta,\varepsilon\to 0$ for given $E$.}

\emph{If the functions $F_{H_{A_1}}$ and $G$ satisfy, respectively,  conditions (\ref{B-D-cond-a}) and (\ref{G-c3})
then continuity bound (\ref{mQCMI-CB-2}) is close-to-tight for large $E$ up to the factor $\,2-2/n\,$ in the main term in both cases $m=n-1$ and $m=n$. This is true if $A_1$ is the $\ell$-mode quantum oscillator and $G=G_{\ell,\omega}$ is the function  defined in (\ref{F-osc}).}
\end{proposition}\smallskip

If $n=2$ and  the functions $F_{H_{A_1}}$ and $G$ satisfy, respectively,  conditions (\ref{B-D-cond-a}) and (\ref{G-c3}) then continuity bound (\ref{mQCMI-CB-2}) is asymptotically tight for large $E$ in both cases $m=1$ and $m=2$. It is natural  to compare accuracy of the
continuity bound given by Proposition \ref{CMI-W-CB} and the continuity bound (\ref{mQCMI-CB-2}) with $n=2$ and $m=1$. The numerical analysis with one-mode quantum oscillator
in the role of system $A_1$ shows that the former continuity bound is essentially sharper than the latter.

\subsubsection{Local continuity conditions.}
For the convenience of formulation of the main results of this subsection introduce the following classes of converging sequences of bipartite states.
\smallskip

\begin{definition}\label{classes-1}
Let $\Upsilon_{\!AB}$ be the class of all converging sequences $\{\omega_n\}\subset\S(\H_{AB})$
such that
\begin{equation}\label{QMI-conv}
 \lim_{n\to+\infty }I(A\!:\!B)_{\omega_n}=I(A\!:\!B)_{\omega_0}<+\infty,
\end{equation}
where  $\omega_0$ is the limit of $\{\omega_n\}$.
\end{definition}
\smallskip

\begin{definition}\label{classes-1C}
Let $\Upsilon_{\!AB|C}$ be the class of all converging sequences $\{\omega_n\}\subset\S(\H_{ABC})$
such that
\begin{equation*}
 \lim_{n\to+\infty }I(A\!:\!B|C)_{\omega_n}=I(A\!:\!B|C)_{\omega_0}<+\infty,
\end{equation*}
where  $\omega_0$ is the limit of $\{\omega_n\}$.
\end{definition}
\smallskip

\begin{proposition}\label{QCMI-lc} \cite{CMI} \emph{A sequence $\{\omega_n\}\subset\S(\H_{AB})$ converging to a state $\omega_0$ belongs to the class $\Upsilon_{\!AB}$ if
\begin{equation}\label{HA-conv}
  \lim_{n\to+\infty }S([\omega_n]_X)=S([\omega_0]_X)<+\infty,\;\, \textrm{ where}\; X\; \textrm{is either}\; A\textrm{ or }B.
\end{equation}
A sequence $\{\omega_n\}\subset\S(\H_{ABC})$ converging to a state $\omega_0$ belongs to the class $\Upsilon_{\!AB|C}$ if
either condition (\ref{HA-conv}) holds or}
\begin{equation*}
  \lim_{n\to+\infty }S([\omega_n]_{XC})=S([\omega_0]_{XC})<+\infty,\;\, \textit{ where}\; X\; \textit{is either}\; A\textit{ or }B.
\end{equation*}
\end{proposition}\smallskip
The first claim of Proposition \ref{QCMI-lc} is proved by using Lemma \ref{L-S} in Section 4.1, the identity
$$
I(A\!:\!B)_{\rho}+I(B\!:\!R)_{\rho}=2S(\rho_{B})
$$
valid for any pure state $\rho$ in $\S(\H_{ABR})$ and the standard arguments of the purification theory.
The second claim is proved by using the first one, Lemma \ref{L-S} and the representations in (\ref{QCMI-rep}).

To formulate  strengthened versions of the conditions in Proposition \ref{QCMI-lc} introduce the following two definitions.\footnote{In these definitions and in what follows speaking about a converging sequence $\{x_n\}$ we assume that
it converges to $x_0$.}

\begin{definition}\label{class-2}
Let $\Upsilon^{*}_{\!AB}$ be the class of all converging sequences $\{\omega_n\}\subset\S(\H_{AB})$
possessing the following property: there exist a system $E$ and a converging sequence $\{\hat{\omega}_n\}$
in $\S(\H_{ABE})$ such that  $[\hat{\omega}_n]_{AB}=\omega_n$ for all $n\geq 0$ and
\begin{equation}\label{HAE-conv}
  \lim_{n\to+\infty }S([\hat{\omega}_n]_{AE})=S([\hat{\omega}_0]_{AE})<+\infty.
\end{equation}
\end{definition}

\begin{definition}\label{class-2C}
Let $\Upsilon^{*}_{\!AB|C}$ be the class of all converging sequences $\{\omega_n\}\subset\S(\H_{ABC})$
possessing the following property: there exist a system $E$ and a converging sequence $\{\hat{\omega}_n\}$
in $\S(\H_{ABCE})$ such that  $[\hat{\omega}_n]_{ABC}=\omega_n$ for all $n\geq 0$ and either (\ref{HAE-conv}) holds or
\begin{equation*}
  \lim_{n\to+\infty }S([\hat{\omega}_n]_{ACE})=S([\hat{\omega}_0]_{ACE})<+\infty.
\end{equation*}
\end{definition}

\begin{remark}\label{sym}
The definitions of the classes $\Upsilon^{*}_{\!AB}$ and $\Upsilon^{*}_{\!AB|C}$ are, in fact, symmetric w.r.t. the subsystems $A$ and $B$. Indeed, if
$\{\hat{\omega}_n\}$ is a converging sequence
in $\S(\H_{ABE})$ such that  $[\hat{\omega}_n]_{AB}=\omega_n$ for all $n\geq 0$ and
(\ref{HAE-conv}) holds  then
  there is a system $E'$ and a converging sequence $\{\tilde{\omega}_n\}$ of pure states
in $\S(\H_{ABEE'})$  such that  $[\tilde{\omega}_n]_{ABE}=\hat{\omega}_n$ for all $n\geq 0$.
Since $S([\tilde{\omega}_n]_{BE'})=S([\tilde{\omega}_n]_{AE})=S([\hat{\omega}_n]_{AE})$, we see that (\ref{HAE-conv}) holds
 with $A$ and $E$ replaced by $B$ and $E'$ and with the sequence $\{\hat{\omega}'_n\doteq[\tilde{\omega}_n]_{ABE'}\}$ instead of $\{\hat{\omega}_n\}$.

The similar arguments show the symmetry of the definition of $\Upsilon^{*}_{\!AB|C}$ w.r.t. the subsystems $A$ and $B$.
\end{remark}\medskip

By using the method described in Section 4.1 and  the lower semicontinuity of the QCMI and of the loss of QCMI
under local channels one can prove the following
\smallskip
\begin{proposition}\label{classes-p}\cite{LSE}  \emph{Let $A$, $B$ and $C$ be arbitrary quantum systems. Then}
\begin{equation}\label{classes+}
  \Upsilon^{*}_{\!AB}\subseteq\Upsilon_{\!AB}\quad \emph{and} \quad\Upsilon^{*}_{\!AB|C}\subseteq\Upsilon_{\!AB|C}.
\end{equation}
\end{proposition}\smallskip

By Remark \ref{sym} any converging sequence $\{\omega_n\}$ satisfying the condition of the first part (correspondingly, the second part)  of Proposition \ref{QCMI-lc}
belongs to the class $\Upsilon^{*}_{\!AB}$ (correspondingly, the class $\Upsilon^{*}_{\!AB|C}$). Hence, continuity condition (\ref{classes+}) is not weaker than the continuity condition given by Proposition \ref{QCMI-lc}. In fact, \emph{it is essentially stronger}.  To see this it suffices to consider a sequence  $\{\omega_n=\rho_n\otimes\sigma\}$ of product
states in $\S(\H_{AB})$, where $\{\rho_n\}$ is a sequence in $\S(\H_A)$ converging to a state $\rho_0$ and $\sigma$ is any state in $\S(\H_B)$. Since
$I(A\!:\!B)_{\omega_n}=0$ for all $n\geq0$, relation (\ref{QMI-conv}) holds trivially, but condition (\ref{HA-conv}) is not valid if
$S(\rho_n)$ does not tend to $S(\rho_0)$ and $S(\sigma)=+\infty$. At the same time, the sequence $\{\omega_n\}$ belongs to the class $\Upsilon^{*}_{\!AB}$. To see this it suffices to take a sequence $\{\bar{\rho}_n\}$ of pure states
in $\S(\H_{AE})$ converging to a pure state $\bar{\rho}_0$ such that  $[\bar{\rho}_n]_{A}=\rho_n$ for all $n\geq 0$. Then the sequence $\{\hat{\omega}_n=\bar{\rho}_n\otimes\sigma\}$ has the properties mentioned in the definition of $\Upsilon^{*}_{\!AB}$.

Nontrivial examples of sequences in $\Upsilon^{*}_{\!AB}$ not satisfying condition (\ref{HA-conv}) can be found in \cite{LSE} (Examples 2 and 3) and in Section 5.4.3 below (Example \ref{EoF-ex}).

It is shown in \cite[Section 5.2]{LSE} and in Sections 5.3.3, 5.4.3 and 5.6.3 below that conditions of local
continuity of many correlation and entanglement measures can be formulated in terms of belonging  of a convergent sequence of states to the class $\Upsilon^{*}_{\!AB}$.

For some time it was unclear whether the classes $\Upsilon_{\!AB}$ and $\Upsilon^*_{\!AB}$
(correspondingly, the classes $\Upsilon_{\!AB|C}$ and $\Upsilon^*_{\!AB|C}$) coincide. Recently Lami and Winter constructed a state $\omega$ in $\S(\H_{AB})$
with finite $I(A\!:\!B)_{\omega}$ such that $S(\hat{\omega}_{AE})=+\infty$ for any extension $\hat{\omega}$ of $\omega$ \cite{L&W}. This example shows that
$\Upsilon^{*}_{\!AB}\varsubsetneq\Upsilon_{\!AB}$ and $\Upsilon^{*}_{\!AB|C}\varsubsetneq\Upsilon_{\!AB|C}$ provided that the systems $A$ and $B$ are infinite-dimensional.
\smallskip

\begin{proposition}\label{DCT-MI}
\emph{The classes $\Upsilon_{\!AB}$ and $\Upsilon_{\!AB|C}$ possess the following properties:}

\noindent A) \emph{If $\{\rho_n\}_{n\geq0}\subset\S(\H_{AB})$ is a converging sequence such that
$c\rho_n\leq \tau_n$ for all $n\geq 0$ and some $c>0$, where $\{\tau_n\}_{n\geq0}$ is a sequence in $\Upsilon_{\!AB}$, then $\{\rho_n\}_{n\geq0}\in \Upsilon_{\!AB}$.}

\noindent B) \emph{If $\{\rho_n\}_{n\geq0}$ and $\{\sigma_n\}_{n\geq0}$ are converging sequences of states in $\S(\H_{AB})$ and $\{p_n\}_{n\geq0}$ is a converging sequence of numbers in $[0,1]$ then
\begin{equation}\label{eq-rel}
\!\{\{\rho_n\}_{n\geq0}\in \Upsilon_{\!AB}\}\wedge\{\{\sigma_n\}_{n\geq0}\in \Upsilon_{\!AB}\}\;\Rightarrow\; \{\{\omega_n\}_{n\geq0}\in \Upsilon_{\!AB}\}
\end{equation}
where $\,\omega_n=p_n\rho_n+(1-p_n)\sigma_n\,$ for all $n\geq0$. If the sequence $\{p_n\}$ converges to $p_0\in(0,1)$ then $"\Leftrightarrow"$ holds in (\ref{eq-rel}).}

\noindent C) \emph{If $\{\omega_n\}_{n\geq0}$ is a sequence in $\Upsilon_{\!AB|C}$ then $\{\Phi_n\otimes\Psi_n\otimes\id_C(\omega_n)\}_{n\geq0}$
is a sequence in $\Upsilon_{\!A'B'|C}$ for any sequences of channels  $\,\Phi_n:A\to A'$ and $\,\Psi_n:B\to B'$ strongly converging to channels $\Phi_0$ and $\Psi_0$ correspondingly.}\footnote{A sequence of channels $\Phi_n$ strongly converges to a channel $\Phi_0$ if $\lim_{n\to+\infty}\Phi_n(\rho)=\Phi_0(\rho)$ for any input state $\rho$ \cite{H-SCI,CSR}.}

\emph{The class $\Upsilon^{*}_{\!AB}$ possesses the above property B. The class $\Upsilon^{*}_{\!AB|C}$ possesses the above property C.}
\end{proposition}\medskip

\emph{Proof.} A) We have to show that the implication (\ref{DCT-II-tau})$\Rightarrow$(\ref{DCT-II-rho}) is valid for the
function $f(\rho)=I(A\!:\!B)_{\rho}$. This function is nonnegative lower semicontinuous on $\S(\H_{AB})$ and satisfies inequalities (\ref{LAA-1})
and (\ref{LAA-2}) with  $a_f=b_f=h_2$. But it is not finite on the set of pure states in $\S(\H_{AB})$, so we can not apply Theorem \ref{DCT-II}A in Section 4.2.3 to this function.
We will prove the above implication by using Theorem \ref{DCT-II}B.

Assume that $\{\rho'_n\}$ and $\{\tau'_n\}$ are  sequences  converging to states $\rho'_0$ and $\tau'_0$
such that $\,c\rho'_n\leq 2\rho_n$, $\,c\rho'_n\leq 2\tau'_n\leq 4\tau_n$ for all $\,n\geq0$ and $\,\sup_n\rank\tau'_n<+\infty$.
If follows from the conditions $\,\sup_n\rank\tau'_n<+\infty\,$ and $\,c\rho'_n\leq 2\tau'_n\leq 4\tau_n\,$ that there exist
\begin{equation*}
\lim_{n\to+\infty} S(\tau'_n)=S(\tau'_0)<+\infty\quad \textrm{and}\quad \lim_{n\to+\infty} S(\rho'_n)=S(\rho'_0)<+\infty
\end{equation*}
and that $f(\tau'_0),f(\rho'_0)<+\infty$. Hence, since the finiteness of $S(\tau'_0),S(\rho'_0),f(\tau'_0)$ and $f(\rho'_0)$ implies the finiteness of $S([\tau'_0]_A)$ and $S([\rho'_0]_A)$ by the second formula in (\ref{MI-e1}), this formula, Proposition 9A in \cite{CMI} and Remark \ref{DCT-II-r} in Section 4.2.3 show that
\begin{equation*}
\limsup_{n\to+\infty}f(\tau'_n)-f(\tau'_0)=2\limsup_{n\to+\infty}S([\tau'_n]_A)-S([\tau'_0]_A)<+\infty
\end{equation*}
and
\begin{equation*}
\limsup_{n\to+\infty}f(\rho'_n)-f(\rho'_0)=2\limsup_{n\to+\infty}S([\rho'_n]_A)-S([\rho'_0]_A)<+\infty.
\end{equation*}
Since
\begin{equation*}
\limsup_{n\to+\infty}S([\rho'_n]_A)-S([\rho'_0]_A)\leq \frac{2}{c} \left(\limsup_{n\to+\infty}S([\tau'_n]_A)-S([\tau'_0]_A)\right),
\end{equation*}
by Theorem \ref{DCT-II}A, the above two expressions imply the validity of condition (\ref{gen-con-II})
with the function $G(x)=2x/c$.

B) We will prove this assertion by using Theorem \ref{convex-m} in Section 4.2.3.

Let $\Psi_m^{\rho}(\rho_n)=P_{\widehat{m}(\rho_0)}^{\rho_n}\rho_n$ and $\Psi_m^{\sigma}(\sigma_n)=P^{\sigma_n}_{\widehat{m}(\sigma_0)}\sigma_n$,
where $P_{\widehat{m}(\rho_0)}^{\rho_n}$ and $P_{\widehat{m}(\sigma_0)}^{\sigma_n}$
are the spectral projectors of $\rho_n$ and $\sigma_n$ defined according to the rule described after (\ref{Psi-m-1+}). The families
of maps $\Psi^{\rho}_m:\{\rho_n\}\cup\{\rho_0\}\mapsto \T_+(\H_{AB})$ and $\Psi^{\sigma}_m:\{\sigma_n\}\cup\{\sigma_0\}\mapsto \T_+(\H_{AB})$ satisfy all the conditions
stated at the begin of Section 4.2.1 with $M_{\rho_0}=\mathbb{N}\cap[m_\rho,+\infty)$ and $M_{\sigma_0}=\mathbb{N}\cap[m_\sigma,+\infty)$, where
$m_\rho$ and $m_\sigma$ are some natural numbers.

Denote by $\rho_n^m$ and $\sigma_n^m$ the states $(\mu_n^m)^{-1}\Psi_m^{\rho}(\rho_n)$ and $(\nu_n^m)^{-1}\Psi_m^{\sigma}(\sigma_n)$ provided that
$\mu_n^m=\Tr\Psi_m^{\rho}(\rho_n)>0$ and $\nu_n^m=\Tr \Psi_m^{\sigma}(\sigma_n)>0$. By Lemma \ref{DTL-L} we have
\begin{equation}\label{mu-nu++}
\lim_{m\to+\infty}\inf_{n\geq0}\mu^m_n=\lim_{m\to+\infty}\inf_{n\geq0}\nu^m_n=1.
\end{equation}

Let $f(\rho)=I(A\!:\!B)_{\rho}$. It follows from (\ref{mu-nu++}) and part A of the proposition  that
\begin{equation}\label{+2-rel}
\lim_{n\to+\infty}f(\rho^m_n)=f(\rho^m_0)<+\infty\;\;\; \textup{and}
\;\; \lim_{n\to+\infty}f(\sigma^m_n)=f(\sigma^m_0)<+\infty
\end{equation}
for all sufficiently large $m$.
Since
\begin{equation}\label{+2-rel+}
\lim_{n\to+\infty}S(\rho^m_n)=S(\rho^m_0)\leq \ln m\;\;\; \textup{and}
\;\; \lim_{n\to+\infty}S(\sigma^m_n)=S(\sigma^m_0)\leq\ln m
\end{equation}
for all $m$ large enough, the limit relations in (\ref{+2-rel}) and in (\ref{+2-rel+}) imply, by the second formula in (\ref{MI-e1}) and the lower semicontinuity of the entropy, that
$$
\lim_{n\to+\infty}S([\rho^m_n]_X)=S([\rho^m_0]_X)<+\infty\quad \textup{and}
\quad \lim_{n\to+\infty}S([\sigma^m_n]_X)=S([\sigma^m_0]_X)<+\infty,
$$
$X=A,B$, for all sufficiently large $m$. By using all the above limit relations and the remark after Corollary 4 in \cite{SSP} we obtain
$$
\lim_{n\to+\infty}S(p_n[\rho^m_n]_X+\bar{p}_n[\sigma^m_n]_X)=S(p_0[\rho^m_0]_X+\bar{p}_0[\sigma^m_0]_X)<+\infty \quad X=A,B,AB.
$$
and hence
$$
\lim_{n\to+\infty}f(p_n\rho^m_n+\bar{p}_n\sigma^m_n)=f(p_0\rho^m_0+\bar{p}_0\sigma^m_0)<+\infty
$$
for any  sequences $\{p_n\}_{n\geq0}$ of numbers in $[0,1]$ converging to $p_0$ and all sufficiently large $m$.

Thus, since $f(\rho)=I(A\!:\!B)_{\rho}$ is a
lower semicontionuous nonnegative function on $\S(\H_{AB})$ satisfying inequalities (\ref{LAA-1}) and  (\ref{LAA-2}) with $a_f=b_f=h_2$,
all the conditions of Theorem \ref{convex-m} are satisfied with the above families $\{\Psi_m^{\rho}\}$ and $\{\Psi_m^{\sigma}\}$.

The last assertion of B follows from part A of the proposition.

C) This assertion directly follows from Proposition 2 in \cite{LSE}.

The validity of property B for the class $\Upsilon^{*}_{\!AB}$ is proved easily by using the definition of this class,  Remark \ref{sym} and the remark after Corollary 4 in \cite{SSP}.

Property C for the class $\Upsilon^{*}_{\!AB|C}$ can be proved in two steps. First, we note that\break $\{\omega'_n=\id_{AC}\otimes\Psi_n(\omega_n)\}_{n\geq0}$
is a sequence in $\Upsilon^{*}_{\!AB'|C}$. Second, by applying the arguments from Remark \ref{sym} to the last sequence we see that
$\{\Phi_n\otimes\Psi_n\otimes\id_C(\omega_n)=\Phi_n\otimes\id_{BC}(\omega'_n)\}_{n\geq0}$
is a sequence in $\Upsilon^{*}_{\!A'B'|C}$.$\square$

To generalize the results of Proposition \ref{DCT-MI} to the multipartite QMI and QCMI introduce the multipartite versions
of the classes defined above.

Denote by $\Upsilon_{\!A_1..A_n}$ the class  of all converging sequences $\{\omega_k\}\subset\S(\H_{A_1..A_n})$
such that
\begin{equation*}
 \lim_{k\to+\infty }I(A_1\!:\!...\!:\!A_n)_{\omega_k}=I(A_1\!:\!...\!:\!A_n)_{\omega_0}<+\infty,
\end{equation*}
where  $\omega_0$ is the limit of $\{\omega_k\}$.

Denote by $\Upsilon_{\!A_1..A_n|C}$ the class  of all converging sequences $\{\omega_k\}\subset\S(\H_{A_1..A_nC})$
such that
\begin{equation*}
 \lim_{k\to+\infty }I(A_1\!:\!...\!:\!A_n|C)_{\omega_k}=I(A_1\!:\!...\!:\!A_n|C)_{\omega_0}<+\infty,
\end{equation*}
where  $\omega_0$ is the limit of $\{\omega_k\}$.

Denote by $\Upsilon^{*}_{\!A_1..A_n}$ the class  of all converging sequences $\{\omega_k\}_{k\geq0}\subset\S(\H_{A_1..A_n})$
possessing the following property: there exist systems $E_1$,..,$E_{n-1}$ and a converging sequence $\{\hat{\omega}_k\}_{k\geq0}$
in $\S(\H_{A_1..A_nE_1..E_{n-1}})$ such that  $[\hat{\omega}_k]_{A_1..A_n}=\omega_k$ for all $k\geq 0$ and
\begin{equation*}
  \lim_{k\to+\infty }S([\hat{\omega}_k]_{A_jE_j})=S([\hat{\omega}_0]_{A_jE_j})<+\infty,\quad j=\overline{1,n-1}.
\end{equation*}

By using a purification it is easy to show that the definition of the class $\Upsilon^{*}_{\!A_1..A_n}$ is symmetric w.r.t. the subsystems $A_1$,..,$A_{n}$.

For a given $(n-1)$-element subset $\pi$ of $\{1,2,..,n\}$ denote by $\Upsilon^{*,\pi}_{\!A_1..A_n|C}$ the class  of all converging sequences $\{\omega_k\}_{k\geq0}\subset\S(\H_{A_1..A_nC})$
possessing the following property: there exist systems $E_1$,..,$E_{n-1}$ and a converging sequence $\{\hat{\omega}_k\}_{k\geq0}$
in $\S(\H_{A_1..A_nCE_1..E_{n-1}})$ such that  $[\hat{\omega}_k]_{A_1..A_nC}=\omega_k$ for all $k\geq 0$ and
$$
  \lim_{k\to+\infty }S([\hat{\omega}_k]_{X_j})=S([\hat{\omega}_0]_{X_j})<+\infty,\;\; \forall j\in\pi,\textrm{ where }X_j\textrm{ is either }A_jE_j\textrm{ or }A_jE_jC,
$$
Since it is not clear how to show that the class $\Upsilon^{*,\pi}_{\!A_1..A_n|C}$ does not depend on $\pi$ if $n>2$ and $\dim\H_C=+\infty$, we define the class
$\Upsilon^{*}_{\!A_1..A_n|C}$ as $\,\bigcup_\pi\Upsilon^{*,\pi}_{\!A_1..A_n|C}$.

Now we can formulate the multipartite versions of Propositions \ref{classes-p} and \ref{DCT-MI}.
\smallskip

\begin{proposition}\label{classes-p-n}\cite{LSE} \emph{Let $A_1,..,A_n$ and $C$ be arbitrary quantum systems. Then}
$$
 \Upsilon^{*}_{\!A_1..A_n}\subseteq\Upsilon_{\!A_1..A_n}\quad \emph{and} \quad\Upsilon^{*}_{\!A_1..A_n|C}\subseteq\Upsilon_{\!A_1..A_n|C}.
$$
\end{proposition}
\smallskip

By using the Lami-Winter example mentioned after Proposition \ref{classes-p} one can show that
$\Upsilon^{*}_{\!A_1..A_n}\subsetneq\Upsilon_{\!A_1..A_n}$ and $\Upsilon^{*}_{\!A_1..A_n|C}\subsetneq\Upsilon_{\!A_1..A_n|C}$
provided that at least two of the subsystems $A_1$,..,$A_n$ are infinite-dimensional.
\smallskip

\begin{proposition}\label{DCT-MI+m} \emph{Let $A_1$,..,$A_n$ and $C$ be arbitrary quantum systems, $n>2$.}

\noindent\emph{The analogs of properties A and B in Proposition \ref{DCT-MI} hold for the class $\Upsilon_{\!A_1..A_n}$.}

\noindent\emph{The analog of property C in Proposition \ref{DCT-MI} holds for the classes  $\Upsilon_{\!A_1..A_n|C}$ and $\Upsilon^{*}_{\!A_1..A_n}$.}

\noindent\emph{The analog of property B in Proposition \ref{DCT-MI} holds for the class $\Upsilon^{*}_{\!A_1..A_n}$.}

\end{proposition}\medskip

\emph{Proof.} The validity of properties $A$, $B$ for the class $\Upsilon_{\!A_1..A_n}$ and property $C$ for the class $\Upsilon_{\!A_1..A_n|C}$
can be derived from the validity of these properties for the classes $\Upsilon_{\!AB}$ and $\Upsilon_{\!AB|C}$
by using the  representation (\ref{n-QCMI-rep}). It suffices to note that the lower semicontinuity of QCMI and Lemma \ref{L-S} in Section 4.1 imply that the local continuity of  the function $\,\omega\mapsto I(A_1\!:\!...\!:\!A_n|C)_{\omega}\,$ \emph{is equivalent to}
the local continuity of all the summands in the r.h.s. of (\ref{n-QCMI-rep}).

The validity of properties $B$ and $C$ for the class $\Upsilon^{*}_{\!A_1..A_n}$ can be shown by natural
generalization of the arguments from the proof of Proposition \ref{DCT-MI}. $\square$

Sometimes it is convenient to use the homogeneous extension of the QMI to the positive cone $\T_+(\H_{A_1...A_n})$ defined as
\begin{equation}\label{QMI-ext}
I(A_1\!:\!...\!:\!A_n)_{\omega}\doteq (\Tr\omega) \,I(A_1\!:\!...\!:\!A_n)_{\omega/\Tr\omega}
\end{equation}
for any nonzero operator  $\omega\in\T_+(\H_{A_1...A_n})$ and $\,I(A_1\!:\!...\!:\!A_n)_{0}=0$. Property $A$ of the class $\Upsilon_{\!A_1..A_n}$ implies the following result that can be called
\textit{dominated convergence theorem for the quantum mutual information}.\smallskip

\begin{corollary}\label{DCT-MI-c1}
\emph{Let $\{\rho_k\}$ and $\{\tau_k\}$ be sequences of operators from $\T_+(\H_{A_1...A_n})$ converging, respectively, to operators $\rho_0$ and $\tau_0$  such that
$\rho_k\leq \tau_k$ for all $k\geq 0$. If
\begin{equation}\label{QMI-conv+m+}
\lim_{k\to+\infty }I(A_1\!:\!...\!:\!A_n)_{\tau_k}=I(A_1\!:\!...\!:\!A_n)_{\tau_0}<+\infty
\end{equation}
then}
\begin{equation*}
 \lim_{k\to+\infty }I(A_1\!:\!...\!:\!A_n)_{\rho_k}=I(A_1\!:\!...\!:\!A_n)_{\rho_0}<+\infty.
\end{equation*}
\end{corollary}\smallskip

\emph{Proof.} If $\rho_0\neq0$ then the claim directly follows from property $A$ of the class $\Upsilon_{\!A_1..A_n}$.

Let $\rho_0=0$. The validity of inequality (\ref{LAA-1}) for the function $f(\omega)=I(A_1\!:\!...\!:\!A_n)_{\omega}$ on $\S(\H_{A_1...A_n})$
with $a_f=h_2$ and the assumption $\,\rho_k\leq \tau_k\,$ for all $k$ imply that
$$
I(A_1\!:\!...\!:\!A_n)_{\rho_k}\leq I(A_1\!:\!...\!:\!A_n)_{\tau_k}-I(A_1\!:\!...\!:\!A_n)_{\tau_k-\rho_k}+(\Tr\tau_k) h_2(\Tr\rho_k/\Tr\tau_k),\quad \forall k.
$$
If $\tau_0\neq0$ then property $A$ of the class $\Upsilon_{\!A_1..A_n}$ shows that condition (\ref{QMI-conv+m+}) implies that
$I(A_1\!:\!...\!:\!A_n)_{\tau_k-\rho_k}$ tends to $I(A_1\!:\!...\!:\!A_n)_{\tau_0}$ and hence $I(A_1\!:\!...\!:\!A_n)_{\rho_k}$
tends to zero as $k\to+\infty$. If $\tau_0=0$ then the above inequality guarantees that $I(A_1\!:\!...\!:\!A_n)_{\rho_k}$
tends to zero, since $\Tr\tau_k\to0$ as $k\to+\infty$ in this case. $\square$\pagebreak

\subsection{One-way classical correlation, its regularization and\\  quantum discord}

\subsubsection{Definitions, AFW-type continuity bounds and their applications.}
The notion of a measure of classical correlations is proposed by Henderson and Vedral in \cite{H&V} as a characteristic  properly describing classical component of correlation  of a state $\omega$ of a bipartite system $AB$.  Henderson and Vedral have found the example of such measure called \emph{one-way classical correlation} that is defined as
\begin{equation}\label{CB-def}
C_B(\omega)=\sup_{\M\in \mathfrak{M}_B}\chi(\{p_i,\omega^i_{A}\}),
\end{equation}
where the supremum  is taken over the set $\mathfrak{M}_B$ of all discrete Positive Operator Values Measures (POVM)
$\M=\{M_i\}$ on the space $\H_B$, $p_i=\Tr(I_A\otimes M_i)\omega$
is the probability of  the $i$-th outcome, $\omega^i_{A}=p_i^{-1}\Tr_B(I_A\otimes M_i)\omega$ is the posteriori state of  system $A$ corresponding to  the $i$-th outcome
(if $p_i=0$ then we assume that the ensemble $\{p_i,\omega^i_{A}\}$ has no state in the $i$-th position).  It is easy to show that the supremum in (\ref{CB-def}) can be taken only over the set $\mathfrak{M}^0_B$ of all  POVM $\M=\{M_i\}$ consisting of one-rank operators \cite{K&W}.

The function $\omega\mapsto C_B(\omega)$ on $\S(\H_{AB})$ is nonnegative, invariant under local unitary trasformations
and  non-increasing under local channels. It is equal to the von Neumann entropy of $\omega_A$  at any pure state $\omega$ and  coincides with the QMI $I(A\!:\!B)$ on the set of q-c states (\ref{qc-states}) \cite{H&V,MI-B,Xi+}.

The one-way classical correlation is nonadditive, its regularization is defined as
\begin{equation*}
  C^{\infty}_B(\omega)=\lim_{m\rightarrow+\infty}m^{-1}C_B(\omega^{\otimes m})=\sup_{m}m^{-1}C_B(\omega^{\otimes m}),
\end{equation*}
where $\omega^{\otimes m}$ is treated as a state of the bipartite quantum system $(A^m)(B^m)$. In the finite-dimensional settings it is proved that
$C^{\infty}_B(\omega)$ coincides with the distillable common randomness (when  the classical communication is in the direction is from B to A) \cite{D&W}.

The \emph{quantum discord}  is the difference between the QMI and the above measure of classical correlations (cf. \cite{Xi+,Str}):
\begin{equation}\label{q-d}
D_B(\omega)\doteq I(A\!:\!B)_{\omega}-C_B(\omega).
\end{equation}
The notion of quantum discord as a quantity describing quantum component of correlations of a state $\omega$ of a bipartite system $AB$
is proposed by Ollivier and Zurek in \cite{O&Z}, where  the quantum discord is defined by the expression similar to (\ref{q-d}) in which $C_B(\omega)$
is defined by formula (\ref{CB-def}) with the supremum over all von Neumann measurements only. One can show that
the Ollivier-Zurek definition coincides with definition (\ref{q-d}) if $B$ is an infinite-dimensional system, but in the finite-dimensional case
these definitions do not coincide. The advantage of the definition (\ref{q-d}) used below is that it provides, due to Naimark's theorem, the invariance of
the quantum discord w.r.t. to the embedding of $\H_{AB}\doteq\H_A\otimes\H_B$ into $\H_{AB'}\doteq\H_A\otimes\H_{B'}$, where $\H_{B'}$ is any space containing $\H_B$ (the Ollivier-Zurek definition does not possess this property) \cite{W-pc}.

Definition (\ref{CB-def}) can be rewritten as follows
\begin{equation}\label{CB-rep}
C_B(\omega)=\sup_{\M\in\mathfrak{M}_B}I(A\!:\!E)_{\id_A\otimes\Psi_{\M}(\omega)},
\end{equation}
where
\begin{equation}\label{Psi-M}
\Psi_{\M}(\rho)=\sum_{i}[\Tr M_i\rho\shs]|i\rangle\langle i|
\end{equation}
is a  channel from $\S(\H_{B})$ to $\S(\H_{E})$ determined by any fixed basis $\{|i\rangle\}$ in a  Hilbert space $\H_E$ (s.t. $\dim\H_E$ is equal to the cardinality of the outcome set of $\M=\{M_i\}$).

For given $\M=\{M_i\}\in \mathfrak{M}_B$ consider the unoptimized classical correlation and unoptimized quantum discord defined for any state $\omega\in\S(\H_{AB})$ as
$$
C_B^{\shs\M}(\omega)=I(A\!:\!E)_{\id_A\otimes\Psi_{\M}(\omega)}\quad \textrm{and} \quad  D_B^{\shs\M}(\omega)=I(A\!:\!B)_{\omega}-I(A\!:\!E)_{\id_A\otimes\Psi_{\M}(\omega)}.
$$
The basic properties of QMI imply that
both these functions are nonnegative and well defined on the dense convex subset of $\S(\H_{AB})$ consisting of states
with finite QMI. The unoptimized quantum discord has operational interpretations described in \cite{NQD,NQD+}.\smallskip

\begin{lemma}\label{BL} \emph{Let $\M=\{M_i\}$ be an arbitrary POVM in $\mathfrak{M}^0_B$.}

A) \emph{The functions  $C_B^{\shs\mathbb{M}}$ and $D_B^{\shs\mathbb{M}}$ satisfy inequalities  (\ref{LAA-1}) and (\ref{LAA-2})  with
$a_f=b_f=h_2$ for any states $\rho$ and $\sigma$ in $\S(\H_{AB})$ with finite quantum mutual information.}

B) \emph{For any state $\omega\in\S(\H_{AB})$ the following inequalities hold}
$$
C_B^{\shs\mathbb{M}}(\omega)\leq \min\{S(\omega_A), S(\omega_B)\},\quad  D_B^{\shs\mathbb{M}}(\omega)\leq \min\{2S(\omega_A), S(\omega_B)\}.
$$
\end{lemma}

\emph{Proof.} A) Let $\rho$ and $\sigma$ be states in $\S(\H_{AB})$ with finite QMI.
If $S(\rho_A),S(\sigma_A)<+\infty$ then $C_B^{\shs\mathbb{M}}(\tau)=S(\tau_A)-S(A|E)_{\id_A\otimes\Psi_{\M}(\tau)}$ and
$D_B^{\shs\mathbb{M}}(\tau)=S(A|E)_{\id_A\otimes\Psi_{\M}(\tau)}-S(A|B)_{\tau}$, $\tau=\rho, \sigma$, where $S(A|X)$ is the extended conditional entropy defined in (\ref{ce-ext}). So,  inequalities  (\ref{LAA-1}) and (\ref{LAA-2})  with
$f=C_B^{\shs\mathbb{M}},D_B^{\shs\mathbb{M}}$ and $\,a_f=b_f=h_2\,$ follow from the properties of the entropy and the extended QCE mentioned in Sections 2 and 5.1.1. Arbitrary states $\rho$ and $\sigma$ in $\S(\H_{AB})$ with finite QMI can be approximated by the sequences of states
$\rho_n=\Pi_n\otimes\id_B(\rho)$ and $\sigma_n=\Pi_n\otimes\id_B(\sigma)$  defined by means of the sequence of channels
$\,\Pi_n(\varrho)=P_n\varrho P_n+[\Tr(I_A- P_n)\varrho]\varrho_0$, $\varrho\in\S(\H_A)$, where $\{P_n\}$ is a sequence of finite-rank projectors in $\B(\H_A)$ strongly converging to the unit operator $I_A$ and $\varrho_0$ is a pure state in $\S(\H_A)$.

Let $\omega=p\rho+(1-p)\sigma$ and $\omega_n=p\rho_n+(1-p)\sigma_n$ for some $p\in(0,1)$. Since $\id_A\otimes\Psi_{\M}(\tau_n)=\Pi_n\otimes\id_B(\id_A\otimes\Psi_{\M}(\tau))$, $\tau=\rho, \sigma,\omega$, for each $n$, the lower semicontinuity of QMI and its monotonicity under local channels imply
\begin{equation}\label{lr}
\lim_{n\rightarrow+\infty}C_B^{\shs\mathbb{M}}(\tau_n)=C_B^{\shs\mathbb{M}}(\tau),\quad  \lim_{n\rightarrow+\infty}D_B^{\shs\mathbb{M}}(\tau_n)=D_B^{\shs\mathbb{M}}(\tau),\quad \tau=\rho, \sigma,\omega.
\end{equation}
Since $S([\rho_n]_A),S([\sigma_n]_A)<+\infty$ for each $n$, the above arguments show that inequalities  (\ref{LAA-1}) and (\ref{LAA-2})  with
$f=C_B^{\shs\mathbb{M}},D_B^{\shs\mathbb{M}}$ and $a_f=b_f=h_2$ hold for the states $\rho_n$ and $\sigma_n$ for any $n$. By the limit
relations in (\ref{lr}) these inequalities  hold for the states $\rho$ and $\sigma$ as well.

B) Since $D_B^{\shs\mathbb{M}}(\omega)\leq I(A\!:\!B)_{\omega}$, the inequality
$D_B^{\shs\mathbb{M}}(\omega)\leq 2S(\omega_A)$ follows from the
upper bound (\ref{MI-UB}). The inequality $C_B^{\shs\mathbb{M}}(\omega)\leq S(\omega_A)$ is due to the
fact that factor $2$ in the upper bound (\ref{MI-UB}) can be removed if $\omega$ is a q-c state \cite{MI-B,Wilde}.
The inequality $C_B^{\shs\mathbb{M}}(\omega)\leq S(\omega_B)$ follows from the same inequality for $C_B$ that can be derived
from the generalized Koashi-Winter relation (\ref{KWS-gen}) in Section 5.5.1, since $S(\omega_B)=S(\omega_{AC})$ for any pure state $\omega$ in $\S(\H_{ABC})$.

To prove the inequality $\,D_B^{\shs\mathbb{M}}(\omega)\leq S(\omega_B)\,$ assume first that $S(\omega)<+\infty$ and note that the assumption $\M\in\mathfrak{M}^0_B$ implies
$$
S(A|E)_{\id_A\otimes\Psi_{\M}(\omega)}=\sum_i p_i S(\omega_i),
$$
where $\omega_i=p_i^{-1}(I_A\otimes \sqrt{M_i})\shs\omega\shs (I_A\otimes \sqrt{M_i})$, $p_i=\Tr (I_A\otimes M_i)\shs\omega$.
If follows that
\begin{equation}\label{D-B-rep}
D_B^{\shs\mathbb{M}}(\omega)=S(\omega_B)-ER(\{I_A\otimes M_i\},\omega),
\end{equation}
where
$\,ER(\{I_A\otimes M_i\},\omega)=S(\omega)-\sum_i p_i S(\omega_i)\,$
is the entropy reduction of the measurement $\{I_A\otimes M_i\}$ at a state $\omega$ \cite{L-ER, Ozawa}.\footnote{In this case we may identify the instrument $\{I_A\otimes \sqrt{M_i}\,(\cdot)\, I_A\otimes \sqrt{M_i}\}$ and the POVM $\{I_A\otimes M_i\}$.} By the  approximation technique used in the proof of part A one can show that representation
(\ref{D-B-rep}) remains valid in the case $\,S(\omega)=S(\omega_A)=+\infty\,$ with the extended definition of $ER$ \cite{QMG,Sh-ER}. Since the measurement $\{I_A\otimes M_i\}$ is efficient, $ER(\{I_A\otimes M_i\},\omega)\geq 0$ by the Lindblad-Ozawa inequality \cite{L-ER,Ozawa}. $\square$

Lemma \ref{BL} implies the well known upper bounds (cf. \cite{Xi+,Xi})
$$
C_B(\omega)\leq \min\{S(\omega_A), S(\omega_B)\}\quad \textrm{and}\quad D_B(\omega)\leq \min\{2S(\omega_A), S(\omega_B)\}.
$$
Within the notation introduced in Section 3.1.2, this lemma shows that
$$
C_B^{\shs\M}\in L_2^1(1,2)\quad\forall \M\in\mathfrak{M}^0_B\quad \textrm{and hence}\quad C_B\in\widehat{L}_2^1(1,2)
$$
in both settings $A_1=A,A_2=B$ and $A_1=B,A_2=A$ and that
$$
D_B^{\shs\M}\in L_2^1(C,2)\quad\forall \M\in\mathfrak{M}^0_B\quad \textrm{and hence}\quad  D_B\in\widehat{L}_2^1(C,2),
$$
where $C=2$ in the setting $A_1=A,A_2=B$ and $C=1$ in the setting $A_1=B,A_2=A$.\pagebreak

Now we can formulate the main result of this subsection.
\smallskip

\begin{proposition}\label{QD-CB-1} \emph{Let $A$ and $B$ be  quantum systems such that either $d_A=\dim\H_A$ or $d_B=\dim\H_B$ is finite and
$\rho$ and $\sigma$ any states in $\S(\H_{AB})$. }\smallskip

\emph{If $\frac{1}{2}\|\rho-\sigma\|_1=\varepsilon$ then the following inequalities are valid
\begin{equation}\label{CB-1}
|C_B(\rho)-C_B(\sigma)|\leq \varepsilon\ln d_A+2g(\varepsilon),\;\;
\end{equation}
\begin{equation}\label{CB-1+}
|C_B(\rho)-C_B(\sigma)|\leq \varepsilon\ln d_B+2g(\varepsilon),\;\;
\end{equation}
\begin{equation}\label{CB-2}
|D_B(\rho)-D_B(\sigma)|\leq 2\varepsilon\ln d_A+2g(\varepsilon),
\end{equation}
\begin{equation}\label{CB-2+}
|D_B(\rho)-D_B(\sigma)|\leq \varepsilon\ln d_B+2g(\varepsilon).\;\;
\end{equation}
If $S(\rho_A)=S(\sigma_A)$ and $\varepsilon\leq 1-1/d_A$ then
\begin{equation}\label{CB-1++}
|C_B(\rho)-C_B(\sigma)|\leq \varepsilon\ln (d_A-1)+h_2(\varepsilon).
\end{equation}
If $\rho_B=\sigma_B$ then inequalities (\ref{CB-1+}) and (\ref{CB-2+}) hold with the term $2g(\varepsilon)$ replaced by $\,g(\varepsilon)$.}\smallskip

\emph{If $F(\rho,\sigma)=1-\delta^2$ then}
\begin{equation}\label{CB-RA}
|C^{\infty}_B(\rho)-C^{\infty}_B(\sigma)|\leq 2\epsilon\ln d_A+g(\epsilon),\qquad\qquad\qquad\qquad\quad
\end{equation}
\begin{equation}\label{CB-RB}
|C^{\infty}_B(\rho)-C^{\infty}_B(\sigma)|\leq 2\epsilon\ln (2d_B)+g(\epsilon),\quad \epsilon=\sqrt{\delta(2-\delta)}.\;
\end{equation}

\emph{Continuity bounds (\ref{CB-1}), (\ref{CB-1+}) and  (\ref{CB-2+}) are tight for large $d_X$, $X=A,B$ (Definition \ref{AT} in Section 3.1.2), continuity bound (\ref{CB-2}) is close to tight
up to the factor $2$ in the main term. Continuity bound (\ref{CB-1++}) is optimal for any given dimension $d_A$.}
\end{proposition}\smallskip

\emph{Proof.}  Inequalities  (\ref{CB-1})-(\ref{CB-2+}) follow directly from  by  Theorem \ref{CB-L} in Section 3.1.2 and the classification of the functions $C_B$ and $D_B$ mentioned before the proposition.

Inequality (\ref{CB-1++}) with $C_B$ replaced by $C_B^{\shs\M}$, where $\mathbb{M}$ is any POVM in $\mathfrak{M}_B$, is  obtained by using Wilde's optimal continuity bound for the QCE of q-c states presented in Proposition \ref{Wilde-CB} in Section 5.1.2, since
$C^{\shs\M}_B(\rho)-C^{\shs\M}_B(\sigma)=S(A|E)_{\id_A\otimes\Psi_{\M}(\sigma)}-S(A|E)_{\id_A\otimes \Psi_{\M}(\rho)}$ provided that $S(\rho_A)=S(\sigma_A)<+\infty$.
By noting that  the r.h.s. of this inequality does not depend on $\M$ we conclude that it holds for $C_B$.

To show that the term $2g(\varepsilon)$ in inequalities (\ref{CB-1+}) and (\ref{CB-2+})  can be replaced by $g(\varepsilon)$ in the case $\rho_B=\sigma_B$
we will apply Corollary \ref{AFW-1-c} in Section 3.1.1 with the family of subsets $\S_{\lambda}=\{\omega\in \S(\H_{AB})\,|\,\omega_B=\lambda\}$, $\lambda\in\S(\H_B)$,
possessing property (\ref{G-Delta-P}).

For any POVM $\M=\{M_i\}$ in $\mathfrak{M}_B$ with a finite number of outcomes we have
$$
C_B^{\shs\mathbb{M}}(\varrho)-C_B^{\shs\mathbb{M}}(\varsigma)=S(E|A)_{\id_A\otimes\Psi_{\M}(\varsigma)}-S(E|A)_{\id_A\otimes \Psi_{\M}(\varrho)}
$$
for any $\varrho,\varsigma\in\S_{\lambda}$. Since the function $f(\rho)=S(E|A)_{\rho}$ satisfies
inequalities  (\ref{LAA-1}) and (\ref{LAA-2})  with
$a_f=0$ and $b_f=h_2$,  Corollary \ref{AFW-1-c} implies that (\ref{CB-1+}) holds with $C_B$ and $2g(\varepsilon)$
replaced, respectively, by $C_B^{\shs\mathbb{M}}$ and $g(\varepsilon)$. This proves the claim for $C_B$, since
the supremum in (\ref{CB-rep}) can be taken over all POVM with a finite number of outcomes (this can be easily shown by using the lower semicontinuity of the QMI).

For any POVM $\M=\{M_i\}$ in $\mathfrak{M}^0_B$ it follows from representation (\ref{D-B-rep}) that
$$
D_B^{\shs\mathbb{M}}(\varrho)-D_B^{\shs\mathbb{M}}(\varsigma)=ER(\{I_A\otimes M_i\},\varsigma)-ER(\{I_A\otimes M_i\},\varrho)
$$
for any $\varrho,\varsigma\in\S_{\lambda}$. By Lemma \ref{BL} the function $f(\omega)=ER(\{I_A\otimes M_i\},\omega)$
satisfies inequalities  (\ref{LAA-1}) and (\ref{LAA-2})  with
$a_f=0$ and $b_f=h_2$. Hence, Corollary \ref{AFW-1-c} implies that (\ref{CB-2+}) holds with $D_B$ and $2g(\varepsilon)$
replaced, respectively, by $D_B^{\shs\mathbb{M}}$ and $g(\varepsilon)$. This proves the claim for $D_B$, since
the supremum in (\ref{CB-rep}) can be taken over all POVM in $\mathfrak{M}^0_B$.

Continuity bounds (\ref{CB-RA}) and (\ref{CB-RB}) are proved below by using the generalized  Koashi-Winter relation (\ref{KWS-gen})
and the continuity bound (\ref{CHI-R+}) for the function $\chi^{\infty}_A$ obtained independently in Section 5.5.1.

Let $\bar{\rho}$ and $\bar{\sigma}$ be purifications in $\S(\H_{ABC})$ of the states $\rho$ and $\sigma$ such that $F(\bar{\rho},\bar{\sigma})=F(\rho,\sigma)$, where $C$ is an appropriate quantum system. Continuity bound (\ref{CB-RA}) directly follows from continuity bound (\ref{CHI-R+}) with $d=\dim\H_A$, since
$F(\bar{\rho}_{AC},\bar{\sigma}_{AC})\geq F(\bar{\rho},\bar{\sigma})=1-\delta^2\,$ by monotonicity of the fidelity
and hence inequality (\ref{F-Tn-ineq}) implies that
\begin{equation}\label{F-ineq}
  \textstyle\frac{1}{2}\|\bar{\rho}_{AC}-\bar{\sigma}_{AC}\|_1\leq \delta.
\end{equation}
Continuity bound (\ref{CB-RB}) is derived from continuity bound (\ref{CHI-R+}) with $d=\dim\H_*$ by using (\ref{F-ineq}), since the rank of the states $\bar{\rho}_{AC}$ and $\bar{\sigma}_{AC}$ does not exceed $\dim\H_B$.

To show that continuity bounds (\ref{CB-1})-(\ref{CB-2+}) are tight (close-to-tight) for large $d_X$ it suffices to
assume that $\rho$ and $\sigma$ are, respectively, the maximally entangled pure state and the chaotic state in $\S(\H_{AB})$, where $A$ and $B$ are $d$-dimensional systems, since in this case
$$
D_B(\rho)-D_B(\sigma)=C_B(\rho)-C_B(\sigma)=\ln d\quad \textrm{and} \quad \textstyle\frac{1}{2}\|\rho-\sigma\|_1=1-1/d^2.
$$
The optimality of the continuity bound (\ref{CB-1++}) follows from the optimality of the upper bound  (\ref{C-UP}) below  derived from (\ref{CB-1++}). $\Box$

\textbf{Note.} Continuity bounds for the quantum discord presented in Proposition \ref{QD-CB-1} essentially refine the
continuity bound obtained in \cite{Xi}, which states (in our notation) that
$$
|D_B(\rho)-D_B(\sigma)|\leq 16\varepsilon\ln d_A+4h_2(2\varepsilon).
$$

Below we consider several applications of the continuity bounds in Proposition \ref{QD-CB-1}

\begin{itemize}
\item \emph{Computable  upper bound for the classical correlation.}
Continuity bound (\ref{CB-1++}) in Proposition \ref{QD-CB-1} gives the following upper bound
\begin{equation}\label{C-UP}
C_B(\rho)\leq \Delta^*(\rho)\ln (d_A-1)+h_2(\Delta^*(\rho)),
\end{equation}
where $\Delta^*(\rho)\doteq \inf\limits_{\sigma\in\S(\H_B)}\textstyle\frac{1}{2}\|\rho-\rho_A\otimes \sigma\|_1\leq \Delta(\rho)\doteq\textstyle\frac{1}{2}\|\rho-\rho_A\otimes \rho_B\|_1$ provided that $\Delta^*(\rho)\leq 1-1/d_A$.

The upper bound (\ref{C-UP}) is \textit{optimal in any dimension}. Consider the q-c state
$\,\rho=\frac{1}{d}\,(|11\rangle\langle11|+...+|dd\rangle\langle dd|)\,$
in $\S(\H_{AB})$, where $A$ and $B$ are $d$-dimensional systems.
Then $\rho_A\otimes \rho_B$ is the chaotic state and it is easy to see that $\Delta(\rho)=1-1/d$. So,
both sides of (\ref{C-UP}) (with $\Delta^*(\rho)$ replaced by $\Delta(\rho)$) are equal to $\ln d$.

In this case $\Delta^*(\rho)=\Delta(\rho)$ (otherwise, the inequality $">"$ would be in  (\ref{C-UP})).

\item \emph{Computable  upper bound for the  quantum discord.} Proposition \ref{QD-CB-1} implies that
\begin{equation*}
 |D_B(\rho)-D_B(\sigma)|\leq \varepsilon \ln d_B+g(\varepsilon)
\end{equation*}
for any states $\rho$ and $\sigma$ in $\S(\H_{AB})$ such that $\frac{1}{2}\|\rho-\sigma\|_1\leq \varepsilon$
and $\rho_B=\sigma_B$. This gives the following upper bound
\begin{equation}\label{D-UP}
D_B(\rho)\leq \Upsilon(\rho)\ln d_B+g(\Upsilon(\rho)),\quad \Upsilon(\rho)=\inf_{\sigma\in\C(\rho)}\textstyle\frac{1}{2}\|\rho-\sigma\|_1,
\end{equation}
where $\C(\rho)$ is the set of q-c states having the form
$\sum_i p_i\varrho_i\otimes |i\rangle\langle i|$, provided that  $\rho_B=\sum_i p_i|i\rangle\langle i|$ and $\{\varrho_i\}$ is an arbitrary collection of states in $\S(\H_A)$.

Let $\rho$ be any maximally entangled pure state (in the case $A\cong B$). By taking the states $\varrho_i=|i\rangle\langle i|$ it is easy to see that $\Upsilon(\rho)\leq 1-1/d$, $d=d_A=d_B$ (while $\Delta(\rho)=1-1/d^2$). So, upper bound (\ref{D-UP}) implies that
\begin{equation*}
D_B(\rho)\leq (1-1/d)\ln d+g(1-1/d).
\end{equation*}
Since $D_B(\rho)=\ln d$ this shows that upper bound (\ref{D-UP}) is tight for large $d$.
\end{itemize}

\subsubsection{Continuity bound under the energy-type constraint.}  In this subsection we present continuity bounds for the functions $C_B$, $D_B$ and $C^{\infty}_B$ under
the energy-type constraint imposed on one of the marginal states $\omega_A$ and $\omega_B$ of a state $\omega$ in $\S(\H_{AB})$.
\smallskip

\begin{proposition}\label{EC-CB} \emph{Let $H$ be a positive operator on the space $\H_X$, where $X$ is either $A$ or $B$, satisfying  conditions (\ref{H-cond+}) and (\ref{star}) and $G$  any function on $\mathbb{R}_+$ satisfying conditions (\ref{G-c1}) and  (\ref{G-c2}). Let $\rho$ and $\sigma$ be  states in $\S(\H_{AB})$ such that $\,\frac{1}{2}\|\shs\rho-\sigma\|_1\leq\varepsilon$
and $\,\Tr H\rho_{X},\Tr H\sigma_{X}\leq E$. Then
\begin{equation}\label{EC-CB-1}
|C_B(\rho)-C_B(\sigma)|\leq \min_{t\in(0,T]}\mathbb{CB}_{\shs t}(E,\varepsilon\,|\,1,2),\quad
\end{equation}
\begin{equation}\label{EC-DB-1}
|D_B(\rho)-D_B(\sigma)|\leq \min_{t\in(0,T]}\mathbb{CB}_{\shs t}(E,\varepsilon\,|\,c_X,2),\;
\end{equation}
where $\mathbb{CB}_{\shs t}(E,\varepsilon\,|\,C,D)$ and $T$  are defined in Theorem \ref{L-1-ca} in Section 3.2.4, $c_A=2$, $c_B=1$.}

\emph{If the functions $F_{H}$ and $G$ satisfy, respectively,  conditions (\ref{B-D-cond-a}) and (\ref{G-c3}) then both continuity bounds in (\ref{EC-CB-1}) and continuity bound (\ref{EC-DB-1}) with $X=B$  are asymptotically tight for large $E$ (Definition \ref{AT} in Section 3.1.2). Continuity bound (\ref{EC-DB-1}) with $X=A$ is close-to-tight
(up to the factor $2$ in the main term). This is true if the system under constraint  is the $\ell$-mode quantum oscillator
and the function $G_{\ell,\omega}$ defined in (\ref{F-osc}) is used in the role of $G$. In this case inequalities
(\ref{EC-CB-1}) and (\ref{EC-DB-1}) are valid with $T$ and $\,\mathbb{CB}_{\shs t}(E,\varepsilon\,|\,C,D)$ replaced by $T_*$ and $\,\mathbb{CB}^*_{\shs t}(E,\varepsilon\,|\,C,D)$  defined in Corollary \ref{L-1-ca-c} in Section 3.2.4.}
\end{proposition}\smallskip

\emph{Proof.} The main assertion of the proposition directly follows from Theorem \ref{L-1-ca} in Section 3.2.4 and the classification of the
functions $C_B$ and $D_B$ before  Proposition \ref{QD-CB-1}.

To prove the last assertion it suffices to note that condition (\ref{as-t-c-1}) with $c_f^-=0,c_f^+=1$ holds for the functions $C_B$ and $D_B$ in both settings $A_1=A,A_2=B$ and $A_1=B,A_2=A$. To show this assume that $\rho=\gamma_A(E)\otimes\gamma_B(E)$, where
$\gamma_X(E)$ is the Gibbs state (\ref{Gibbs}) in $\S(\H_X)$ determined by the operator $H$ and a given arbitrary $E>0$, and $\sigma$ is a pure state
in $\S(\H_{AB})$ such that $\sigma_A=\gamma_A(E)$ and $\sigma_B=\gamma_B(E)$, then
$$
C_B(\rho)=D_B(\rho)=0\quad\textrm{and}\quad C_B(\sigma)=D_B(\sigma)=F_H(E),
$$
where $F_H(E)=S(\gamma_X(E))$ is the function defined in (\ref{F-def}). $\square$

By using Theorem \ref{L-1-ip} in Section 3.2.3 with $m=1$, the generalised Koashi-Winter relation (\ref{KWS-gen}) and the continuity bounds
for the functions $\chi_A$ and $\chi_A^{\infty}$ proved independently in Section 5.5.2 one can obtain the following
\smallskip

\begin{proposition}\label{CB-CB-2} \emph{Let $H$ be a positive operator on the space $\H_{X}$, where $X$ is either $A$ or $B$, satisfying condition (\ref{H-cond+}) and $F_{H}$ the function defined in (\ref{F-def}). Let $\rho$ and $\sigma$ be states in $\S(\H_{AB})$ such that $\,\Tr H\rho_{X},\,\Tr H\sigma_{X}\leq E\,$ and $F(\rho,\sigma)\geq 1-\delta^2\geq0$. Then
\begin{equation}\label{CB-CB-2+}
    |C_B(\rho)-C_B(\sigma)|\leq \delta F_{H}\!\!\left[\frac{2E}{\delta^2}\right]+2g(\delta),\qquad\qquad\qquad\;\;
\end{equation}
\begin{equation}\label{DB-CB-2+}
    |D_B(\rho)-D_B(\sigma)|\leq c_X\delta F_{H}\!\!\left[\frac{2E}{\delta^2}\right]+2g(\delta),\qquad\qquad\quad\;\,
\end{equation}
\begin{equation}\label{CB-CB-2++}
    |C_B^{\infty}(\rho)-C_B^{\infty}(\sigma)|\leq 2\epsilon F_{H}\!\!\left[\frac{2E}{\epsilon^2}\right]+2\epsilon \ln(3-c_X)+g(\epsilon),
\end{equation}
where $\,\epsilon=\sqrt{\delta(2-\delta)}$, $c_A=2$, $c_B=1$ and $\,g(x)$ is the function defined in (\ref{g-def}).}
\end{proposition}\smallskip

\emph{Proof.} Inequalities (\ref{CB-CB-2+}) and (\ref{DB-CB-2+}) directly follow from Theorem \ref{L-1-ip} with $m=1$ and the  classification of the
functions $C_B$ and $D_B$ before  Proposition \ref{QD-CB-1}.

Let $\bar{\rho}$ and $\bar{\sigma}$ be pure states in $\S(\H_{ABC})$ such that
$\bar{\rho}_{AB}=\rho$, $\bar{\sigma}_{AB}=\sigma$ and $F(\bar{\rho},\bar{\sigma})\geq 1-\delta^2$, where $C$ is an appropriate quantum system.

Inequality (\ref{CB-CB-2++}) in the case $X=A$ directly follows, due to estimate (\ref{F-ineq}), from the generalised Koashi-Winter relation (\ref{KWS-gen})
and Proposition \ref{CHI-CB-2} in Section 5.5.2.

If $X=B$ then the condition $\,\Tr H\rho_{B},\,\Tr H\sigma_{B}\leq E\,$ implies existence of positive operators $H_{\rho}$ and $H_{\sigma}$ on
the space $\H_{AC}$ isometrically equivalent to the operator $H$ such that $\,\Tr H_{\rho}\bar{\rho}_{AC},\,\Tr H_{\sigma}\bar{\sigma}_{AC}\leq E$. This follows from the fact that
the states $\bar{\rho}_{AC}$ and $\bar{\sigma}_{AC}$ have the same positive parts of the spectrum as the states $\rho_{B}$ and $\sigma_{B}$ correspondingly. By Lemma \ref{H-lemma} below there a positive operator
$H_*$ on $\H_{AC}$ satisfying condition (\ref{H-cond+}) such that
$$
\Tr H_*\bar{\rho}_{AC},\,\Tr H_*\bar{\sigma}_{AC}\leq E\quad \textrm{and}\quad F_{H_*}(E)=F_H(E)+\ln2.
$$
This observation,  the generalised Koashi-Winter relation (\ref{KWS-gen}) and  Proposition \ref{CHI-CB-2} in Section 5.5.2 imply, due to estimate (\ref{F-ineq}), inequality (\ref{CB-CB-2++}) in the case $X=B$. $\Box$\smallskip\pagebreak

\begin{lemma}\label{H-lemma}  \emph{Let $H_1$ and $H_2$ be positive operators on a Hilbert space
$\H$ satisfying condition (\ref{H-cond+}) which are isometrically  equivalent in the following sense
$$
 H_1|\varphi\rangle=V^*H_2V|\varphi\rangle,\quad  H_2|\psi\rangle=VH_1V^*|\psi\rangle,\quad \forall \varphi\in\D(H_1),\psi\in\D(H_2),
$$
for some partial isometry $V$ isometrically mapping $\D(H_1)$ onto $\D(H_2)$.
Then there is a positive operator $H_0$ on $\H$ satisfying condition (\ref{H-cond+}) such that
\begin{equation}\label{EA-1}
  \D(\sqrt{H_i})\subseteq\D(\sqrt{H_0}),\quad \langle\varphi|H_0|\varphi\rangle\leq\langle\varphi|H_i|\varphi\rangle,\;\;\forall\varphi\in\D(\sqrt{H_i}),
\end{equation}
\begin{equation}\label{EA-2}
F_{H_0}(E)=F_{H_i}(E)+\ln2,\quad  i=1,2,
\end{equation}
where $F_{H_i}$ is the function defined in (\ref{F-def}) with $H=H_i$.}
\end{lemma}

\emph{Proof.} Since the operators $H_1$ and $H_2$ are isometrically  equivalent and satisfy condition (\ref{H-cond+}) they have
the representations
\begin{equation*}
H_i=\sum_{k=0}^{+\infty} E_k |\tau^i_k\rangle\langle\tau^i_k|,\quad i=1,2,
\end{equation*}
where
$\mathcal{T}_i\doteq\left\{\tau^i_k\right\}_{k=0}^{+\infty}$ is an orthonormal
system of vectors and $\left\{E_k\right\}_{k=0}^{+\infty}$ is a \emph{nondecreasing} sequence tending to $+\infty$.  In this case $\D(\sqrt{H_i})=\{ \varphi\in\H_{\mathcal{T}_i}\,| \sum_{k=0}^{+\infty} E_k |\langle\tau^i_k|\varphi\rangle|^2<+\infty\}$, where $\H_{\mathcal{T}_i}$
is the closed subspace of $\H$ generated by the system $\mathcal{T}_i$, $i=1,2$.

Let $P_n^i=\sum_{k=0}^{n} |\tau^i_k\rangle\langle\tau^i_k|$ and $\bar{P}_n^i=I_{\H}-P_n^i$. Then it is easy to see that
\begin{equation}\label{SHR}
\langle\varphi|H_i|\varphi\rangle=E_0\langle\varphi|\varphi\rangle+\sum_{n=0}^{+\infty}(E_{n+1}-E_n)\langle\varphi|\bar{P}_n^i|\varphi\rangle\quad \forall\varphi\in\D(\sqrt{H}_i),
\end{equation}
and that the series in (\ref{SHR}) does not converge if $\varphi$ does not lie in $\D(\sqrt{H}_i)$, $i=1,2$.

Let $\{P^0_n\}_{n=0}^{\infty}$ be any sequence of orthogonal projectors on $\H$ such that
\begin{equation*}
P^0_n\geq P^i_n, \;\; i=1,2,\quad P^0_{n+1}>P^0_n,\quad \mathrm{rank}P^0_n=2(n+1)\quad \forall n\geq0.
\end{equation*}
Then there exists an orthonormal system $\mathcal{T}_0\doteq\left\{\tau^0_k\right\}_{k=0}^{+\infty}$ of vectors in $\H$ such that $P_n^0=\sum_{k=0}^{n} |\tau^0_{2k}\rangle\langle\tau^0_{2k}|+|\tau^0_{2k+1}\rangle\langle\tau^0_{2k+1}|$ for all $n=0,1,2,..$. Consider the positive operator
$$
H_0=\sum_{k=0}^{+\infty} E_k (|\tau^0_{2k}\rangle\langle\tau^0_{2k}|+|\tau^0_{2k+1}\rangle\langle\tau^0_{2k+1}|)
$$
defined on the set $\D(H_0)=\{ \varphi\in\H_{\mathcal{T}_0}\,| \sum_{k=0}^{+\infty} E^2_k (|\langle\tau^0_{2k}|\varphi\rangle|^2+|\langle\tau^0_{2k+1}|\varphi\rangle|^2)<+\infty\}$.
By noting that (\ref{SHR}) and the remark after it hold for $i=0$ with $\bar{P}^0_n=I_{\H}-P^0_n$, we see that
$\varphi\in\D(\sqrt{H_i})\,$ if and only if $\;\sum_{n=0}^{+\infty}\langle\varphi|\bar{P}_n^i|\varphi\rangle(E_{n+1}-E_n)<+\infty$, $i=0,1,2$.
Thus, all the relations in (\ref{EA-1}) are valid, since $\bar{P}_n^0\leq\bar{P}_n^i$, $i=1,2$ by the construction.

To prove (\ref{EA-2}) note that
$F_{H_i}(E)=\beta_i(E) E+\ln \Tr e^{-\beta_i(E) H_i}$, where $\beta_i(E)$ is defined by the equation
$\Tr H_ie^{-\beta H_i}=E\Tr e^{-\beta H_i}$, $i=0,1,2$ \cite{W,EC}. Since the spectrum of
$H_0$ has the form $\{E_0,E_0,E_1,E_1,E_2,E_2,... \}$, the solutions of the above equation for $H_0$
and for $H_i$, $i=1,2$ coincide, i.e. $\beta_0=\beta_1=\beta_2$.  Using this it is easy to prove  (\ref{EA-2}).

To show that the operator $H_0$ satisfies  condition (\ref{H-cond+}) it suffices to use (\ref{EA-2}) and the
equivalence of (\ref{H-cond+}) and (\ref{H-cond+a}). $\Box$

\subsubsection{Local continuity conditions.} It follows from Proposition \ref{QD-CB-1} that the functions $C_B$, $D_B$ and $C^{\infty}_B$
are uniformly continuous on $\S(\H_{AB})$ provided that either $\dim\H_A$ or $\dim\H_B$ is finite. It is easy to see that these functions
are not continuous on $\S(\H_{AB})$ if $\dim\H_A=\dim\H_B=+\infty$. Sufficient
conditions of local continuity of the functions $C_B$, $D_B$ and $C^{\infty}_B$ are presented in Propositions \ref{CB-DB} and \ref{CB-DB-R} below.
\smallskip

\begin{proposition}\label{CB-DB}  A) \emph{The function $C_B$ is
lower semicontinuous on the set $\S(\H_{AB})$.}

B) \emph{If $\,\{\omega_n\}$ is a sequence in  $\,\mathfrak{S}(\mathcal{H}_{AB})$ converging to a state $\omega_0$ such that
\begin{equation}\label{I-cont}
 \lim_{n\to+\infty }\!I(A\!:\!B)_{\omega_n}=I(A\!:\!B)_{\omega_0}<+\infty
\end{equation}
then
\begin{equation}\label{CB-DB-conv}
  \lim_{n\to+\infty }\!C_B(\omega_n)=C_B(\omega_0)<+\infty,\;\;\lim_{n\to+\infty }\!D_B(\omega_n)=D_B(\omega_0)<+\infty.
\end{equation}
Moreover, for arbitrary  sequences of channels $\,\Phi_n:A\to A'$ and $\,\Psi_n:B\to B'$ strongly converging\footnote{A sequence of channels $\Phi_n$ strongly converges to a channel $\Phi_0$ if $\lim_{n\to+\infty}\Phi_n(\rho)=\Phi_0(\rho)$ for any input state $\rho$ \cite{H-SCI,CSR}.} to channels
$\,\Phi_0$ and $\Psi_0$ both limit relations in (\ref{CB-DB-conv}) hold with $\omega_n$ replaced by $\Phi_n\!\otimes\!\Psi_n(\omega_n)$ for all $n\geq0$.}
\end{proposition}\smallskip

Proposition \ref{CB-DB}B has a clear physical interpretation: it states that \emph{local continuity of the total correlation
implies local continuity of its classical and quantum components.}\smallskip

\begin{example}\label{CB-DB-ex}
Let $\omega$ be a state in $\S(\H_{AB})$ such that $I(A\!:\!B)_{\omega}$ is finite. Let $\{\Phi_t:A\rightarrow A\}_{t\in \mathbb{R}_+}$ and $\{\Psi_t:B\rightarrow B\}_{t\in \mathbb{R}_+}$ be arbitrary strongly continuous families of quantum channels (for instance, quantum dynamical semigroups). The last assertion of Proposition \ref{CB-DB} and
the monotonicity of the QMI imply that
the nonnegative functions
$$
t\mapsto C_B(\Phi_t\otimes\Psi_t(\omega))\quad \textrm{and} \quad t\mapsto D_B(\Phi_t\otimes\Psi_t(\omega))
$$
are continuous on $\mathbb{R}_+$ and bounded above by $I(A\!:\!B)_{\omega}$.
\end{example}\medskip

Proposition \ref{classes-p} in Section 5.2.2 and Proposition \ref{CB-DB} imply the following\smallskip

\begin{corollary}\label{CB-DB-c} \emph{If $\,\{\omega_n\}$ is a sequence
in $\,\S(\H_{AB})$ converging to a state $\omega_0$ that belongs to the class $\Upsilon^{*}_{\!AB}$ (Definition \ref{class-2} at the begin of Section 5.2.3.)
then both limit relations in (\ref{CB-DB-conv}) hold.}
\end{corollary}\smallskip

By using Corollary \ref{CB-DB-c} one can prove  both limit relations in (\ref{CB-DB-conv}) for the sequence
considered in Example 3 in \cite{LSE} (now the condition $S([\rho_0]_X)<+\infty$ can be removed).

One can conjecture that the analogue of the continuity condition in Proposition \ref{CB-DB}B holds for the function $C_B^{\infty}$. However, it was possible to prove only the following\smallskip\pagebreak

\begin{proposition}\label{CB-DB-R}  A) \emph{The function $C_B^{\infty}$ is
lower semicontinuous on the set $\S(\H_{AB})$.}

B) \emph{If $\,\{\omega_n\}$ is a sequence in  $\,\mathfrak{S}(\mathcal{H}_{AB})$ converging to a state $\omega_0$ such that
either
\begin{equation}\label{CSA-cont}
 \lim_{n\to+\infty }S([\omega_n]_A)=S([\omega_0]_A)<+\infty
\end{equation}
or
\begin{equation}\label{CSB-cont}
 \lim_{n\to+\infty }S([\omega_n]_B)=S([\omega_0]_B)<+\infty\quad and \quad S(\omega_0)<+\infty
\end{equation}
then}
\begin{equation}\label{CB-R-conv}
  \lim_{n\to+\infty }C^{\infty}_B(\omega_n)=C^{\infty}_B(\omega_0)<+\infty.
\end{equation}
\end{proposition}\smallskip

Note that each of the conditions (\ref{CSA-cont}) and (\ref{CSB-cont}) implies (\ref{I-cont}) by Proposition \ref{QCMI-lc} in Section 5.2.3, but one can easily
construct a converging sequence $\{\omega_n\}$ satisfying (\ref{I-cont}) such that neither (\ref{CSA-cont}) nor (\ref{CSB-cont})
is valid (see Section 5.2.3).

\emph{Proof of Proposition \ref{CB-DB}.} A) The lower semicontinuity of $C_B$ follows from representation (\ref{CB-rep}) and the
lower semicontinuity of QMI.

B) We start by showing that the first limit relation in (\ref{CB-DB-conv}) holds for any
sequence $\{\omega_n\}$ of states in $\S(\H_{AB})$ converging to a state $\omega_0$ such that (\ref{CSA-cont}) is valid.

Let $\{P_k\}$ be a sequence of finite rank projectors in $\B(\H_A)$ strongly converging to the unit operator $I_A$ and $\tau$ is a fixed pure state in $\S(\H_A)$.
Consider the sequence of channels $\Phi_k(\rho)=P_k\rho P_k+[\Tr(I_A-P_k)\rho\shs]\tau$ from $A$ to itself. It follows from Proposition \ref{QD-CB-1} that the function
$\omega\mapsto C_B(\Phi_k\otimes\id_B(\omega))$ is continuous on the
set $\S(\H_{AB})$ for each $k$. Thus, to prove the first limit relation in (\ref{CB-DB-conv}) it suffices to show that
\begin{equation*}
\lim_{k\to+\infty}\sup_{n\geq0} |C_B(\omega_n)-C_B(\Phi_k\otimes\id_B(\omega_n))|=0.
\end{equation*}
This can be done by using Lemma \ref{CB-R-ap}  below, since condition (\ref{CSA-cont}) implies that
\begin{equation*}
\lim_{k\to+\infty}\sup_{n\geq0} |S([\omega_n]_A)-S(P_k[\omega_n]_AP_k)|=0.
\end{equation*}
Indeed, this follows from the standard Dini's lemma, since the function
$\rho\mapsto S(P\rho P)\doteq [\Tr P\rho] S([\Tr P\rho]^{-1}P\rho P)$ is continuous on $\S(\H)$
for any finite rank projector $P$ and $S(\rho)=\sup_P S(P\rho P)$, where the supremum is over all finite rank projectors \cite{L-2}.

Assume now that $\{\omega_n\}$ is a sequence of
states in $\S(\H_{AB})$ converging to a state $\omega_0$ such that (\ref{I-cont}) holds.
Let $\Psi_m(\omega_n)=P_m^{\omega_n}\omega_n$ for any $m$ and $n\geq0$, where $P_m^{\omega_n}$ is the spectral projector of
$\omega_n$ corresponding to its $m$ maximal eigenvalues (taken the multiplicity into account). It is shown at the end of Section 4.2.1
that the family $\{\Psi_m\}$ of maps from $\,\A=\{\omega_n\}\cup\{\omega_0\}\,$ into $\T_+(\H_{AB})$ satisfies all the conditions stated before  Lemma \ref{DTL-L}. In this case the set $M_{\omega_0}$
consists of all natural numbers $m$ such that either $\lambda^{\omega_0}_{m+1}<\lambda^{\omega_0}_{m}$ or $\lambda^{\omega_0}_{m}=0$ ($\{\lambda^{\omega_0}_{k}\}_k$ is a sequence of eigenvalues of $\omega_0$ taken in the non-increasing order). Let $\mathbb{N}_{\A}$ be the subset of $\mathbb{N}$ defined in (\ref{N-def}).

For each $n\geq 0$ and $m\in \mathbb{N}_{\A}$ denote by $\omega^m_n$ and $\mu^m_n$ the state  proportional to
the operator $\Psi_m(\omega_n)$ and the positive number $\Tr\Psi_m(\omega_n)$ correspondingly.
By property A of the class $\Upsilon_{\!AB}$ stated in Proposition \ref{DCT-MI} in Section 5.2.3 condition (\ref{I-cont}) implies that
$$
 \lim_{n\to+\infty }I(A\!:\!B)_{\omega^m_n}=I(A\!:\!B)_{\omega^m_0}<+\infty\quad \forall m\in M'_{\omega_0}\doteq M_{\omega_0}\cap\mathbb{N}_{\A}.
$$
Since $\,\lim_{n\to+\infty }S(\omega^m_n)=S(\omega^m_0)\leq\ln m\,$ for all $m\in M'_{\omega_0}$, the above limit relation shows, by the second formula in (\ref{MI-e1}) and the lower semicontinuity of the entropy, that
$$
\lim_{n\to+\infty }S([\omega^m_n]_A)=S([\omega^m_0]_A)<+\infty\quad \textrm{and} \quad \lim_{n\to+\infty }S([\omega^m_n]_B)=S([\omega^m_0]_B)<+\infty
$$
for all $m\in M'_{\omega_0}$. So, by the observation at the begin of this proof we have
\begin{equation}\label{tmp-r}
  \lim_{n\to+\infty }C_B(\omega^m_n)=C_B(\omega^m_0)<+\infty\quad \forall m\in M'_{\omega_0}.
\end{equation}

The function $f(\omega)=I(A\!:\!B)_{\omega}$ is nonnegative and lower semicontinuous on $\S(\H_{AB})$. It satisfies inequalities (\ref{LAA-1}) and (\ref{LAA-2}) with $a_f=b_f=h_2$ (inequalities (\ref{MI-LAA-1}) and (\ref{MI-LAA-2})).

Representation (\ref{CB-rep}) and inequality (\ref{MI-LAA-1}) with trivial system $C$ imply that
\begin{equation*}
\mu_0^m C_B(\omega^m_0)\leq C_B(\omega_0)+h_2(1-\mu_0^m)\quad \forall m\in\mathbb{N}_{\A}.
\end{equation*}
It follows from this inequality that
\begin{equation}\label{tmp-r+}
  \limsup_{m\to+\infty }\mu_0^m C_B(\omega^m_0)\leq C_B(\omega_0).
\end{equation}
By using representation (\ref{CB-rep}) and inequality (\ref{MI-LAA-2}) with trivial system $C$ it is easy to show that the nonnegative function $g(\omega)=C_B(\omega)$ satisfies
inequality (\ref{LAA-2}) with $b_g=h_2$.

Thus, since $g(\omega)\leq f(\omega)$ for all $\omega$ in $\S(\H_{AB})$, by applying  Corollary \ref{DTL-c} in Section 4.2.1 to the above functions $f$ and $g$ with the use of (\ref{tmp-r}) and (\ref{tmp-r+})
we obtain
\begin{equation*}
  \limsup_{m\to+\infty }C_B(\omega_n)\leq C_B(\omega_0).
\end{equation*}
This shows the validity of the first limit relation in (\ref{CB-DB-conv}) by part A of the proposition.  The second  limit relation in (\ref{CB-DB-conv})
obviously follows from the first one and (\ref{I-cont}).

The last assertion  follows from Proposition \ref{DCT-MI} in Section 5.2.2. $\Box$\smallskip

\emph{Proof of Proposition \ref{CB-DB-R}.} A) The lower semicontinuity of $C^{\infty}_B$ follows from its definition and the
lower semicontinuity of $C_B$ mentioned in Proposition \ref{CB-DB}.

B) Assume that (\ref{CSA-cont}) holds. Let $\{\Phi_k\}$ be the sequence of channels introduced in the proof of part B
of Proposition \ref{CB-DB}. Since the function
$\omega\mapsto C^{\infty}_B(\Phi_k\otimes\id_B(\omega))$ is continuous on the
set $\S(\H_{AB})$ for each $k$ by Proposition \ref{QD-CB-1}, to prove (\ref{CB-R-conv}) one can repeat
the arguments from the first paragraph of the proof of part B of Proposition \ref{CB-DB}
based on using of Lemma \ref{CB-R-ap}  below and the standard Dini's lemma.

Assume that (\ref{CSB-cont}) holds. Let $\{\bar{\omega}_n\}$ be a sequence
of pure states in $\S(\H_{ABC})$ converging to a pure state $\bar{\omega}_0$ such that
$[\bar{\omega}_n]_{AB}=\omega_n$, $\forall n\geq0$, where $C$ is an appropriate quantum system.
It follows from the limit relation in (\ref{CSB-cont}) that the similar relation holds with $\omega_n$ replaced by $[\bar{\omega}_n]_{AC}$.
Since $S([\bar{\omega}_0]_{C})=S(\omega_0)<+\infty$, the validity of (\ref{CB-R-conv}) follows from Proposition \ref{CHI-LC} in Section 5.5.3 below (proved independently) and the generalized Koashi-Winter relations (\ref{KWS-gen}). $\Box$\pagebreak

\begin{lemma}\label{CB-R-ap} \emph{Let $\Phi(\rho)=P\rho P+[\Tr(I_A-P)\rho\shs]\tau$ be a channel $A\to A$, where $P$ is an orthogonal
projector in $\B(\H_A)$ and $\tau$ is a fixed state in $\S(\H_A)$. Then
\begin{equation}\label{CB-delta}
0\leq C_B(\omega)-C_B(\Phi\otimes \id_B(\omega))\leq (S(\omega_A)-S(P\omega_A P))+h_2(c_P),\;\;
\end{equation}
\begin{equation}\label{CB-delta+}
0\leq C^{\infty}_B(\omega)-C^{\infty}_B(\Phi\otimes \id_B(\omega))\leq 2(S(\omega_A)-S(P\omega_A P))+h_2(c_P)
\end{equation}
for any state $\omega$ in $\S(\H_{AB})$ such that $S(\omega_A)<+\infty$ and $c_P\doteq\Tr P\omega_A>0$,
where $S(P\omega_A P)\doteq c_PS(c_P^{-1}P\omega_A P)$.}
\end{lemma}\smallskip

\emph{Proof.} The first inequalities in (\ref{CB-delta}) and (\ref{CB-delta+}) follow from the monotonicity of $C_B$ and $C^{\infty}_B$ under local channels \cite{H&V}.

Let $m$ be a natural number, $\mathbb{M}=\{M_i\}$ an arbitrary POVM  on $\H^{\otimes m}_{B}$, $\Psi_{\M}$  a  channel from $\S(\H^{\otimes m}_{B})$ to $\S(\H_{E})$ defined in (\ref{Psi-M}) and $C_B^{\mathbb{M}}$ the function defined after (\ref{Psi-M}).

We will estimate the difference $\,\Delta^m(\omega)\doteq C_B^{\mathbb{M}}(\omega^{\otimes m})-C_B^{\mathbb{M}}((\Phi\otimes\id_B(\omega))^{\otimes m})\,$
by using the telescopic technique (cf.\cite{L&S}).  Let $\sigma^0_m=\id_{A^m}\otimes\Psi_{\M}(\omega^{\otimes m})$ and
$$
\sigma^k_m=[\Phi_1\otimes...\otimes\Phi_k\otimes\id_{A_{k+1}..A_m}]\otimes \id_{B^m}(\sigma^0_m),\quad k=\overline{1,m},
$$
where $A_i$ is the $i$-th copy of system $A$ and $\Phi_i$ denotes the copy of $\Phi$ acting from $A_i$ to itself. Then
$C_B^{\mathbb{M}}(\omega^{\otimes m})=I(A^m\!:\!E)_{\sigma^0_m}$ and $C_B^{\mathbb{M}}((\Phi\otimes\id_B(\omega))^{\otimes m})=I(A^m\!:\!E)_{\sigma^m_m}$. So, we have
\begin{equation}\label{Delta-m}
 \Delta^m(\omega)\leq \sum_{k=1}^m \delta_k,\quad \textrm{where}\quad  \delta_k=I(A^m\!:\!E)_{\sigma^{k-1}_m}-I(A^m\!:\!E)_{\sigma^k_m}\geq0.
\end{equation}
By noting that $\Tr_{A_k}\sigma^{k-1}_m=\Tr_{A_k}\sigma^{k}_m$ we obtain
$\delta_k=I(A_k\!:\!E|A^c_k)_{\sigma^{k-1}_m}-I(A_k\!:\!E|A^c_k)_{\sigma^k_m}$,
where $A^c_k=A^m\setminus A_k$. Since
$$
\sigma^k_m=\id_{A^c_kE}\otimes\Phi_k(\sigma^{k-1}_m)=I_{A^c_kE}\otimes P_k\cdot\sigma^{k-1}_m\cdot I_{A^c_kE}\otimes P_k+\tau_k\otimes \Tr_{A_k}\sigma^{k-1}_m\cdot I_{A^c_kE}\otimes \bar{P}_k,
$$
where $P_k$ and $\tau_k$ are the copies  of $P$ and $\tau$ acting on $\H_{A_k}$, $\bar{P}_k=I_{A_k}-P_k$, by using inequality (\ref{MI-LAA-1})
we obtain
$$
\begin{array}{c}
\delta_k\leq I(A_k\!:\!E|A^c_k)_{\sigma^{k-1}_m}-I(A_k\!:\!E|A^c_k)_{I_{A^c_kE}\otimes P_k\cdot\sigma^{k-1}_m\cdot I_{A^c_kE}\otimes P_k}+h_2(\Tr P_k[\sigma^{k-1}_m]_{A_k})\\\\
\leq 2(S(\omega_A)-S(P\omega_AP))+h_2(\Tr P\omega_A),
\end{array}
$$
where it is assumed that $\,f(\varrho)=[\Tr\varrho] f(\varrho/\Tr\varrho)\,$  for any positive trace class operator $\varrho$, $f=I(A_k\!:\!E|A^c_k),S$, and the second
inequality follows from Lemma 9 in \cite{SE}. This and (\ref{Delta-m}) imply that
$$
C_B^{\mathbb{M}}(\omega^{\otimes m})-C_B^{\mathbb{M}}((\Phi\otimes\id_B(\omega))^{\otimes m})\leq 2m(S(\omega_A)-S(P\omega_AP))+mh_2(\Tr P\omega_A).
$$
Since the r.h.s. of this inequality does not depend on $\mathbb{M}$, it implies (\ref{CB-delta+}).

To prove (\ref{CB-delta}) it suffices to apply the above arguments in the case $m=1$ and to use Lemma 25 in \cite{CHI} instead of Lemma 9 in \cite{SE}. $\Box$

\subsection{Entanglement of
Formation and its regularization}

\subsubsection{Definitions and the Nielsen-Winter-Wilde continuity bounds.} The Entanglement of
Formation (EoF) is one of the basic entanglement measures in bipartite quantum systems \cite{Bennett,ESP,4H,P&V,Wilde-new}. In a finite-dimensional bipartite system $AB$ the EoF is defined as the convex roof extension to the set $\S(\H_{AB})$ of the function $\omega\mapsto S(\omega_{A})$ on the set $\mathrm{ext}\shs\S(\H_{AB})$ of pure states in $\S(\H_{AB})$ , i.e.
\begin{equation}\label{ef-def}
  E_{F}(\omega)=\inf_{\sum_kp_k\omega_k=\omega}\sum_k p_kS([\omega_k]_{A}),
\end{equation}
where the infimum is over all finite ensembles $\{p_k, \omega_k\}$ of pure states in $\S(\H_{AB})$ with the average state $\omega$ \cite{Bennett}.
In this case $E_F$ is a uniformly continuous function on $\S(\H_{AB})$  possessing basic properties of an entanglement measure \cite{4H,P&V,Wilde-new}.
In fact, definition (\ref{ef-def}) and all these properties hold regardless of the dimension of system $B$ (which may be infinite).

The EoF is nonadditive (this follows, due to Shor's theorem in \cite{Shor}, from the nonadditivity of the minimal output entropy \cite{Hastings}), its regularization is defined as follows
\begin{equation*}
  E^{\infty}_F(\omega)=\lim_{m\rightarrow+\infty}m^{-1}E_F(\omega^{\otimes m})=\inf_{m}m^{-1}E_F(\omega^{\otimes m}),
\end{equation*}
where $\omega^{\otimes m}$ is treated as a state of the bipartite quantum system $(A^m)(B^m)$. In the finite-dimensional settings it is proved that
$E^{\infty}_F$ coincides with the entanglement cost $E_C$ \cite{E-C}. Generalization
of this result to the infinite-dimensional case  is an interesting open question (requiring, in particular,
finding  adequate definition of $E_C$ in this case).

The first continuity bound for the EoF in a finite-dimensional bipartite system $AB$ was obtained by Nielsen (his technique was used in the proof of Theorem \ref{L-1-ip}) \cite{Nielsen}.
Then Winter improved this continuity bound in \cite{W-CB} by using Nielsen's arguments and the continuity bound for the QCE of q-c-states  presented in Proposition \ref{AFW-CB}. Finally, Wilde used the optimal continuity bound for the QCE  of q-c-states  presented in Proposition \ref{Wilde-CB} to obtain
the most accurate continuity bound for the EoF to date.

Continuity bound for the regularized EoF is obtained by Winter by means of the "quantum coupling lemma" presented in \cite[Proposition 5]{W-CB}.
\smallskip

\begin{proposition}\label{EF-CB-1} \cite{Wilde-CB,W-CB} \emph{Let $AB$ be a bipartite quantum system such that $d=\min\{\dim\H_A,\dim\H_B\}<\infty$. Then
\begin{equation}\label{EF-CB-1+}
  |E_F(\rho)-E_F(\sigma)|\leq \delta\ln(d-1)+h_2(\delta),\quad \delta=\sqrt{\varepsilon(2-\varepsilon)},
\end{equation}
for any states $\rho$ and $\sigma$ in $\S(\H_{AB})$  such that $\,\frac{1}{2}\|\rho-\sigma\|_1\leq \varepsilon\leq 1-d^{-1}\sqrt{2d-1}\,$ and
\begin{equation*}
  |E^{\infty}_F(\rho)-E^{\infty}_F(\sigma)|\leq 2\delta\ln d+g(\delta),\quad \delta=\sqrt{\varepsilon(2-\varepsilon)},
\end{equation*}
for arbitrary states $\rho$ and $\sigma$ in $\S(\H_{AB})$  such that $\frac{1}{2}\|\rho-\sigma\|_1\leq \varepsilon\leq1$.}\smallskip

\emph{Inequality (\ref{EF-CB-1+}) holds  provided that $F(\rho,\sigma)\doteq \|\sqrt{\rho}\sqrt{\sigma}\|_1^2 \geq 1-\delta^2\geq d^{-2}(2d-1)$.}
\end{proposition}\medskip

The last statement follows from the proof of inequality (\ref{EF-CB-1+}) presented in \cite{Wilde-CB,W-CB}, since the condition $F(\rho,\sigma)\geq 1-\delta^2\geq d^{-2}(2d-1)$
implies, due to relation (\ref{F-Tn-eq}), that for any purification $\bar{\rho}$ of $\rho$ there is a purification $\bar{\sigma}$ of $\sigma$ such that
$\frac{1}{2}\|\bar{\rho}-\bar{\sigma}\|_1\leq \delta\leq 1-1/d$. Note that continuity bound (\ref{EF-CB-1+}) with $\delta=\sqrt{1-F(\rho,\sigma)}$ may be
\emph{significantly sharper} than the same bound with $\delta=\sqrt{\varepsilon(2-\varepsilon)}$, especially, if the states $\rho$ and $\sigma$ are nearly pure.
Indeed, if $\rho$ and $\sigma$ are pure states than inequality (\ref{EF-CB-1+}) with $\delta=\sqrt{1-F(\rho,\sigma)}\,$ implies the inequality
\begin{equation*}
  |S(\rho_A)-S(\sigma_A)|\leq \varepsilon\ln(\dim\H_A-1)+h_2(\varepsilon),\quad \varepsilon=\textstyle\frac{1}{2}\|\rho-\sigma\|_1,
\end{equation*}
which differs from the Audenaert optimal bound for $|S(\rho_A)-S(\sigma_A)|$ presented in \cite{Aud} by replacing $\frac{1}{2}\|\rho_A-\sigma_A\|_1$ with $\frac{1}{2}\|\rho-\sigma\|_1$ (that may be close to $\frac{1}{2}\|\rho_A-\sigma_A\|_1$).

\subsubsection{Continuity bound under the energy-type constraint.} If both systems $A$ and $B$ are infinite-dimensional then  there are two versions $E_{F,d}$ and $E_{F,c}$ of the EoF defined, respectively, by means of discrete and continuous convex roof extensions
\begin{equation}\label{E_F-def-d}
E_{F,d}(\omega)=\!\inf_{\sum_k\!p_k\omega_k=\omega}\sum_kp_kS([\omega_k]_A)\quad
\end{equation}
\begin{equation}\label{E_F-def-c}
E_{F,c}(\omega)=\!\inf_{\int\omega'\mu(d\omega')=\omega}\int\! S(\omega'_A)\mu(d\omega'),
\end{equation}
where the  infimum in (\ref{E_F-def-d}) is over all countable ensembles $\{p_k, \omega_k\}$ of pure states in $\S(\H_{AB})$ with the average state $\omega$ and the  infimum in (\ref{E_F-def-c}) is over all Borel probability measures on the set of pure states in $\S(\H_{AB})$ with the barycenter $\omega$ (the last infimum is always attained) \cite[Section 5]{EM}. It follows from definitions that $E_{F,d}(\omega)\geq E_{F,c}(\omega)$.

Physically, the discrete version of EoF is more preferable, but the continuous version has better analytical properties: it is proved in \cite[Section 5]{EM} that
the function $E_{F,c}(\omega)$ is lower semicontinuous on $\S(\H_{AB})$ and does not increase under generalized selective LOCC operations. Moreover,
the assumption $E_{F,d}\neq E_{F,c}$ leads to serious problems with the discrete version, in particular, it is not clear how to prove that $E_{F,d}$ vanishes on countably non-decomposable separable states \cite[Remark 6]{EM}. The conjecture of coincidence of $E_{F,d}$ and $E_{F,c}$ on $\S(\H_{AB})$ is an interesting open question. One can prove that this coincidence is equivalent to the lower semicontinuity of $E_{F,d}$ on $\S(\H_{AB})$.

Thus, at the moment we have two regularizations
\begin{equation}\label{E-F-r-def+d}
  E^{\infty}_{F,d}(\omega)=\lim_{m\rightarrow+\infty}m^{-1}E_{F,d}(\omega^{\otimes m})=\inf_{m}m^{-1}E_{F,d}(\omega^{\otimes m}),
\end{equation}
\begin{equation}\label{E-F-r-def+c}
  E^{\infty}_{F,c}(\omega)=\lim_{m\rightarrow+\infty}m^{-1}E_{F,c}(\omega^{\otimes m})=\inf_{m}m^{-1}E_{F,c}(\omega^{\otimes m}).
\end{equation}
One of the interesting open questions consists in  clarifying the relationship of the quantities $E^{\infty}_{F,d}$ and $E^{\infty}_{F,c}$ with the entanglement cost $E_C$ (provided that the appropriate definition of $E_C$ in the infinite-dimensional settings is found).

In \cite{EM} it is shown that $E_{F,d}(\omega)=E_{F,c}(\omega)$ for any state $\omega$ in $\S(\H_{AB})$ such that $\min\{S(\omega_{A}),S(\omega_{B})\}<+\infty$.
So, if we restrict attention to states in the set
$$
\S_{\!A}(\H_{AB})=\{\omega\in\S(\H_{AB})\,|\, S(\omega_{A})<+\infty\}
$$
we may forget about possible non-coincidence of $E_{F,d}$ and $E_{F,c}$. Since $S(\omega^{\otimes m}_{A^m})<+\infty$ for any state $\omega$ in $\S_{\!A}(\H_{AB})$ and any $m$, the same statements are valid for the reqularizations $E^{\infty}_{F,d}$ and $E^{\infty}_{F,c}$. So, we will denote by $E_F(\omega)$ and $E^{\infty}_{F}(\omega)$ the values  $E_{F,d}(\omega)=E_{F,c}(\omega)$ and $E^{\infty}_{F,d}(\omega)=E^{\infty}_{F,c}(\omega)$ for any state $\omega$ in $\S_{\!A}(\H_{AB})$ correspondingly. We will follow
this notation in Propositions \ref{E_F-WCB} and \ref{EF-CB-2} below, since the conditions imposed therein on the states $\rho$ and $\sigma$ imply that
these states belong to $\S_{\!A}(\H_{AB})$.

The following proposition gives Winter-type continuity bound for the EoF and its regularization under the energy-type constraint.
\smallskip

\begin{proposition}\label{E_F-WCB} \emph{Let $H$ be a positive operator on the space $\H_A$ satisfying conditions (\ref{H-cond}) and (\ref{star}). Let $\rho$ and $\sigma$ be states in $\S(\H_{AB})$ such that $\Tr H\rho_A,\Tr H\sigma_A\leq E$ and
$\,\frac{1}{2}\|\rho-\sigma\|_1\leq \varepsilon<1$. Let $\eta=\sqrt{\varepsilon(2-\varepsilon)}$. Then
\begin{equation}\label{E_F-WCB+}
\left|E_F(\rho)-E_F(\sigma)\right|\leq(\varepsilon'+2\delta)F_H\!\left(E/\delta\right)
+g(\varepsilon')+2h_2(\delta),
\end{equation}
\begin{equation}\label{E_F-WCB++}
\left|E^{\infty}_F(\rho)-E^{\infty}_F(\sigma)\right|\leq(2\varepsilon'+4\delta)F_H\!\left(E/\delta\right)
+g(\varepsilon')+2h_2(\delta),
\end{equation}
$\delta=(\varepsilon'-\eta)/(1+\varepsilon')$, for any $\varepsilon'\in(\eta,1]$, where $F_H$ is the function defined in (\ref{F-def}), $h_2(x)$ is the binary entropy and $g(x)$ is the function defined in (\ref{g-def}).}
\end{proposition}\smallskip

\begin{remark}\label{E_F-WCB-r}
The "free parameter" $\varepsilon'$  can be used to optimize continuity bounds (\ref{E_F-WCB+}) and (\ref{E_F-WCB++})
for given values of $E$ and $\varepsilon$. The faithfulness of this continuity bounds follows from the equivalence of
(\ref{H-cond}) and (\ref{H-cond-a}), it can be shown by taking $\,\varepsilon'=\sqrt{\eta}$.
\end{remark}\smallskip

\emph{Proof.} Continuity bound (\ref{E_F-WCB+}) is an improved version of the continuity bound for $E_F$ presented
in Proposition 22 in \cite{SE}. The improvement consists in using in the proof  more sharp continuity bound for the QCMI from  Proposition \ref{CMI-W-CB} in Section 5.2.2.

Continuity bound (\ref{E_F-WCB++}) can be obtained by repeating all the arguments from the proof of Corollary 4  in \cite{W-CB} with the use
of Winter's continuity bound for the QCE under the energy constraint (presented in Proposition \ref{Winter-CB} in Section 5.1.2) at the final step.
We have only to note that in this case one should use $\epsilon$-optimal decomposition of the state $\sigma^{AB}\otimes\Omega^{A'B'}$
introduced in that proof  because of possible nonexistence of the optimal decomposition. $\Box$

Another continuity bound for the function $E_F$ can be obtained by noting that this function  belongs to the class $N_{2,3}^1(1,1)$ in terms of Section 3.1.2 (in the settings $A_1=A$, $A_2=B$). Indeed,
this follows from the fact that the function $\rho\mapsto \frac{1}{2}I(A\!:\!B|E)_{\rho}$ belongs to the class $L_{3}^1(1,1)$ (in the settings $A_1=A$, $A_2=B$, $A_3=E$), since
it is easy to see that
\begin{equation*}
E_F(\omega)=\textstyle\frac{1}{2}\displaystyle\inf_{\hat{\omega}\in\mathfrak{M}_3(\omega)}I(A\!:\!B|E)_{\hat{\omega}},
\end{equation*}
where $\mathfrak{M}_3(\omega)$ is the set of all q-c states $\hat{\omega}=\sum_k p_k \omega_k\otimes |k\rangle\langle k|$
such that $\sum_k p_k \omega_k=\omega$ and $\{\omega_k\}$ is a set of pure states in $\S(\H_{AB})$.

By applying Theorem \ref{L-1-ip} in Section 3.2.3 and Winter's arguments from \cite{W-CB} we obtain
\smallskip

\begin{proposition}\label{EF-CB-2} \emph{Let $H$ be a positive operator on the space $\H_{A}$ satisfying condition (\ref{H-cond+}) and $F_{H}$ the function defined in (\ref{F-def}). Let $\,g(x)$ be the function defined in (\ref{g-def}). Then
\begin{equation}\label{EF-CB-2+}
    |E_F(\rho)-E_F(\sigma)|\leq \delta F_{H}\!\!\left[\frac{2E}{\delta^2}\right]+g(\delta),\;\;
\end{equation}
\begin{equation}\label{EF-CB-2++}
    |E^{\infty}_F(\rho)-E^{\infty}_F(\sigma)|\leq 2\delta F_{H}\!\!\left[\frac{2E}{\delta^2}\right]+g(\delta)
\end{equation}
for any states $\rho$ and $\sigma$ in $\S(\H_{AB})$ such that $\,\Tr H\rho_{A},\,\Tr H\sigma_{A}\leq E\,$ and
\begin{equation}\label{cl-cond}
  \varepsilon(2-\varepsilon)\leq\delta^2\leq1,\quad \textit{where} \quad  \varepsilon=\textstyle\frac{1}{2}\|\rho-\sigma\|_1.
\end{equation}
Inequality (\ref{EF-CB-2+}) remains valid  if  condition (\ref{cl-cond}) is replaced by the weaker condition $F(\rho,\sigma)\doteq \|\sqrt{\rho}\sqrt{\sigma}\|_1^2 \geq 1-\delta^2\geq0$.}
\end{proposition}\medskip

\textbf{Note:} the r.h.s. of (\ref{EF-CB-2+}) and (\ref{EF-CB-2++}) tend to zero as $\,\delta\to 0\,$ due to the equivalence of (\ref{H-cond+}) and (\ref{H-cond+a}).
\smallskip

\emph{Proof.} Continuity bound (\ref{EF-CB-2+}) and the last claim directly follow from Theorem \ref{L-1-ip} with $m=1$ and the observations before the proposition.

Continuity bound (\ref{EF-CB-2++}) can be obtained by repeating all the arguments from the proof of Corollary 4 in \cite{W-CB} with the use
of $\epsilon$-optimal decomposition of the state $\sigma^{AB}\otimes\Omega^{A'B'}$ (because of possible nonexistence of the optimal decomposition). Indeed, in
the proof of Corollary 4 Winter use the states $\psi$ and $\Theta$ such that $F(\psi,\Theta)\geq(1-\varepsilon)^2$. By monotonicity of the fidelity under
quantum channels we have $F(\tilde{\rho},\tilde{\sigma})\geq(1-\varepsilon)^2$ and hence $1-F(\tilde{\rho},\tilde{\sigma})\leq \varepsilon(2-\varepsilon)\leq\delta^2$, where $\tilde{\rho}$ and $\tilde{\sigma}$ are the states introduced by Winter later. Thus,
by applying Proposition \ref{Sh-CB} in Section 5.1.2 at the last step of that proof we obtain
$$
E_F(\rho\otimes\Omega)-E_F(\sigma\otimes\Omega)\leq S(A|A'X)_{\tilde{\rho}}-S(A|A'X)_{\tilde{\sigma}}+\epsilon\leq 2\delta F_{H}\!\!\left[\frac{2E}{\delta^2}\right]+g(\delta)+\epsilon.\quad \Box
$$
It is easy to show that the continuity bounds for the functions $E_F$ and $E^{\infty}_F$ given by Proposition \ref{EF-CB-2}
are more accurate than the continuity bounds given by Proposition \ref{EF-CB-1} if the function $F_{H}$ has a logarithmic
growth, in particular, if $H$ is the Hamiltonian of a multi-mode quantum oscillator.  At the same time, the continuity bounds in Proposition \ref{EF-CB-1} are
faithful if the operator $H$ satisfies the Gibbs condition (\ref{H-cond}) which is  weaker than the condition (\ref{H-cond+}) required in Proposition \ref{EF-CB-2}. Hence, they imply the uniform continuity and
the asymptotic continuity of the functions $E_F$ and $E^{\infty}_F$ on a wider set of states.

\subsubsection{Local continuity conditions.} By Proposition \ref{EF-CB-1} the EoF and its regularization  are uniformly continuous on the set
$\S(\H_{AB})$ provided that at least one of the systems $A$ and $B$ is finite-dimensional. If both systems $A$ and $B$ are infinite-dimensional then these
functions are discontinuous and may take the  value $+\infty$. Analytical properties of the functions $E_{F,d}$, $E_{F,c}$, $E^{\infty}_{F,d}$ and $E^{\infty}_{F,c}$
defined in (\ref{E_F-def-d})-(\ref{E-F-r-def+c}) are described in  the following
\smallskip

\begin{proposition}\label{EF-2} A) \emph{The function $E_{F,c}$ is lower semicontinuous on the set $\S(\H_{AB})$. The function $E_{F,d}$ is finite and coincides with
the function $E_{F,c}$ on the set
\begin{equation*}
\S_*(\H_{AB})\doteq\left\{\omega\in\S(\H_{AB})\,|\,\min\{S(\omega_{A}),S(\omega_{B})\}<+\infty\right\}.
\end{equation*}
For any sequence $\{\omega_n\}\subset\S(\H_{AB})$ converging to a state $\omega_0$ in $\S_*(\H_{AB})$ it holds}
\begin{equation}\label{EF-lsc}
\liminf_{n\to+\infty}E_{F,d}(\omega_n)\geq E_{F,d}(\omega_0).
\end{equation}

\emph{The functions $E^{\infty}_{F,d}$ and $E^{\infty}_{F,c}$  coincide on the set $\S_*(\H_{AB})$, on which they  are finite and lower semicontinuous.} \emph{Moreover,
\begin{equation}\label{EFR-lsc}
\liminf_{n\to+\infty}E^{\infty}_{F,d}(\omega_n)\geq E^{\infty}_{F,d}(\omega_0),\quad\liminf_{n\to+\infty}E^{\infty}_{F,c}(\omega_n)\geq E^{\infty}_{F,c}(\omega_0)
\end{equation}
for any sequence $\{\omega_n\}\subset\S(\H_{AB})$ converging to a state $\omega_0$ in $\S_*(\H_{AB})$.}\smallskip

\noindent B) \emph{Let $\{\omega_n\}$ be a sequence
in $\S(\H_{AB})$ converging to a state $\omega_0$  that belongs to the class $\Upsilon^{*}_{\!AB}$ (see Definition \ref{class-2} at the begin of Section 5.2.3.).}

\emph{If $\omega_0\in\S_*(\H_{AB})$ then}
\begin{equation}\label{EoF-conv+}
  \lim_{n\to+\infty }E_{F,d}(\omega_n)=\lim_{n\to+\infty }E_{F,c}(\omega_n)=E_{F,d}(\omega_0)=E_{F,c}(\omega_0)<+\infty,
\end{equation}
\begin{equation}\label{EoF-conv++}
  \lim_{n\to+\infty }E^{\infty}_{F,d}(\omega_n)=\lim_{n\to+\infty }E^{\infty}_{F,c}(\omega_n)=E^{\infty}_{F,d}(\omega_0)=E^{\infty}_{F,c}(\omega_0)<+\infty.
\end{equation}

\emph{If $\{\Phi_n:A\to A'\}$ and $\{\Psi_n:B\to B'\}$ are sequences of quantum channels strongly converging\footnote{A sequence of channels $\Phi_n$ strongly converges to a channel $\Phi_0$ if $\lim_{n\to+\infty}\Phi_n(\rho)=\Phi_0(\rho)$ for any input state $\rho$ \cite{H-SCI,CSR}.} to channels $\Phi_0$ and $\Psi_0$ such that the state $\Phi_0\otimes\Psi_0(\omega_0)$ belongs to the set $\S_*(\H_{A'B'})$ then
(\ref{EoF-conv+}) and (\ref{EoF-conv++}) hold with $\omega_n$ replaced by $\Phi_n\!\otimes\!\Psi_n(\omega_n)$ for all $n\geq0 $.}
\end{proposition}\smallskip

\emph{Proof.} A) The lower semicontinuity of the function $E_{F,c}$ on $\S(\H_{AB})$ and its coincidence with $E_{F,d}$ on the set $\S_*(\H_{AB})$ are  proved in \cite{EM} by using the stability and the $\mu$-compactness of the set of quantum states along
with the lower semicontinuity of the von Neumann entropy. These properties imply the limit  relation (\ref{EF-lsc}), since  $E_{F,d}(\omega)\geq E_{F,c}(\omega)$ for any state
$\omega$ in $\S(\H_{AB})$.

The coincidence of the functions $E^{\infty}_{F,d}$ and $E^{\infty}_{F,c}$ on the set $\S_*(\H_{AB})$ follows from their definitions and the above observations, since the state $\omega^{\otimes m}$
belongs to the set $\S_*(\H_{(A^m)(B^m)})$ for any state $\omega$ in $\S_*(\H_{AB})$ and any natural $m$.

Consider the functions
$$
f_A(\omega)=\sup_{\Phi\in\F_A}E^{\infty}_{F,c}(\Phi\otimes \id_B(\omega))\quad \textrm{and} \quad f_B(\omega)=\sup_{\Psi\in\F_B}E^{\infty}_{F,c}(\id_A\otimes\Psi(\omega))
$$
where $\F_X$ is the set of all channels from $X$ to itself with finite-dimensional output. By Proposition \ref{EF-CB-1} the functions  $\omega \mapsto E^{\infty}_{F,c}(\Phi\otimes\id_B(\omega))$ and $\omega\mapsto E^{\infty}_{F,c}(\id_A\otimes\Psi(\omega))$ are continuous on $\S(\H_{AB})$.
Hence, the function $f_m=\max\{f_A,f_B\}$ is lower semicontinuous  on $\S(\H_{AB})$. Thus, since
\begin{equation*}
f_m(\omega)\leq E^{\infty}_{F,c}(\omega)\leq E^{\infty}_{F,d}(\omega),\quad \forall\omega\in\S(\H_{AB}),
\end{equation*}
by monotonicity of  $E^{\infty}_{F,c}$ under local channels, to prove the limit  relations in (\ref{EFR-lsc}) it suffices to show  that
\begin{equation}\label{rq-r-EF}
f_m(\omega)=E^{\infty}_{F,c}(\omega)=E^{\infty}_{F,d}(\omega)\quad \forall\omega\in\S_*(\H_{AB}).
\end{equation}

Assume that $\omega$ is a state in $\S(\H_{AB})$ such that $S(\omega_A)<+\infty$. By Proposition 4 in \cite{EC} there is a positive operator $H$
satisfying the Gibbs condition (\ref{H-cond}) such that $\Tr H\omega_A<+\infty$. Let $P_k$ be the spectral projector of $H$ corresponding
to its $k$ maximal eigenvalues (taking the multiplicity into account). Consider the sequence of channels
$\Phi_k(\rho)=P_k\rho P_k+[\Tr(I_A-P_k)\rho\shs]\tau$, $\rho\in\S(\H_A)$, where $\tau$ is a pure state in $\S(\H_A)$ corresponding to the minimal eigenvalue of $H$. Then it is easy to show
that  $\Tr H\Phi_k(\omega_A)\leq \Tr H\omega_A<+\infty$ for all $k$. Thus, by applying continuity bound for $E^{\infty}_{F}$ from Proposition \ref{E_F-WCB}
we conclude that $E^{\infty}_{F,c}(\Phi_k\otimes\id_B(\omega))$ tends to $E^{\infty}_{F,c}(\omega)$ as $k\to+\infty$.
Since $\Phi_k\in\F_A$ for all $k$, this implies (\ref{rq-r-EF}).

B) Show first that limit relations (\ref{EoF-conv+}) and (\ref{EoF-conv++}) hold
provided that
\begin{equation}\label{HA-conv+}
  \lim_{n\to+\infty }S([\omega_n]_X)=S([\omega_0]_X)<+\infty,\;\, \textrm{ where}\; X\; \textrm{is either}\; A\textrm{ or }B,
\end{equation}
(one of these implications is proved in \cite[Proposition 8]{EM}). If $X=A$ then this can be done by using Lemma \ref{L-S} in Section 4.1, part A of this proposition and the representations
$$
E_{F,d}(\omega)=E_{F,c}(\omega)=S(\omega_A)-\chi_{A}(\omega),\quad E^{\infty}_{F,d}(\omega)=E^{\infty}_{F,c}(\omega)=S(\omega_A)-\chi^{\infty}_{A}(\omega)
$$
valid for any state $\omega$ in $\S(\H_{AB})$ with finite $S(\omega_A)$, where $\chi_{_A}(\omega)$ and $\chi^{\infty}_{A}(\omega)$ are the constrained Holevo capacity
of the partial trace channel $\Phi_A:\omega\mapsto \omega_A$ at a state $\omega$ and its regularization
(see Section 5.5 below). It suffices to note that the functions $\chi_{A}$ and $\chi^{\infty}_{A}$ are
lower semicontinuous on $\S(\H_{AB})$ by Proposition \ref{CHI-LC} in Section 5.5.3.

Let $\{\omega_n\}\subset\S(\H_{AB})$ be a converging sequence satisfying the conditions of part B of the proposition. Since $\omega_0\in\S_*(\H_{AB})$, it is shown in the proof of part A that
$$
E_{F,d}(\omega_0)=E_{F,c}(\omega_0)<+\infty\quad \textrm{and} \quad E^{\infty}_{F,d}(\omega_0)=E^{\infty}_{F,c}(\omega_0)=f_m(\omega_0)<+\infty,
$$
where $f_m$ is the function introduced in that proof. Since the functions $E_{F,c}$ and $f_m$ are lower semicontinuous on $\S(\H_{AB})$,
$$
E_{F,c}(\omega_n)\leq E_{F,d}(\omega_n)\quad \textrm{and} \quad f_m(\omega_n)\leq E^{\infty}_{F,c}(\omega_n)\leq E^{\infty}_{F,d}(\omega_n)\quad \forall n,
$$
to prove (\ref{EoF-conv+}) and (\ref{EoF-conv++}) it suffices to show that
\begin{equation}\label{EoF-usc}
  \limsup_{n\to+\infty }E_{F,d}(\omega_n)\leq E_{F,d}(\omega_0)\quad \textrm{and} \quad\limsup_{n\to+\infty }E^{\infty}_{F,d}(\omega_n)\leq E^{\infty}_{F,d}(\omega_0).
\end{equation}
Both of these relations are proved similarly by using such common properties of $E_{F,d}$ and $E^{\infty}_{F,d}$ as convexity and monotonicity under local channels (the convexity of $E^{\infty}_{F,d}$ follows from Proposition 13 in \cite{DHR}).
So, in what follows we will assume that $E^*_{F,d}$ is either $E_{F,d}$ or $E^{\infty}_{F,d}$.  We will also use essentially the implication (\ref{HA-conv+})$\,\Rightarrow\,$(\ref{EoF-conv+}),(\ref{EoF-conv++}) mentioned before. We will refer it as
"continuity condition (\ref{HA-conv+})" for brevity.

By definition of the class $\Upsilon^*_{AB}$ there is a  sequence $\{\hat{\omega}_n\}$
in $\S(\H_{ABE})$ converging to a state $\hat{\omega}_0$ such that  $[\hat{\omega}_n]_{AB}=\omega_n$ for all $n\geq 0$ and limit relation (\ref{HAE-conv}) holds.

Let $\Psi_m(\hat{\omega}_n)=P_m^{\hat{\omega}_n}\hat{\omega}_n$ for any $m$ and $n\geq0$, where $P_m^{\hat{\omega}_n}$ is the spectral projector of
$\hat{\omega}_n$ corresponding to its $m$ maximal eigenvalues (taken the multiplicity into account). It is shown at the end of Section 4.2.1
that the family $\{\Psi_m\}$  of maps from $\,\A=\{\hat{\omega}_n\}\cup\{\hat{\omega}_0\}\,$ into $\T_+(\H_{ABE})$ satisfies all the conditions stated before Lemma \ref{DTL-L}. In this case the set $M_{\hat{\omega}_0}$
consists of all natural numbers $m$ such that either $\lambda^{\hat{\omega}_0}_{m+1}<\lambda^{\hat{\omega}_0}_{m}$ or $\lambda^{\hat{\omega}_0}_{m}=0$ ($\{\lambda^{\hat{\omega}_0}_{k}\}_k$ is a sequence of eigenvalues of $\hat{\omega}_0$ taken in the non-increasing order).

Denote by $\hat{\omega}^m_n$ and $\mu^m_n$ the state  proportional to
 the operator $\Psi_m(\hat{\omega}_n)$ and the positive number $\Tr\Psi_m(\hat{\omega}_n)$ correspondingly (we will assume that $m\in \mathbb{N}_{\A}$, see (\ref{N-def})).

The function $f(\hat{\omega})=S(\hat{\omega}_{AE})$ is nonnegative concave and lower semicontinuous on $\S(\H_{ABE})$. By  condition (\ref{HAE-conv}) we have
\begin{equation*}
  \lim_{n\to+\infty }f(\hat{\omega}_n)=f(\hat{\omega}_0)<+\infty.
\end{equation*}
The function $g(\hat{\omega})=E^*_{F,d}(\hat{\omega}_{AB})$ is nonnegative and convex on $\S(\H_{ABE})$. Moreover,
\begin{equation*}
g(\hat{\omega})= E^*_{F,d}(\hat{\omega}_{A|B})\leq E^*_{F,d}(\hat{\omega}_{AE|B})\leq S(\hat{\omega}_{AE})=f(\hat{\omega})\quad \forall \hat{\omega} \in\S(\H_{AEB}),
\end{equation*}
where the first inequality follows from the monotonicity of $E^*_{F,d}$ under local channels, while the second one is due to the obvious upper bound for $E^*_{F,d}$ \cite{Bennett,P&V,EM}.

Thus, to prove  (\ref{EoF-usc}) it suffices, by Corollary \ref{DTL-c} in Section 4.2.1, to show that
\begin{equation}\label{d-l-r+}
 \lim_{m\to+\infty}g(\hat{\omega}^m_0)=g(\hat{\omega}_0)\quad \textrm{and} \quad \lim_{n\to+\infty }g(\hat{\omega}^m_n)=g(\hat{\omega}^m_0)<+\infty\
\end{equation}
for all $m$ in $M_{\hat{\omega}_0}$. The first limit relation
in (\ref{d-l-r+}) follows from the continuity condition (\ref{HA-conv+}), since the assumed finiteness of $S([\hat{\omega}_0]_{X})$, where $X$ is either $A$ or $B$, implies that $S([\hat{\omega}_0^m]_{X})$ tends to
$S([\hat{\omega}_0]_{X})$ as $m\to+\infty$ by Simon's dominated convergence theorem \cite[the Appendix]{Ruskai} (because $\mu_0^m[\hat{\omega}_0^m]_{X}\leq [\hat{\omega}_0]_{X}$ for all $m$ and $\mu_0^m$ tends to $1$ as $m\to+\infty$).

To prove the second limit relation in (\ref{d-l-r+}) note first that
\begin{equation}\label{2-l-r+1}
\lim_{n\to+\infty }S(\hat{\omega}^m_n)=S(\hat{\omega}^m_0)<+\infty\quad \forall m\in M_{\hat{\omega}_0}
\end{equation}
and
\begin{equation}\label{2-l-r+2}
\lim_{n\to+\infty }S([\hat{\omega}^m_n]_{AE})=S([\hat{\omega}^m_0]_{AE})<+\infty\quad \forall m\in M_{\hat{\omega}_0}.
\end{equation}
The limit relation (\ref{2-l-r+1}) follows from the convergence of $\hat{\omega}^m_n$ to $\hat{\omega}^m_0$  as $n\rightarrow+\infty$, since $\rank\shs \hat{\omega}^m_n\leq m$ for all $n\geq0$. The limit relation (\ref{2-l-r+2}) follows from relation (\ref{HAE-conv}) by Corollary 4 in \cite{SSP}, since $\hat{\omega}^m_n$ tends to $\hat{\omega}^m_0$ as $n\rightarrow+\infty$, $\mu^m_n$ tends to $1$ as $m\rightarrow+\infty$ uniformly on $n$ (by Lemma \ref{DTL-L} in Section 4.2.1) and $\,\mu^m_n[\hat{\omega}^m_n]_{AE}\leq[\hat{\omega}_n]_{AE}\,$ for all $m$ and $n$.

By Proposition \ref{QCMI-lc} in Section 5.2.3 the limit relation (\ref{2-l-r+2}) implies that
\begin{equation*}
  \lim_{n\to+\infty}I(AE\!:\!B)_{\hat{\omega}_n^m}=I(AE\!:\!B)_{\hat{\omega}_0^m}<+\infty.
\end{equation*}
Hence, both limit
relations (\ref{2-l-r+1}) and (\ref{2-l-r+2}) show, by the second formula in (\ref{MI-e1}), that
\begin{equation*}
\lim_{n\to+\infty }S([\hat{\omega}^m_n]_{B})=S([\hat{\omega}^m_0]_{B})<+\infty.
\end{equation*}
By applying  the continuity condition (\ref{HA-conv+}), we obtain the second limit relation in (\ref{d-l-r+}).

The last claim of part B  follows from the main claim  and the last assertion of Proposition \ref{DCT-MI} in Section 5.2.3. $\square$

The continuity condition given by Proposition \ref{EF-2}B is essentially  stronger than the continuity condition (\ref{HA-conv+}). This  can be illustrated
by the following\smallskip
\begin{example}\label{EoF-ex}
Let $\{\rho_n\}$ and $\{\sigma_n\}$ be sequences of states
in $\S(\H_{AB})$ converging to states $\rho_0$ and $\sigma_0$ such that
$$
\lim_{n\to+\infty }S([\rho_n]_A)=S([\rho_0]_A)<+\infty\quad \textrm{and}\quad \lim_{n\to+\infty }S([\sigma_n]_B)=S([\sigma_0]_B)<+\infty.
$$
Proposition \ref{EF-2} and the last assertion of Proposition \ref{DCT-MI} in Section 5.2.3 allow us to prove that (\ref{EoF-conv+}) and (\ref{EoF-conv++})
hold with $\,\omega_n=p_n\rho_n+(1-p_n)\sigma_n$, $n\geq0$, for arbitrary sequence $\{p_n\}\subset[0,1]$ converging to $p_0$ provided that
either $S([\rho_0]_B)$ or $S([\sigma_0]_A)$ is finite.
This assertion can not be proved by using continuity condition (\ref{HA-conv+}).
\end{example}

By reformulating Example 3 in \cite{LSE} one can get another nontrivial example that shows efficiency of the continuity condition in Proposition \ref{EF-2}B.

\subsection{The constrained Holevo capacity of a partial trace and its regularization}

\subsubsection{Definitions and the AFW-type continuity bounds.} For an arbitrary state $\omega$ of a bipartite system $AC$ consider the quantity
\begin{equation*}
\chi_A(\omega)=\sup_{\sum_k p_k\omega_k=\omega}\sum_{k}p_k D([\omega_k]_A\|\shs\omega_A)=\sup_{\hat{\omega}\in\mathfrak{M}_2(\omega)}I(A\!:\!E)_{\hat{\omega}},
\end{equation*}
where the first supremum is over all countable ensembles $\{p_k, \omega_k\}$ of  states in $\S(\H_{AB})$ with the average state $\omega$ and
the second supremum is over all extensions of $\omega$ having the form (\ref{m-2}).  The quantity $\chi_A(\omega)$ is the constrained Holevo capacity
$\bar{C}(\Phi_A,\omega)$ of the partial trace channel $\Phi_A:\varpi\mapsto \varpi_A$ at a state $\omega$ defined in (\ref{CHI-QC-def}). It is shown in \cite{H-Sh-2}  that
\begin{equation*}
\chi_A(\omega)=\sup_{\int\omega'\mu(d\omega')=\omega}\int\!D(\omega'_A\|\shs\omega_A)\mu(d\omega'),
\end{equation*}
where the supremum is over all Borel probability measures $\mu$ on the set $\S(\H_{AC})$ with the barycenter $\omega$ (this supremum is attained if either $S(\omega)$ or $S(\omega_A)$ is finite \cite{H-Sh-2,Sh-H-ED}).

The function $\chi_A(\omega)$ is nonadditive (this follows, due to Shor's theorem in \cite{Shor}, from the nonadditivity of the minimal output entropy \cite{Hastings}).  Its regularization is defined as
\begin{equation*}
\chi^{\infty}_A(\omega)=\sup_m m^{-1}\chi_{A^m}(\omega^{\otimes m})=\lim_{m\to+\infty} m^{-1}\chi_{A^m}(\omega^{\otimes m}),
\end{equation*}
where $\omega^{\otimes m}$ is treated as a state of the bipartite system $(A^m)(C^m)$. It is easy to see that
\begin{equation*}
\chi_A(\omega)\leq\chi^{\infty}_A(\omega)\leq \min\{S(\omega),S(\omega_A)\}
\end{equation*}
for arbitrary state $\omega\in\S(\H_{AC})$ and that
\begin{equation}\label{CHI-1}
\chi_A(\omega)=S(\omega_A)-E_{F,d}(\omega)=S(\omega_A)-E_{F,c}(\omega),
\end{equation}
\begin{equation}\label{CHI-1R}
\chi^{\infty}_A(\omega)=S(\omega_A)-E^{\infty}_{F,d}(\omega)=S(\omega_A)-E^{\infty}_{F,c}(\omega)
\end{equation}
for any state $\omega\in\S(\H_{AC})$ such that $S(\omega_A)<+\infty$, where $E_{F,d}(\omega)$ and $E_{F,c}(\omega)$ are the discrete and continuous versions of
the EoF defined in (\ref{E_F-def-d}) and (\ref{E_F-def-c}) with $C=B$, $E^{\infty}_{F,d}(\omega)$ and $E^{\infty}_{F,c}(\omega)$ are their regularizations
defined in (\ref{E-F-r-def+d}) and (\ref{E-F-r-def+c}).

By updating the arguments from the proof of Theorem 1 in \cite{K&W} one can obtain
the generalised Koashi-Winter relations
\begin{equation}\label{KWS-gen}
C_B(\omega_{AB})=\chi_A(\omega_{AC}),\quad C_B^{\infty}(\omega_{AB})=\chi^{\infty}_A(\omega_{AC})
\end{equation}
valid for any pure state $\omega$ in $\S(\H_{ABC})$, where $C_B$ and $C_B^{\infty}$ are the one way classical correlation and its regularization defined in Section 5.3. If $S(\omega_A)<+\infty$ then the equalities in (\ref{KWS-gen}) are reduced to the standard Koashi-Winter relations due to representations (\ref{CHI-1}) and (\ref{CHI-1R}).
The advantage of the relations in (\ref{KWS-gen}) consists in their \emph{independence of the condition} $S(\omega_A)<+\infty$. The same arguments from \cite{K&W} show that for
any mixed state $\omega$ in $\S(\H_{ABC})$ both relations in (\ref{KWS-gen}) hold with $"="$ replaced by $"\leq"$. Using these relations it is easy to establish
the following \emph{monotonicity property}:
\begin{equation}\label{CHI-M}
  \chi_A(\omega)\leq\chi_A(I_A\otimes\Phi(\omega)),\quad\chi^{\infty}_A(\omega)\leq\chi^{\infty}_A(I_A\otimes\Phi(\omega))
\end{equation}
for any channel $\Phi:C\to C$ and any state $\omega\in\S(\H_{AC})$ that can be also directly deduced from the definitions of $\chi_A$ and $\chi^{\infty}_A$.

Since $I(A\!:\!E)_{\omega}\leq S(\omega_A)$ for any state $\omega$ in $\S(\H_{ACE})$ that is separable w.r.t. the decomposition $(AC)(E)$ \cite{MI-B,Wilde},  the function $\S(\H_{ACE})\ni\omega\mapsto I(A\!:\!E)_{\omega}$
belongs to the class $L^1_3(1,2|\S_{\rm qc})$ in the settings $A_1=A$, $A_2=C$, $A_3=E$, where $\S_{\rm qc}$ is the set of all q-c states having the form (\ref{m-2}) with $n=3$ (we use the notation of Section 3.1.2).
Hence, the function $\chi_A$ belongs to the class $N_{2,2}^1(1,2)$ in the settings $A_1=A$, $A_2=C$.

Since $I(A\!:\!E)_{\omega}\leq I(AC\!:\!E)_{\omega}\leq S(\omega_{AC})$ for any state $\omega$ in $\S(\H_{ACE})$ which is separable w.r.t. the decomposition $(AC)(E)$ \cite{MI-B,Wilde}, $\S(\H_{ACE})\ni\omega\mapsto I(A\!:\!E)_{\omega}$ is a function
from the class $L^1_2(1,2|\S_{\rm qc})$  in the settings $A_1=AC$, $A_2=E$, where $\S_{\rm qc}$ is the set of all q-c states having the form (\ref{m-2}) with $n=2$.
It follows that the function $\chi_A$ belongs to the class $N^1_{1,2}(1,2)$ in the settings $A_1=AC$.
\smallskip

\begin{proposition}\label{CHI-1-CB} \emph{Let $\rho$ and $\sigma$ be arbitrary states in $\S(\H_{AC})$ and $\H_*$  the minimal subspace containing the supports of  $\rho$ and $\sigma$. Assume that $d=\min\left\{\dim\H_A, \dim\H_*\right\}<+\infty$.
Let $\varepsilon=\textstyle\frac{1}{2}\|\rho-\sigma\|_1$, $F(\rho,\sigma)=\|\sqrt{\rho}\sqrt{\sigma}\|_1^2$ and $g(x)$ be the function defined in (\ref{g-def}).}

\emph{If  $F(\rho,\sigma)\geq 1-\delta^2$ then}
\begin{equation}\label{CHI-R}
|\chi_A(\rho)-\chi_A(\sigma)|\leq \delta\ln d+2g(\delta).
\end{equation}

\emph{If $\,\varepsilon(2-\varepsilon)\leq\delta^2$ then (\ref{CHI-R}) holds without  factor $2$ in the right hand side and}
\begin{equation}\label{CHI-R+}
|\chi^{\infty}_A(\rho)-\chi^{\infty}_A(\sigma)|\leq 2\delta\ln d+g(\delta).
\end{equation}
\end{proposition}\smallskip

\emph{Proof.} The first assertion of the proposition directly follows from the observations before the proposition and Corollary \ref{CB-N} in Section 3.1.2.

Let $\,\varepsilon(2-\varepsilon)\leq\delta^2$. Inequality (\ref{CHI-R+}) can be proved  by using Winter's arguments from the proof of the continuity bound for $E_F^{\infty}$ in Corollary 4 in \cite{W-CB}. One should repeat all the steps of that proof with the systems $C,E$ instead of $B,X$, the function $-\chi_A$ instead of $E_F$ and the function  $-I(A\!:\!E)$ instead of $S(A|X)$. At the final step we obtain
\begin{equation*}
\begin{array}{rl}
 \chi_A(\sigma\otimes\Omega)-\chi_A(\rho\otimes\Omega) \!\! & \leq I(AA'\!:\!E)_{\tilde{\sigma}}-I(AA'\!:\!E)_{\tilde{\rho}} \\\\
 & = I(A\!:\!E|A')_{\tilde{\sigma}}-I(A\!:\!E|A')_{\tilde{\rho}} \leq 2\delta\ln d+g(\delta),
\end{array}
\end{equation*}
where the identity $\,I(AA'\!:\!E)=I(A\!:\!E|A')+I(A'\!:\!E)\,$  and the coincidence of $\tilde{\rho}_{A'E}$ and $\tilde{\sigma}_{A'E}$ were used. The last inequality follows
from Proposition \ref{CMI-CB-S} in Section 5.2.1, since $\frac{1}{2}\|\tilde{\rho}-\tilde{\sigma}\|_1\leq \delta$ and $\tilde{\rho}_{A'E}=\tilde{\sigma}_{A'E}$.

To prove the possibility to remove the factor $2$ in  (\ref{CHI-R}) assume that $\chi_A(\rho)\geq\chi_A(\sigma)$. By Proposition 5 in \cite{W-CB} there exist a
system $R$, a pure state $\bar{\rho}$ in $\S(\H_{ACR})$ and a state $\bar{\sigma}$ in $\S(\H_{ACR})$ such that
$\bar{\rho}_{AC}=\rho$, $\bar{\sigma}_{AC}=\sigma$, $\bar{\rho}_{R}=\bar{\sigma}_{R}$ and $F(\bar{\rho},\bar{\sigma})\geq(1-\varepsilon)^2$.

For given $\epsilon>0$ let $\hat{\rho}$ be an extension of $\rho$ having form (\ref{m-2}) such that
\begin{equation*}
I(A\!:\!E)_{\hat{\rho}}\geq \chi_A(\rho)-\epsilon.
\end{equation*}
By the Schrodinger-Gisin–Hughston–Jozsa–Wootters theorem \cite{Schr,Gisin,HJW} there is a POVM $\{M_k\}$
in $R$ such that $p_k\rho_k=\mathrm{Tr}_R [I_{AC}\otimes M_k]\shs\bar{\rho}\,$ for all $k$.  So, if
$\Lambda(\varrho)=\sum_k [\mathrm{Tr} M_k\varrho]|k\rangle\langle k|$
is a q-c channel from $R$ to $E$ then  $\hat{\rho}=\id_{AC}\otimes \Lambda(\bar{\rho})$.

Since $\hat{\sigma}=\id_{AC}\otimes \Lambda(\bar{\sigma})$ is an extension of $\sigma$ having form (\ref{m-2}),
we have $I(A\!:\!E)_{\hat{\sigma}}\leq \chi_A(\sigma)$ and hence
\begin{equation}\label{LS-tmp}
\chi_A(\rho)-\chi_A(\sigma)\leq I(A\!:\!E)_{\hat{\rho}}-I(A\!:\!E)_{\hat{\sigma}}+\epsilon \leq \delta\ln d+g(\delta)+\epsilon,
\end{equation}
where the last inequality follows
from Proposition \ref{CMI-CB-S} in Section 5.2.1, since $\,\frac{1}{2}\|\hat{\rho}-\hat{\sigma}\|_1\leq \frac{1}{2}\|\bar{\rho}-\bar{\sigma}\|_1\leq\sqrt{1-F(\bar{\rho},\bar{\sigma})}\leq \sqrt{\varepsilon(2-\varepsilon)}\leq\delta\,$ and $\,\hat{\rho}_{E}=\hat{\sigma}_{E}\,$. $\Box$

\subsubsection{Continuity bounds under the energy-type constraint.} Continuity bounds for the function $\chi_A$ and $\chi_A^{\infty}$ under two forms of energy-type constraints are presented in the following\smallskip

\begin{proposition}\label{CHI-CB-2} \emph{Let $H$ be a positive operator on the space $\H_{X}$, where $X$ is either $A$ or $AC$  that satisfies condition (\ref{H-cond+}) and $F_{H}$ the function defined in (\ref{F-def}). Let $\rho$ and $\sigma$ be any states in $\S(\H_{AC})$ such that $\,\Tr H\rho_{X},\,\Tr H\sigma_{X}\leq E\,$. Let $\varepsilon=\textstyle\frac{1}{2}\|\rho-\sigma\|_1$, $F(\rho,\sigma)=\|\sqrt{\rho}\sqrt{\sigma}\|_1^2$ and $\,g(x)$ be the function defined in (\ref{g-def}).}

\emph{If $F(\rho,\sigma)\geq 1-\delta^2\geq0$ then}
\begin{equation*}
    |\chi_A(\rho)-\chi_A(\sigma)|\leq \delta F_{H}\!\!\left[\frac{2E}{\delta^2}\right]+2g(\delta).\qquad\qquad\qquad\;\;
\end{equation*}

\emph{If $\,\varepsilon(2-\varepsilon)\leq\delta^2\leq1$ then}
\begin{equation}\label{CHI-CB-2++}
    |\chi_A^{*}(\rho)-\chi_A^{*}(\sigma)|\leq 2\delta F_{H}\!\!\left[\frac{2E}{\delta^2}\right]+g(\delta),\quad \chi_A^{*}=\chi_A,\chi_A^{\infty}.
\end{equation}
\end{proposition}\smallskip

\emph{Proof.} As noted before Proposition \ref{CHI-1-CB} the function $\chi_A$ belongs to the class $N_{2,2}^1(1,2)$ in the settings $A_1=A$, $A_2=C$ and to the class $N_{1,2}^1(1,2)$ in the settings $A_1=AC$. So,
the first claim of this proposition follows from Theorem \ref{L-1-ip} in Section 3.2.3 with $m=1$.

Let $\,\varepsilon(2-\varepsilon)\leq\delta^2\leq1$. To prove inequality (\ref{CHI-CB-2++}) with $\chi_A^{*}=\chi_A$ one can repeat all the steps
from the last part of the proof of Proposition \ref{CHI-1-CB}. At the final step we obtain
\begin{equation*}
\chi_A(\rho)-\chi_A(\sigma)\leq I(A\!:\!E)_{\hat{\rho}}-I(A\!:\!E)_{\hat{\sigma}}+\epsilon \leq 2\delta F_{H}\!\!\left[\frac{2E}{\delta^2}\right]+g(\delta)+\epsilon,
\end{equation*}
instead of (\ref{LS-tmp}), where the last inequality follows
from Proposition \ref{CMI-ECCB-S} in Section 5.2.2, since $\,F(\hat{\rho},\hat{\sigma})\geq F(\bar{\rho},\bar{\sigma})\geq 1-\delta^2\,$ by monotonicity of the fidelity and $\,\hat{\rho}_{E}=\hat{\sigma}_{E}\,$.

Inequality (\ref{CHI-CB-2++}) with $\chi_A^{*}=\chi_A^{\infty}$ can be proved  by using Winter's arguments form the proof of the continuity bound for $E_F^{\infty}$ in Corollary 4 in \cite{W-CB}. One should repeat all the steps of that proof with the systems $C,E$ instead of $B,X$, the function $-\chi_A$ instead of $E_F$, the function $-I(A\!:\!E)$ instead of $S(A|X)$ and the use of a $\epsilon$-optimal decomposition of the state $\sigma^{AB}\otimes\Omega^{A'B'}$ (since  the optimal decomposition may not exist in this case). At the final step we obtain
\begin{equation}\label{f-step}
\begin{array}{l}
 \chi_A(\sigma\otimes\Omega)-\chi_A(\rho\otimes\Omega) \leq  I(AA'\!:\!E)_{\tilde{\sigma}}-I(AA'\!:\!E)_{\tilde{\rho}}+\epsilon\\\\
 \quad=I(A\!:\!E|A')_{\tilde{\sigma}}-I(A\!:\!E|A')_{\tilde{\rho}}+\epsilon \leq \displaystyle 2\delta F_{H}\!\!\left[\frac{2E}{\delta^2}\right]+g(\delta)+\epsilon,
\end{array}
\end{equation}
where the identity $\,I(AA'\!:\!E)=I(A\!:\!E|A')+I(A'\!:\!E)\,$ and the coincidence of $\tilde{\rho}_{A'E}$ and $\tilde{\sigma}_{A'E}$  were used and the last inequality follows
from Proposition \ref{CMI-ECCB-S} in Section 5.2.2, since $F(\tilde{\rho},\tilde{\sigma})\geq (1-\varepsilon)^2\geq 1-\delta^2$ and $\tilde{\rho}_{A'E}=\tilde{\sigma}_{A'E}$.
It is easy to see that all the terms in (\ref{f-step}) are finite in both cases $X=A$, $X=AC$, since $\,\Tr H\rho_{X},\,\Tr H\sigma_{X}\leq E$.
$\Box$

\subsubsection{Local continuity conditions.} It follows from Proposition \ref{CHI-1-CB} that the functions $\chi_A$ and $\chi_A^{\infty}$
are uniformly continuous on the set $\S(\H_{AC})$ provided that $\dim\H_A$ is finite. It is easy to see that these functions
are not continuous on $\S(\H_{AC})$ if $\dim\H_A=+\infty$, sufficient
conditions of their local continuity are presented in the following
\smallskip

\begin{proposition}\label{CHI-LC} A) \emph{The functions $\chi_A$ and $\chi_A^{\infty}$ are concave and
lower semicontinuous on the set $\S(\H_{AC})$.}

B) \emph{Let$\,\{\omega_n\}$ be a sequence in  $\,\mathfrak{S}(\mathcal{H}_{AC})$ converging to a state $\omega_0$.}

\emph{If
\begin{equation}\label{CHS-A-cont}
 \lim_{n\to+\infty }S([\omega_n]_A)=S([\omega_0]_A)<+\infty
\end{equation}
then}
\begin{equation}\label{CHI-conv}
  \lim_{n\to+\infty }\chi^*_A(\omega_n)=\chi^*_A(\omega_0)<+\infty,\quad \chi^*_A=\chi_A,\chi^{\infty}_A.
\end{equation}

\emph{If
\begin{equation}\label{CHS-cont}
 \lim_{n\to+\infty }S(\omega_n)=S(\omega_0)<+\infty
\end{equation}
then (\ref{CHI-conv}) holds with $\chi^*_A=\chi_A$. If, in addition,
$S([\omega_0]_C)<+\infty$ then (\ref{CHI-conv}) holds with $\chi^*_A=\chi^{\infty}_A$.}
\end{proposition}\smallskip

\emph{Proof.} A) The concavity and lower semicontinuity of the function $\chi_A$ follow from Proposition 3 in \cite{cmp-2}.
The concavity of $\chi_A^{\infty}$ can be shown by using concavity and superadditivity of $\chi_A$ with the technique
presented in the proof of Proposition 13 in \cite{DHR} (applied to the convex subadditive function $E=-\chi_A$).

The lower semicontinuity of $\chi_A$ and $\chi_A^{\infty}$ follows from the
lower semicontinuity of $C_B$ and $C_B^{\infty}$ (Propositions \ref{CB-DB},\ref{CB-DB-R}) and the generalized Koashi-Winter relation (\ref{KWS-gen}).\smallskip

B) Let $\{\bar{\omega}_n\}$ be a sequence
of pure states in $\S(\H_{ABC})$ converging to a pure state $\bar{\omega}_0$ such that
$[\bar{\omega}_n]_{AC}=\omega_n$, $\forall n\geq0$, where $B$ is an appropriate quantum system.

Assume that (\ref{CHS-A-cont}) holds. Then  Propositions \ref{QCMI-lc}, \ref{CB-DB} and \ref{CB-DB-R} imply that
\begin{equation}\label{CB-tmp}
  \lim_{n\to+\infty }C_B^*([\bar{\omega}_n]_{AB})=C_B^*([\bar{\omega}_0]_{AB})<+\infty,\quad C_B^*=C_B,C_B^{\infty}.
\end{equation}
So, limit relations in
(\ref{CHI-conv}) follow from the generalized Koashi-Winter relations (\ref{KWS-gen}).

Assume that (\ref{CHS-cont}) holds. Then the similar relation holds with $\omega_n$ replaced by $[\bar{\omega}_n]_{B}$ and hence (\ref{CB-tmp}) holds with $C_B^*=C_B$
by Propositions \ref{QCMI-lc} and \ref{CB-DB}. Thus, the limit relation in
(\ref{CHI-conv}) with $\chi^*_A=\chi_A$ follows from the generalized Koashi-Winter relation (\ref{KWS-gen}).

Assume for a moment that $\dim\H_C<+\infty$. Then $\omega\mapsto\Tr_C\omega$ is channel from $AC$ to $A$ with finite Choi rank. By Theorem 1 in \cite{PCE} condition (\ref{CHS-cont}) implies the validity of (\ref{CHS-A-cont}). So, in this case the limit relation in
(\ref{CHI-conv}) with $\chi^*_A=\chi^{\infty}_A$  holds  by the proved part
of the proposition.

To replace the condition $\dim\H_C<+\infty$ by the weaker condition $S([\omega_0]_C)<+\infty$ take any
sequence $\{P_k\}$ of finite rank projectors in $\B(\H_C)$ strongly converging to the unit operator $I_C$ and consider the sequence of channels $\,\Phi_k(\rho)=P_k\rho P_k+[\Tr(I_C-P_k)\rho\shs]\tau$ from $C$ to itself, where $\tau$ is a fixed pure state in $\S(\H_C)$.

Introduce the function $\hat{\chi}^{\infty}_A(\omega)\doteq\inf_{k\in\mathbb{N}} \chi^{\infty}_A(\id_A\otimes\Phi_k(\omega))$. The monotonicity
property (\ref{CHI-M}) implies that
\begin{equation}\label{CHI-UB}
  \hat{\chi}^{\infty}_A(\omega)\geq \chi^{\infty}_A(\omega)\quad \forall \omega\in\S(\H_{AC}).
\end{equation}
It follows from the above observation (the case $\dim\H_C<+\infty$) that
the limit relation in
(\ref{CHI-conv}) with $\chi^*_A=\chi^{\infty}_A$  holds with $\omega_n$ replaced by $\id_A\otimes\Phi_k(\omega_n)$ for each $k$.
Hence
\begin{equation}\label{CHI-USC}
  \limsup_{n\to+\infty}\hat{\chi}^{\infty}_A(\omega_n)\leq\hat{\chi}^{\infty}_A(\omega_0).
\end{equation}
Since the function $\chi^{\infty}_A$ is lower semicontinuous on $\S(\H_{AC})$, it follows from
(\ref{CHI-UB}) and  (\ref{CHI-USC}) that to prove the limit relation in
(\ref{CHI-conv}) with $\chi^*_A=\chi^{\infty}_A$ it suffices to show that
$\hat{\chi}^{\infty}_A(\omega_0)=\chi^{\infty}_A(\omega_0)$. This can be done by showing that
\begin{equation*}
  \lim_{k\to+\infty}\chi^{\infty}_A(\id_A\otimes\Phi_k(\omega_0))=\chi^{\infty}_A(\omega_0).
\end{equation*}
This relation follows from the proved part of the proposition,
since $[\id_A\otimes\Phi_k(\omega_0)]_A=[\omega_0]_A$ for all $k$ and $\,S([\omega_0]_A)\leq S(\omega_0)+S([\omega_0]_C)<+\infty\,$ by the assumption. $\Box$

\subsection{Squashed entanglement and c-squashed entanglement}

\subsubsection{Definitions and the AFW-type continuity bounds.} The squashed entanglement is one of the basic entanglement measures in composite (bipartite and multipartite) quantum systems. For a state $\omega$ of a finite-dimensional bipartite system $AB$ it is defined as
\begin{equation}\label{se-def}
  E_{sq}(\omega)=\textstyle\frac{1}{2}\displaystyle\inf_{\hat{\omega}\in\mathfrak{M}_1(\omega)}I(A\!:\!B|E),
\end{equation}
where  $E$ is a system of any dimension and $\mathfrak{M}_1(\omega)$ is the set of all extensions of $\omega$ to a state in $\S(\H_{ABE})$ \cite{C&W,Tucci}.

The \emph{c-squashed entanglement} is an important  entanglement measure in composite quantum systems. For a state $\omega$ of a finite-dimensional bipartite system $AB$ it is defined as
\begin{equation}\label{c-se-def}
  E^c_{sq}(\rho_{AB})=\textstyle\frac{1}{2}\displaystyle\inf_{\hat{\omega}\in\mathfrak{M}_2(\omega)}I(A\!:\!B|E),
\end{equation}
where $E$ is a system of any dimension and $\mathfrak{M}_2(\omega)$ is the set of all extensions of $\omega$  having the form (\ref{m-2}) with $A_1=A$, $A_2=B$ and $A_3=E$ \cite{N&R,Tucci,Y&C,Wak,Sz}.

Definitions (\ref{se-def}) and (\ref{c-se-def})  can be generalized to any state $\omega$ of an infinite-dimensional bipartite system $AB$ assuming that $I(A\!:\!B|E)$ is the extended QCMI defined by the equivalent expressions (\ref{QCMI-e1}) and (\ref{QCMI-e2}) \cite{SE,CBM}. The main analytical problem of this case is the discontinuity of the
functions $E_{sq}$ and $E^c_{sq}$ on $\S(\H_{AB})$  as well as the unproven equality of these functions to zero at any countably-non-decomposable separable state \cite[Remark 10]{SE}.

The definitions of the $n$-partite squashed entanglement and c-squashed entanglement are direct generalizations
of the bipartite definitions \cite{Y&C,AHS}. If $\omega$ is any state in $\S(\H_{A_1...A_n})$ then
\begin{equation*}
  E_{sq}(\omega)=\textstyle\frac{1}{2}\displaystyle\inf_{\hat{\omega}\in\mathfrak{M}_1(\omega)}I(A_1\!:\!...\!:\!A_n|E)_{\hat{\omega}},
\end{equation*}
where $\mathfrak{M}_1(\omega)$ is the set of all extensions of $\omega$ to a state in $\S(\H_{A_1...A_nE})$,  and
\begin{equation*}
  E^c_{sq}(\omega)=\textstyle\frac{1}{2}\displaystyle\inf_{\hat{\omega}\in\mathfrak{M}_2(\omega)}I(A_1\!:\!...\!:\!A_n|E)_{\hat{\omega}},
\end{equation*}
where $\mathfrak{M}_2(\omega)$ is the set of all extensions of $\omega$ having the form (\ref{m-2}) with $A_{n+1}=E$.

We may equivalently define the c-squashed entanglement
as the $\sigma$-convex hull (mixed convex roof) of the QMI, i.e.
\begin{equation*}
E^c_{sq}(\omega)=\textstyle\frac{1}{2}\displaystyle\inf_{\sum_i p_i\omega_i=\omega}\sum_i p_i I(A_1\!:\!...\!:\!A_n)_{\omega_i},
\end{equation*}
where the infimum is over all countable ensembles $\{p_i,\omega_i\}$ of states in $\S(\H_{A_1...A_n})$ with the average state $\omega$. This follows from
\begin{lemma}\label{sl} \emph{Let $\Lambda$ be a quantum operation from $\T(\H_{A_1...A_n})$ to itself. Then\smallskip
\begin{equation}\label{sl+}
\sum_i p_i I(A_1\!:\!...\!:\!A_n)_{\Lambda(\omega_i)}=I(A_1\!:\!...\!:\!A_n|E)_{\Lambda\otimes\id_E(\hat{\omega})},
\end{equation}
where $\hat{\omega}=\sum_i p_i \omega_i\otimes |i\rangle\langle i|$ is a q-c state determined by any ensemble $\{p_i,\omega_i\}$ of states
in $\S(\H_{A_1...A_n})$ and it is assumed that  $\,I(A_1\!:\!...\!:\!A_n|E)_{\sigma}=(\Tr\sigma)I(A_1\!:\!...\!:\!A_n|E)_{\sigma/\Tr\sigma}$ for any positive
operator $\sigma\neq0$ in $\T(\H_{A_1...A_nE})$ and that $\,I(A_1\!:\!...\!:\!A_n|E)_{0}=0$.}
\end{lemma}\smallskip

\emph{Proof.} Equality (\ref{sl+}) is directly proved if all the states $c_i^{-1}\Lambda(\omega_i)$, $c_i\doteq\Tr\Lambda(\omega_i)>0$, have finite marginal entropies and the Shannon entropy of the
probability distribution $\{p_i\}$ is finite. To show the validity of (\ref{sl+}) in general case one can use the approximation property
stated in \cite[Proposition 5]{CMI}. $\square$\smallskip

Continuity bounds for the squashed entanglement and c-squashed entanglement in the finite-dimensional case
were originally obtained using the Alicki-Fannes continuity bound for the conditional entropy \cite{C&W,Y&C}. Moreover, it was the continuity bounds for the squashed entanglement that was the main motivation of the research of Alicki and Fannes \cite{A&F}. In this section we present improved versions of these results.

In Section 5.2.1 it is mentioned that the function $\omega\mapsto I(A_1\!:\!...\!:\!A_n|A_{n+1})_{\omega}$
belongs to the class $L_{n+1}^{n-1}(2,n)$ (in the notation introduced in Section 3.1.2). Thus, the $n$-partite squashed entanglement and c-squashed entanglement belong, respectively,
to the classes $N_{n,1}^{n-1}(1,n/2)$ and $N_{n,2}^{n-1}(1,n/2)$ and, hence, directly applying Corollary
\ref{CB-N} in Section 3.1.2 to these functions we obtain the following two propositions.
\smallskip\pagebreak

\begin{proposition}\label{SE-2-CB} \emph{If $\,d=\min\{\dim\H_A,\dim\H_B\}<+\infty\,$ then
\begin{equation*}
 |E^*_{sq}(\rho)-E^*_{sq}(\sigma)|\leq \delta\ln d+g(\delta),\quad E^*_{sq}=E_{sq},E^c_{sq},
\end{equation*}
for any states $\rho$ and $\sigma$ in $\S(\H_{AB})$ such that either $F(\rho,\sigma)\geq 1-\delta^2$ or $\,\varepsilon(2-\varepsilon)\leq\delta^2$, where $\,\varepsilon=\textstyle\frac{1}{2}\|\rho-\sigma\|_1$, $F(\rho,\sigma)=\|\sqrt{\rho}\sqrt{\sigma}\|_1^2\,$
and $g(x)$ is the function defined in (\ref{g-def}).}
\end{proposition}\medskip

\begin{proposition}\label{SE-n-CB} \cite{CBM} \emph{If $\,d_k\doteq\dim\H_{A_k}<+\infty\,$ for $\,k=\overline{1,n-1}\,$ then
\begin{equation*}
|E^*_{sq}(\rho)-E^*_{sq}(\sigma)|\leq \delta\sum_{k=1}^{n-1}\ln d_k+(n/2)g(\delta),\quad E^*_{sq}=E_{sq},E^c_{sq},
\end{equation*}
for any states $\rho$ and $\sigma$ in $\S(\H_{A_1..A_n})$ such that either $F(\rho,\sigma)\geq 1-\delta^2$ or $\,\varepsilon(2-\varepsilon)\leq\delta^2$, where $\,\varepsilon=\textstyle\frac{1}{2}\|\rho-\sigma\|_1$, $F(\rho,\sigma)=\|\sqrt{\rho}\sqrt{\sigma}\|_1^2\,$
and $g(x)$ is the function defined in (\ref{g-def}).}
\end{proposition}

\subsubsection{Continuity bound under the energy-type constraint.}
As mentioned before, the bipartite squashed entanglement and c-squashed entanglement belong, respectively,
to the classes $N_{2,1}^{1}(1,1)$ and $N_{2,2}^{1}(1,1)$. So, directly applying Theorem
\ref{L-1-ip} in Section 3.2.3 with $m=1$ to these functions we obtain the following
\smallskip

\begin{proposition}\label{SE-2-CB-EC} \emph{Let $H$ be a positive operator on the space $\H_{A}$ that satisfy condition (\ref{H-cond+})
and $F_{H}$ the function defined in (\ref{F-def}). Let $\delta\in(0,1]$ and $\,g(x)$ be the function defined in (\ref{g-def}). Then
\begin{equation}\label{SE-2-CB-EC+}
    |E^*_{sq}(\rho)-E^*_{sq}(\sigma)|\leq \delta F_{H}\!\!\left[\frac{2E}{\delta^2}\right]+g(\delta),\quad E^*_{sq}=E_{sq},E^c_{sq},
\end{equation}
for any states $\rho$ and $\sigma$ in $\S(\H_{AB})$ such that $\,\Tr H\rho_{A},\Tr H\sigma_{A}\leq E\,$ and}
$$
  \textit{either}\quad F(\rho,\sigma)\doteq\|\sqrt{\rho}\sqrt{\sigma}\|^2_1\geq 1-\delta^2\quad \textit{or} \quad \varepsilon(2-\varepsilon)\leq\delta^2,\quad \textit{where} \quad  \varepsilon=\textstyle\frac{1}{2}\|\rho-\sigma\|_1.
$$
\end{proposition}\smallskip

The r.h.s of (\ref{SE-2-CB-EC+}) tends to zero as $\delta\to 0$ due to the equivalence of (\ref{H-cond+}) and (\ref{H-cond+a}).
Note that continuity bound  (\ref{SE-2-CB-EC+}) coincides with the continuity bound (\ref{EF-CB-2+}) for the EoF.

There is another continuity bound for the bipartite squashed entanglement under the energy-type constraint on one of the subsystems obtained in \cite[Proposition 22]{SE} by
using  the Winter-type continuity bound for the QCMI presented in Proposition \ref{CMI-W-CB} in Section 5.2.1.  This
continuity bound coincides with the continuity bound (\ref{E_F-WCB+}) for the EoF, it is faithful if the operator $H$ satisfies the Gibbs condition (\ref{H-cond}) which is slightly weaker than condition (\ref{H-cond+}). Nevertheless, one can show that it is less accurate than continuity bound (\ref{SE-2-CB-EC+}) if the function $F_{H}$ has a logarithmic
growth, in particular, if $H$ is the Hamiltonian of a multi-mode quantum oscillator.

Since the function $\omega\mapsto I(A_1\!:\!...\!:\!A_n|A_{n+1})_{\omega}$
belongs to the classes  $L_{n+1}^{n-1}(2,n)$ and $L_{n+1}^{n}(2-2/n,n)$, the $n$-partite squashed entanglement and c-squashed entanglement lie
in $N_{n,1}^{n-1}(1,n/2)\cup N_{n,1}^{n}(1-1/n,n/2)$ and in $N_{n,2}^{n-1}(1,n/2)\cup N_{n,2}^{n}(1-1/n,n/2)$ correspondingly. Hence by  applying Theorem
\ref{L-1-ip} in Section 3.2.3 one can obtain continuity bounds for these functions under the following two forms of energy-type constraint:
\begin{itemize}
  \item the constraint on the marginal states $\omega_{A_1},...,\omega_{A_{n-1}}$ of a state $\omega$;
  \item the constraint on the marginal states $\omega_{A_1},...,\omega_{A_{n}}$ of  a state $\omega$.
\end{itemize}
These forms of constraints correspond to the cases $m=n-1$ and $m=n$ in the following
\smallskip

\begin{proposition}\label{SE-m-CB-EC} \emph{Let $n\geq2$ be arbitrary. Let $H_{A_1},...,H_{A_{m}}$ be
positive operators on the spaces $\H_{A_1},...,\H_{A_{m}}$ satisfying  condition (\ref{H-cond+}), where either $m=n-1$ or $m=n$, and $F_{H_{m}}$  the function defined in (\ref{F-H-m}). Let $\delta\in(0,1]$, $C_m=(n-1)/m$ and $\,g(x)$ be the function defined in (\ref{g-def}). Then
$$
 |E^*_{sq}(\rho)-E^*_{sq}(\sigma)|\leq C_m\delta\shs F_{H_{m}}\!\!\left[\frac{2mE}{\delta^2}\right]+(n/2)g(\delta),\quad E^*_{sq}=E_{sq},E^c_{sq},
$$
for any states $\rho$ and $\sigma$  in $\,\S(\H_{A_1..A_n})$ s.t. $\,\sum_{k=1}^{m}\Tr H_{A_k}\rho_{A_k},\,\sum_{k=1}^{m}\Tr H_{A_k}\sigma_{A_k}\leq mE\,$
and either $F(\rho,\sigma)\doteq\|\sqrt{\rho}\sqrt{\sigma}\|^2_1\geq 1-\delta^2$ or $\,\varepsilon(2-\varepsilon)\leq\delta^2$, where  $\,\varepsilon=\textstyle\frac{1}{2}\|\rho-\sigma\|_1$.}

\emph{If the operators $H_{\!A_1}$,..,$H_{\!A_{m}}$ are unitary equivalent to each other then the above inequality can be written as
$$
 |E^*_{sq}(\rho)-E^*_{sq}(\sigma)|\leq C_m\delta\shs mF_{H_{A_1}}\!\!\left[\frac{2E}{\delta^2}\right]+(n/2)g(\delta),\quad E^*_{sq}=E_{sq},E^c_{sq},
$$
where $F_{H_{A_1}}$ is the function defined in (\ref{F-def}) with $H=H_{A_1}$.}
\end{proposition}\smallskip

The continuity bounds given by Proposition \ref{SE-m-CB-EC} for the $n$-partite squashed entanglement and c-squashed entanglement
essentially improve the continuity bounds for these quantities obtained in \cite[Proposition 20]{CBM}.

\subsubsection{Local continuity conditions.} By Proposition \ref{SE-2-CB} the functions $E_{sq}$ and $E^c_{sq}$ are uniformly continuous on the set
$\S(\H_{AB})$ provided that at least one of the systems $A$ and $B$ is finite-dimensional. If both systems $A$ and $B$ are infinite-dimensional then these
functions are discontinuous and may take the  value $+\infty$. In this case we have the following
\smallskip

\begin{proposition}\label{SE-2} A) \emph{The functions $E_{sq}$ and $E^c_{sq}$ are finite and lower semicontinuous on the set}
\begin{equation*}
\S_*(\H_{AB})\doteq\left\{\omega\in\S(\H_{AB})\,|\,\min\{S(\omega_{A}),S(\omega_{B})\}<+\infty\right\}.
\end{equation*}
\emph{Moreover,
\begin{equation*}
\liminf_{n\to+\infty}E^*_{sq}(\omega_n)\geq E^*_{sq}(\omega_0),\quad E^*_{sq}=E_{sq},E^c_{sq},
\end{equation*}
for any sequence $\{\omega_n\}\subset\S(\H_{AB})$ converging to a state $\omega_0$ in $\S_*(\H_{AB})$.}\smallskip

\noindent B) \emph{If  $\{\omega_n\}$ is a sequence of states converging to a state $\omega_0$ such that
\begin{equation}\label{MI-cont}
\lim_{n\to+\infty}I(A\!:\!B)_{\omega_n}=I(A\!:\!B)_{\omega_0}<+\infty
\end{equation}
and $\omega_0\in\S_*(\H_{AB})$ then
\begin{equation}\label{SE-cont}
\lim_{n\to+\infty}E^*_{sq}(\omega_n)=E^*_{sq}(\omega_0)<+\infty, \quad E^*_{sq}=E_{sq},E^c_{sq}.
\end{equation}
Moreover, for arbitrary  sequences of channels $\Phi_n:A\to A'$ and $\Psi_n:B\to B'$ strongly converging\footnote{A sequence of channels $\Phi_n$ strongly converges to a channel $\Phi_0$ if $\lim_{n\to+\infty}\Phi_n(\rho)=\Phi_0(\rho)$ for any input state $\rho$ \cite{H-SCI,CSR}.} to channels
$\,\Phi_0$ and $\Psi_0$ such that the state $\Phi_0\otimes\Psi_0(\omega_0)$ lies in $\S_*(\H_{A'B'})$
condition (\ref{MI-cont}) implies that}
\begin{equation*}
\lim_{n\to+\infty}E^*_{sq}(\Phi_n\otimes\Psi_n(\omega_n))=E^*_{sq}(\Phi_0\otimes\Psi_0(\omega_0))<+\infty, \quad E^*_{sq}=E_{sq},E^c_{sq}.
\end{equation*}
\end{proposition}\smallskip

\emph{Proof.}  Part A and the main assertion of part B are proved in \cite{SE} for $E^*_{sq}=E_{sq}$. Below we will prove them for $E^*_{sq}=E^c_{sq}$
by modifying the arguments from \cite{SE}.

A) Let $\{P^n_A\}\subset\B(\H_A)$ and
$\{P^n_B\}\subset\B(\H_B)$ be any sequences of finite rank projectors strongly
converging to the unit operators $I_A$ and $I_B$. Assuming that $I(A\!:\!B|E)_\sigma=(\Tr\sigma) I(A\!:\!B|E)_{\sigma/\Tr\sigma}$ for a positive operator  $\sigma\in\T(\H_{ABE})$ not equal to $0$ and that $I(A\!:\!B|E)_0=0$
introduce the sequences of functions
$$
f_n(\omega)=\textstyle\frac{1}{2}\displaystyle\inf_{\hat{\omega}\in\mathfrak{M}_2(\omega)}\shs
I(A\!:\!B|E)_{Q^n_A \hat{\omega}Q^n_A}\quad\textrm{and}\quad \displaystyle
g_n(\omega)=\textstyle\frac{1}{2}\displaystyle\inf_{\hat{\omega}\in\mathfrak{M}_2(\omega)}\shs
I(A\!:\!B|E)_{Q^n_B\hat{\omega} Q^n_B},
$$
where $Q^n_A=P^n_A\otimes I_{BE}$, $Q^n_B=I_{AE}\otimes P^n_B$ and the infima are over the set $\mathfrak{M}_2(\omega)$ of all extensions $\shs\hat{\omega}$ of the state
$\omega$ having the form (\ref{m-2}). By Lemma \ref{sl} the functions $f_n$  and $g_n$ coincide with the $\sigma$-convex hull
of the functions $\omega\mapsto I(A\!:\!B)_{P^n_A\otimes I_{B}\cdot\omega\cdot P^n_A\otimes I_{B}}$ and $\omega\mapsto I(A\!:\!B)_{I_{A}\otimes P^n_B \cdot\omega\cdot I_{A}\otimes P^n_B}$ which are continuous on $\S(\H_{AB})$ by Proposition \ref{CMI-CB} in Section 5.2.1. By Proposition 5 in \cite{EM} the functions $f_n$  and $g_n$ are continuous on $\S(\H_{AB})$.

Now we repeat the arguments from the proof of Proposition 8C in \cite{SE} to show that
\begin{equation}\label{sup-h-n}
E^c_{sq}(\omega)=E_*(\omega)\doteq\sup_n\max\{f_n(\omega),g_n(\omega)\},
\end{equation}
for any $\omega\in \S_*(\H_{AB})$. Since $E_*$ is a lower semicontinuous
function on $\S(\H_{AB})$ and $E^c_{sq}(\omega)\geq E_*(\omega)$ for any state $\omega\in \S(\H_{AB})$
by monotonicity of the QCMI under local operations, claim A  for $E^*_{sq}=E^c_{sq}$ follows from (\ref{sup-h-n}).\smallskip

B) For each natural $m$ consider the function
\begin{equation*}
 E^c_{sq,m}(\omega)=\textstyle\frac{1}{2}\displaystyle\inf_{\sum_{i=1}^m p_i\omega_i=\omega}\sum_{i=1}^m p_i I(A\!:\!B)_{\omega_i},
\end{equation*}
where the infimim is over all ensembles $\{p_i,\omega_i\}$ \emph{consisting of $m$ states} in $\S(\H_{AB})$ with the average state $\omega$.
By using inequality (\ref{MI-LAA-1}) it is easy to show that
\begin{equation}\label{SE-imp}
E^c_{sq}(\omega)=\inf_m E^c_{sq,m}(\omega)\quad\textrm{provided that}\quad I(A\!:\!B)_{\omega}<+\infty.
\end{equation}
We will show that condition (\ref{MI-cont}) implies that
\begin{equation}\label{SE-m-usc}
\limsup_{n\to+\infty}E^c_{sq,m}(\omega_n)\leq E^c_{sq,m}(\omega_0)\quad \forall m.
\end{equation}
Let $\varepsilon>0$ be arbitrary and $\{p^0_i,\omega^0_i\}$ an ensemble of $m$ states with the average state $\omega_0$ such that
$\sum_i p^0_i I(A\!:\!B)_{\omega^0_i}<E^c_{sq,m}(\omega_0)+\varepsilon$. By Lemma 3 in \cite{cmp-2} there exists a sequence
$\{\{p^n_i,\omega^n_i\}_i\}_n$ of ensembles consisting of $m$ states such that
$$
\sum_{i=1}^m p^n_i\omega^n_i=\omega_n\quad \forall n\quad \textrm{and} \quad \lim_{n\to+\infty} p^n_i\omega^n_i =p^0_i\omega^0_i,\quad i=\overline{1,m}.
$$
Since $p^n_i\omega^n_i\leq \omega_n$ for all $n$ and $i$,  condition (\ref{MI-cont}) implies, by Corollary \ref{DCT-MI-c1} in Section 5.2.3, that
\begin{equation*}
 \lim_{n\to+\infty} p^n_i I(A\!:\!B)_{\omega^n_i}=p^0_i I(A\!:\!B)_{\omega^0_i}<+\infty,\quad i=\overline{1,m}.
\end{equation*}
It follows that
$$
\limsup_{n\to+\infty}E^c_{sq,m}(\omega_n)\leq \lim_{n\to+\infty} \sum_{i=1}^m p^n_i I(A\!:\!B)_{\omega^n_i}=\sum_{i=1}^m p^0_i I(A\!:\!B)_{\omega^0_i}<E^c_{sq,m}(\omega_0)+\varepsilon.
$$
This implies limit relation (\ref{SE-m-usc}). It follows from  (\ref{SE-imp}) and (\ref{SE-m-usc})  that
\begin{equation}\label{SE-usc}
\limsup_{n\to+\infty}E^c_{sq}(\omega_n)\leq E^c_{sq}(\omega_0).
\end{equation}
Since $\omega_0$ is a state in $\S_*(\H_{AB})$, by the proof of part A we have
\begin{equation*}
  E_*(\omega_n)\leq E^c_{sq}(\omega_n)\quad\forall n\quad \textrm{and} \quad E_*(\omega_0)=E^c_{sq}(\omega_0)<+\infty,
\end{equation*}
where $E_*$ is the function defined in (\ref{sup-h-n}). Thus, by using the lower semicontinuity of
$E_*$ and limit relation (\ref{SE-usc}) we obtain (\ref{SE-cont}) with $E^*_{sq}=E^c_{sq}$.

The last assertion of part B for both functions   $E_{sq}$ and $E^c_{sq}$
follows from the main assertion and Proposition \ref{DCT-MI}C in Section 5.2.3. $\square$

Proposition \ref{SE-2} and Proposition \ref{classes-p} in Section 5.2.3 implies the following\smallskip

\begin{corollary}\label{SE-2-c}
\emph{If $\{\omega_n\}$ is a sequence
in $\S(\H_{AB})$  belonging to the class $\Upsilon^{*}_{\!AB}$ (see Definition \ref{class-2} at the begin of Section 5.2.3)
which converges to a state $\omega_0$ such that either $S([\omega_0]_A)$ or $S([\omega_0]_B)$ is finite then both limit relations in (\ref{SE-cont}) hold.}
\end{corollary}\smallskip

Example 3 in \cite[Section 5.2]{LSE} demonstrates how to apply this continuity condition.

The multipartite version of Proposition \ref{SE-2} is presented below. All its claims for the function $E_{sq}$ is obtained in \cite[Proposition 5]{LSE}.
The proofs of these claims for the function $E^c_{sq}$ can be obtained by the same arguments combined with the observations from the proof of  Proposition \ref{SE-2}.
\smallskip\pagebreak

\begin{proposition}\label{SE-n} A) \emph{The functions $E_{sq}$ and $E^c_{sq}$ are finite and lower semicontinuous on the set}
\begin{equation*}
\S_*(\H_{A_1...A_n})\doteq\left\{\omega\in\S(\H_{A_1...A_n})\,|\,S(\omega_{A_i})<+\infty, i=\overline{1,n}\right\}.
\end{equation*}
\emph{Moreover,
\begin{equation*}
\liminf_{k\to+\infty}E^*_{sq}(\omega_k)\geq E^*_{sq}(\omega_0),\quad E^*_{sq}=E_{sq},E^c_{sq},
\end{equation*}
for any sequence $\{\omega_k\}\subset\S(\H_{A_1...A_n})$ converging to a state $\omega_0$ in $\S_*(\H_{A_1...A_n})$.}

\noindent B) \emph{If  $\{\omega_k\}\subset\S(\H_{A_1...A_n})$ is a sequence converging to a state $\omega_0$ such that
\begin{equation}\label{MI-cont-n}
\lim_{k\to+\infty}I(A_1\!:\!...\!:\!A_n)_{\omega_k}=I(A_1\!:\!...\!:\!A_n)_{\omega_0}<+\infty
\end{equation}
and $\omega_0\in\S_*(\H_{A_1...A_n})$ then
\begin{equation}\label{SE-cont-n}
\lim_{k\to+\infty}E^*_{sq}(\omega_k)=E^*_{sq}(\omega_0)<+\infty, \quad E^*_{sq}=E_{sq},E^c_{sq}.
\end{equation}
Moreover, for any  sequences of channels $\,\Phi_k^{A_1}:A_1\to A_1'$,...,$\,\Phi_k^{A_n}:A_n\to A_n'$  strongly converging to channels
$\,\Phi_0^{A_1}$,...,$\Phi_0^{A_n}$ correspondingly s.t. the state $\Phi_0^{A_1}\otimes...\otimes\Phi_0^{A_n}(\omega_0)$ lies in $\S_*(\H_{A'_1...A'_n})$
condition (\ref{MI-cont-n}) implies that}
$$
\lim_{k\to+\infty}E^*_{sq}(\Phi_k^{A_1}\otimes...\otimes\Phi_k^{A_n}(\omega_k))=E^*_{sq}(\Phi_0^{A_1}\otimes...\otimes\Phi_0^{A_n}(\omega_0))<+\infty, \quad E^*_{sq}=E_{sq},E^c_{sq}.
$$
\end{proposition}\smallskip

By applying Proposition \ref{classes-p-n} in Section 5.2.3 we obtain the following\smallskip

\begin{corollary}\label{SE-n-c}
\emph{If $\{\omega_n\}$ is a sequence
in $\S(\H_{A_1...A_n})$  belonging to the class $\Upsilon^{*}_{\!A_1..A_n}$ (defined at the end of Section 5.2.3.)
which converges to a state $\omega_0$ such that $S([\omega_0]_{A_i})$ is finite for all $\,i=\overline{1,n}$ then both limit relations in (\ref{SE-cont-n}) hold.}
\end{corollary}

\section{Concluding remarks}

\subsection{On other applications}

In Sections 3 and 4 of this review we describe general methods of quantitative and qualitative continuity analysis of characteristics of
composite quantum systems which are defined (explicitly or implicitly) via the von Neumann entropy. The methods are quite universal but the limited size of the review did not allow us to consider applications of these methods to all basic characteristics of composite quantum systems.

The methods of proving uniform continuity bounds described in Section 3  can be \emph{directly applied} to any characteristics
of composite quantum systems belonging to the classes $\widehat{L}^m_n(C,D)$ and $N^m_{n,s}(C,D)$ introduced in Section 3.1.2.
The incomplete list of such characteristics (not considered in Section 5 of this review) includes:
\begin{itemize}
  \item the relative entropy of entanglement, the relative entropy distance to the PPT states and the Rains bound in bipartite quantum systems \cite{ER-1,ER-2,E-PPT,Rains}, all these characteristics  belong to the classes $L^1_2(1,1)$ and $L^2_2(1/2,1)$ \cite{L&Sh};
  \item the relative entropy of entanglement and the relative entropy distance to the set of $\pi$-separable states in $n$-partite quantum system \cite{ER-1,Sz,Lami+}; all these characteristics belong to the classes $L^{n-1}_n(1,1)$ and $L^{n}_n(1-1/n,1)$ \cite{L&Sh};
  \item the unconditional and conditional dual total correlation in $n$-partite quantum system \cite{Han78} (also called secrecy monotones \cite{Y&C,CMS02}), they belong to the classes $L^{n-1}_n(2,n)$
  and $L^{n-1}_{n+1}(2,n)$ correspondingly (this can be shown by using the observations before Proposition 7 in \cite{CMI});
  \item the conditional entanglement of mutual information in $n$-partite quantum system \cite{CondEnt,YHW}, it belongs to the class $N^n_{n,1}(2,n+1)$ \cite[Section IV.C]{CBM};
  \item the information gain of a quantum measurement with and without quantum side information \cite{QMG,IGSI,IGSI+}, they belong to the classes $L^{1}_2(1,1)$
  and $L^{1}_{1}(1,1)$ correspondingly (if $A_1$ is a measured subsystem).
\end{itemize}

The continuity bounds for the relative entropy of entanglement, the relative entropy distance to the PPT states and the Rains bound in bipartite quantum systems are obtained in \cite[Section V.B]{L&Sh}. The continuity bounds for the relative entropy of entanglement and the relative entropy distance to the set of $\pi$-separable states in $n$-partite quantum systems are presented in \cite[Section V.C]{L&Sh}.
The continuity bounds for the conditional entanglement of mutual information are obtained in \cite[Section IV.C]{CBM}. Note that
continuity bound (70)  in  \cite[Proposition 22]{CBM} can be essentially improved by using
Theorem \ref{L-1-ip} in Section 3.2.3 (which is a strengthened version of Theorem 5 in \cite{CBM}).

The methods of proving local continuity conditions described in Section 4 can be applied to the characteristic mentioned before
as well as to many other characteristics satisfying one of the inequalities (\ref{LAA-1}) and  (\ref{LAA-2}) or both of these inequalities.
A generalized version of the Dini-type lemma (Proposition \ref{DTL} in Section 4.2) and its applications to the characteristics not  considered in this review are presented in \cite{DTL}.

\subsection{On continuity bounds under the spectrum constraint}

A significant part of this review was devoted to the problem of obtaining continuity bounds for
characteristics of a composite quantum system $A_1...A_n$, which hold for states $\rho$ and $\sigma$ satisfying the conditions
\begin{equation}\label{e-t-const}
\Tr H\rho_X\leq E,\quad \Tr H\sigma_X\leq E,
\end{equation}
where $X$ is a particular subsystem of $A_1...A_n$ and $H$ is a positive operator satisfying either (case A) the Gibbs condition (\ref{H-cond})
or (case B) its strengthened version (\ref{H-cond+}). Having such a continuity bound, one can get continuity bound
for the same characteristic, which holds for states $\rho$ and $\sigma$ satisfying the conditions
\begin{equation}\label{sp-const}
\sum_{k=0}^{+\infty} c_k\lambda_k^{\rho_X}\leq E,\quad \sum_{k=0}^{+\infty} c_k\lambda_k^{\sigma_X}\leq E,
\end{equation}
where  $\{\lambda_k^{\rho_X}\}$ and $\{\lambda_k^{\sigma_X}\}$ are sequences of eigenvalues of the states
$\rho_X$ and $\sigma_X$ taken in the non-increasing order and $\{c_k\}$ is a nondecreasing sequence of nonnegative numbers
satisfying the condition
\begin{equation}\label{C-cond}
\sum_{k=0}^{+\infty}e^{-\beta c_k}<+\infty\quad \forall \beta>0
\end{equation}
in case A and the condition
\begin{equation*}
\lim_{\beta\to0^+}\left[\sum_{k=0}^{+\infty}e^{-\beta c_k}\right]^{\beta}=1
\end{equation*}
in case B. This can be done by applying the following observation easily proved by using Lemma \ref{H-lemma} in Section 5.3.2.\footnote{I am grateful to G.G.Amosov for the question that stimulated the appearance of this observation.}
\smallskip

\begin{proposition}\label{L5-C} \emph{Let $\rho$ and $\sigma$ be states in $\S(\H_{A_1..A_n})$ satisfying
condition (\ref{sp-const}), where $X$ is a particular subsystem of $A_1...A_n$. Then there is a positive operator $H$ on the space $\H_X$
with the spectrum $\{c_0,c_0,c_1,c_1,c_2,c_2,...\}$ such that condition (\ref{e-t-const}) holds.}
\end{proposition}\smallskip

Thus, if we have a continuity bound for some characteristic under constraint (\ref{e-t-const})
then by replacing the function $F_H(E)$ in the r.h.s. of this continuity bound by the function $F_{D\{c_k\}}(E)+\ln 2$, where $D\{c_k\}$ is a positive operator
with the spectrum $\{c_0,c_1,c_2,...\}$, we obtain a  continuity bound for
this characteristic under constraint (\ref{sp-const}).

For example, by using Winter's continuity bound for the quantum conditional entropy presented in Proposition \ref{Winter-CB} in Section 5.1.2
one can obtain the following\smallskip

\begin{proposition}\label{Winter-CB++} \emph{Let $\{c_k\}_{k=0}^{+\infty}$ be a sequence of nonnegative numbers  satisfying condition (\ref{C-cond}) such that $c_0=0$. Let $\rho$ and $\sigma$ be states in $\S(\H_{AB})$ such that (\ref{sp-const}) holds with $X=A$ and
$\,\frac{1}{2}\|\rho-\sigma\|_1\leq \varepsilon<1$. Then
\begin{equation*}
\left|S(A|B)_{\rho}-S(A|B)_{\sigma}\right|\leq(2\varepsilon'+4\delta)\left(F_{D\{c_k\}}\!\left(E/\delta\right)+\ln2\right)
+g(\varepsilon')+2h_2(\delta),
\end{equation*}
$\delta=(\varepsilon'-\varepsilon)/(1+\varepsilon')$, for any $\varepsilon'\in(\varepsilon,1]$, where $S(A|B)$ is the QCE defined in (\ref{ce-ext}), $F_{D\{c_k\}}$ is the function defined in (\ref{F-def}) with $H=D\{c_k\}\doteq\sum_{k=0}^{+\infty}c_k|k\rangle\langle k|$ and $g(x)$ is the function defined in (\ref{g-def}).}
\end{proposition}\smallskip

The function $F_{D\{c_k\}}$ mentioned above is defined by the expressions
$$
F_{D\{c_k\}}(E)=\min_{\beta\in(0,+\infty)}\left\{\beta E+\ln\sum_{k=0}^{+\infty}e^{-\beta c_k}\right\}=\beta(E)E+\ln\sum_{k=0}^{+\infty}e^{-\beta(E) c_k},
$$
where $\beta(E)$ is a unique solution of the equation $\sum_{k=0}^{+\infty}c_ke^{-\beta c_k}=E\sum_{k=0}^{+\infty}e^{-\beta c_k}$ \cite{W,EC}.
For example, if $\{c_0,c_1,c_2,...\}=\{0,1,2,...\}$ (the spectrum of the number operator $N$ in
one-mode quantum oscillator) then $F_{D\{c_k\}}(E)=F_{N}(E)=(E+1)h_2(1/(E+1))$.

By Proposition 1 in \cite{EC} condition (\ref{C-cond}) implies that $F_{D\{c_k\}}(E)=o(E)$ as $E\to+\infty$. This guarantees the faithfulness
of the continuity bound given by Proposition \ref{Winter-CB++}.

\bigskip

\textbf{Acknowledgements.}
I am grateful to A.S.Holevo, G.G.Amosov, S.N.Philippov and other participants of the seminar
"Quantum probability, statistic, information" (the Steklov
Mathematical Institute) for useful discussion. I am very grateful to A.Winter who stimulated my interest in the problem of quantitative continuity analysis of characteristics of quantum systems and channels and helped me a lot in understanding many complex issues in this field.
I am also grateful to N.Datta, M.Wilde and L.Lami for valuable communication and to A.T.Rezakhani for the relevant reference. Special thanks to A.Winter and L.Lami
for the help to show the mismatch of the classes $\Upsilon_{\!AB}$ and $\Upsilon^{*}_{\!AB}$ introduced in Section 5.2.3.

This work was funded by Russian Federation represented by the Ministry of Science
and Higher Education (grant number 075-15-2020-788).


\end{document}